\documentclass[final,1p,times]{elsarticle}
\usepackage[table]{xcolor} 
\usepackage{longtable}
\usepackage{booktabs} 
\usepackage{tabularx} 

\usepackage{array}
\usepackage{listings} 
\usepackage{upquote}
\usepackage{amsmath}
\usepackage{comment}
\usepackage[colorlinks=true, linkcolor=blue, citecolor=blue, urlcolor=blue]{hyperref}

\definecolor{codegray}{gray}{0.9}
\definecolor{codegreen}{rgb}{0,0.6,0}
\definecolor{codepurple}{rgb}{0.58,0,0.82}
\definecolor{backcolour}{rgb}{0.98,0.98,0.98}
\lstdefinestyle{mystyle}{
    language=TeX,
    basicstyle=\small\ttfamily,
    breaklines=true,
    keywordstyle=\bfseries,
    morekeywords={AND, OR},
}
\lstnewenvironment{query}[1][]
    {\lstset{style=mystyle,#1}}
    {}

\usepackage[utf8]{inputenc} 
\usepackage[T1]{fontenc}    
\usepackage{hyperref}       
\usepackage{url}            
\usepackage{booktabs}       
\usepackage{amsfonts}       
\usepackage{nicefrac}       
\usepackage{microtype}      
\usepackage{lipsum}		
\usepackage{multicol}
\usepackage{graphicx}
\usepackage{doi}
\usepackage{dirtytalk}
\usepackage{amsmath}
\usepackage{xcolor}
\usepackage{fancyvrb}
\usepackage{chronosys}
\usepackage{caption}
\usepackage{pdflscape}  

\begin{document}
\begin{frontmatter}
\title {A Comprehensive Review of Recommender Systems: Transitioning from Theory to Practice}





\makeatletter
\def\@fnsymbol#1{\ensuremath{\ifcase#1\or *\or \dagger\or \ddagger\or
  \mathsection\or \mathparagraph\or \|\else\@ctrerr\fi}}
\makeatother

\author[inst1]{Shaina Raza\corref{cor1}\fnref{equal}}
\ead{shaina.raza@vectorinstitute.ai}

\author[inst2]{Mizanur Rahman\corref{cor1}\fnref{equal}}
\ead{mizanur.york@gmail.com}

\author[inst1]{Safiullah Kamawal\fnref{equal}}
\ead{safiullah.kamawal@vectorinstitute.ai}

\author[inst1]{Armin Toroghi}
\ead{armin.toroghi@vectorinstitute.ai}

\author[inst1]{Ananya Raval}
\ead{ananya.raval@vectorinstitute.ai}

\author[inst2]{Farshad Navah}
\ead{farshad.navah@gmail.com}

\author[inst1]{Amirmohammad Kazemeini}
\ead{amirmohammad.kazemeini@vectorinstitute.ai}

\affiliation[inst1]{organization={Vector Institute},
            addressline={},
            city={Toronto},
            state={ON},
            postcode={},
            country={Canada}}

\affiliation[inst2]{organization={Independent Researcher},
            addressline={},
            city={Toronto},
            state={ON},
            postcode={},
            country={Canada}}

\cortext[cor1]{Corresponding author.}
\fntext[equal]{Equal contribution.}
\fntext[Disclaimer]{The ideas and discussions produced in this work are strictly those of the authors and do not represent the points of view of the institutions the authors belong to.}


\begin{abstract}
Recommender Systems (RS) play an integral role in enhancing user experiences by providing personalized item suggestions. This survey reviews the progress in RS inclusively from 2017 to 2024, effectively connecting theoretical advances with practical applications. We explore the development from traditional RS techniques like content-based and collaborative filtering to advanced methods involving deep learning, graph-based models, reinforcement learning, and large language models. We also discuss specialized systems such as context-aware, review-based, and fairness-aware RS. The primary goal of this survey is to bridge theory with practice. It addresses challenges across various sectors, including e-commerce, healthcare, and finance, emphasizing the need for scalable, real-time, and trustworthy solutions. Through this survey, we promote stronger partnerships between academic research and industry practices. The insights offered by this survey aim to guide industry professionals in optimizing RS deployment and to inspire future research directions, especially in addressing emerging technological and societal trends. The survey resources are available in the public GitHub repository \url{https://github.com/VectorInstitute/Recommender-Systems-Survey}  .

\end{abstract}


\begin{keyword}
Recommender Systems \sep Graph-based Recommender Systems \sep Knowledge-based Systems \sep  Multimodal Recommender Systems
\sep  Large Language Models  \sep  Personalization
\sep  Industry Applications \sep  Explainable AI \sep Transparency
\sep Fairness \sep  Deep Learning \sep  Survey\sffamily
\end{keyword}

\end{frontmatter}
\begin{multicols}{2}   
{\scriptsize
\tableofcontents
}
\end{multicols}

\section{Introduction}
Recommender Systems (RS) are a type of information filtering system designed to predict and suggest items or content—such as products, movies, music, or articles—that a user might be interested in. These predictions are based on the user's past behavior, preferences, or the behavior of similar users \cite{aggarwal2016recommender}. 
The main goal of any RS is to enhance user experience, increase engagement, and facilitate decision-making processes \cite{berkovsky2008mediation}. This is applicable across various domains, including e-commerce, entertainment, and social media. RS hold significant roles in both theoretical research (academics) and practical applications (industry).

The importance of RS has grown exponentially with the advent of big data and advancements in artificial intelligence \cite{adomavicius2005toward,ricci2021recommender}. As users interact with digital platforms, they generate vast amounts of data that can be leveraged to make precise and personalized recommendations. This ability to tailor suggestions not only improves user satisfaction but also increases the likelihood of users discovering new and relevant content \cite{amatriain2016past}. In e-commerce, for example, RS can drive significant sales by suggesting products that align with users' preferences, while in entertainment, they enhance user engagement by recommending shows or music that match users' tastes. Additionally, RS are now being integrated into new and emerging fields such as personalized education \cite{khanal2020systematic}, where they help tailor learning experiences to individual student needs, and healthcare \cite{etemadi2023systematic}, where they assist in suggesting personalized treatment plans and health interventions. The development of large language models (LLMs) further enhances RS by enabling them to understand and process vast amounts of text data,  leading to more sophisticated and context-aware recommendations \cite{fan2023recommender}.

In academia, RS are the subject of extensive research aimed at understanding user behavior and decision-making processes \cite{adomavicius2005toward}. This research utilizes sophisticated data analytics and Artificial intelligence (AI) techniques. The ACM Recommender Systems Conference (RecSys) \cite{ACMRecSys}, along with related scholarly journals and venues,  highlights emerging technologies and their potential impact across various sectors, including entertainment, e-learning, and academic publishing. Recent academic advancements have focused on integrating reinforcement learning and LLMs into RS, leading to more accurate and dynamic recommendation capabilities.

In the industry, RS enhance customer satisfaction and drive revenue growth by providing tailored suggestions \cite{amatriain2016past}. Major corporations such as Amazon, Netflix, and Spotify integrate RS into their operations, significantly contributing to their business models. For instance, Amazon reports that 35\% of its revenue comes from its RS \cite{mackenzie2013retailers}, while Netflix attributes revenues of approximately \$33.7 billion and its success in customer retention significantly to its RS \cite{businessofapps_netflixstats}. The global market for recommendation engines, as per Precision Reports \cite{PrecisionReports}, is forecasted to witness substantial growth from 2023 to 2030, highlighting their increasing importance in business strategies. Privacy-preserving algorithms and bias mitigation are also becoming key areas of focus for industry practitioners.

This survey focuses on the theory of RS and their transition to practical applications, aiming to bridge the gap between academic research and industry practices. It highlights how theoretical advancements can be effectively implemented in real-world scenarios.

\textbf{Necessity of this Survey}
Previous surveys often concentrate solely on the theoretical aspects of RS, exploring methods and algorithmic foundations to improve prediction accuracy and personalization \cite{burke2011recommender,ricci2021recommender}. Conversely, practical-focused research or applications typically views RS as essential tools for enhancing user engagement, retention, and business growth \cite{amatriain2015recommender,melville2010recommender}. There is a need for the collaboration between academia and industry to address both technical challenges and real-world demands, which in turn enhances user satisfaction and business value. This interdependence highlights the growing importance of such partnerships.


\textbf{Difference with Existing Surveys}
Unlike previous surveys that often focus solely on the theoretical or practical aspects, our survey uniquely covers the integration of theoretical advancements with practical applications, offering a comprehensive overview that addresses both academic and industry perspectives. Furthermore, we identify emerging trends and future research directions, such as the integration of explainable AI in RS to ensure transparency and user trust.

{\scriptsize
\begin{longtable}{|c|p{2.5cm}|p{0.8cm}|p{0.8cm}|c|p{2.5cm}|p{0.8cm}|p{0.8cm}|}
\caption{Overview of Related Surveys Ordered by Date of Publication and Comparison Criteria}
\label{tab:comparison} \\
\hline
\textbf{Survey} & \textbf{Topic} & \textbf{Theory} & \textbf{Practise} & \textbf{Survey} & \textbf{Topic} & \textbf{Theory} & \textbf{Practise} \\ 
\hline
\endfirsthead

\hline
\multicolumn{8}{|c|}{\textbf{Table \thetable: Continuation of Survey List}} \\
\hline
\textbf{Survey} & \textbf{Topic} & \textbf{Theory} & \textbf{Practise} & \textbf{Survey} & \textbf{Topic} & \textbf{Theory} & \textbf{Practise} \\ 
\hline
\endhead

\hline
\multicolumn{8}{|r|}{{Continued on next page}} \\ 
\hline
\endfoot

\hline
\endlastfoot

\cite{DEBIASIO2024101352}  & Economics & \text{\sffamily X} & \checkmark & \cite{THAKKAR2021114800} & Stock market & \text{\sffamily X} & \checkmark \\ \hline
\cite{behera2020personalized} & Digital marketing & \checkmark & \checkmark & \cite{sharaf2022survey} & Finance & \checkmark & \text{\sffamily X} \\ \hline
\cite{deldjoo2020recommender} & Multimedia content & \checkmark & \text{\sffamily X} & \cite{chaudhari2020comprehensive} & Travel & \text{\sffamily X} & \checkmark \\ \hline
\cite{de2021health} & Health & \checkmark & \checkmark & \cite{etemadi2023systematic} & Health & \checkmark & \text{\sffamily X} \\ \hline
\cite{ali2023deep} & Health & \checkmark & \text{\sffamily X} & \cite{saha2020review} & Health & \checkmark & \text{\sffamily X} \\ \hline
\cite{cheung2019recommender} & Health & \text{\sffamily X} & \checkmark & \cite{sun2023development} & Health & \checkmark & \text{\sffamily X} \\ \hline
\cite{Tarus2017KnowledgebasedRA} & E-learning & \checkmark & \text{\sffamily X} & \cite{khanal2020systematic} & E-learning & \checkmark & \checkmark \\ \hline
\cite{klavsnja2015recommender} & E-Learning & \text{\sffamily X} & \checkmark & \cite{portugal2018use} & Machine learning & \checkmark & \text{\sffamily X} \\ \hline
\cite{guo2020survey} & Knowledge integration & \checkmark & \text{\sffamily X} & \cite{zhang2020explainable} & Explainability & \checkmark & \text{\sffamily X} \\ \hline
\cite{kulkarni2020context} & Context awareness & \checkmark & \text{\sffamily X} & \cite{raza2019progress} & Context awareness & \checkmark & \text{\sffamily X} \\ \hline
\cite{wei2017collaborative} & Collaborative filtering & \checkmark & \text{\sffamily X} & \cite{chen2018survey} & Collaborative filtering & \checkmark & \text{\sffamily X} \\ \hline
\cite{ccano2017hybrid} & Hybrid methods & \checkmark & \text{\sffamily X} & \cite{quadrana2018sequence} & Sequence awareness & \checkmark & \text{\sffamily X} \\ \hline
\cite{wang2021survey} & Session integration & \checkmark & \text{\sffamily X} & \cite{ludewig2021empirical} & Session integration & \checkmark & \text{\sffamily X} \\ \hline
\cite{jannach2021survey} & Conversation integration & \checkmark & \text{\sffamily X} & \cite{afchar2022explainability} & Music & \checkmark & \checkmark \\ \hline
\cite{schedl2018current} & Music & \checkmark & \text{\sffamily X} & \cite{afsar2022reinforcement} & Reinforcement learning & \checkmark & \text{\sffamily X} \\ \hline
\cite{deldjoo2021survey} & Adversarial methods & \checkmark & \text{\sffamily X} & \cite{srifi2020recommender} & Review texts & \checkmark & \text{\sffamily X} \\ \hline
\cite{wu2022graphneural} & Graph neural network & \checkmark & \text{\sffamily X} & \cite{gao2023survey} & Graph Neural network & \checkmark & \text{\sffamily X} \\ \hline
\cite{wu2021comprehensive} & Graph Neural network & \checkmark & \text{\sffamily X} & \cite{batmaz2019review} & Deep learning & \checkmark & \text{\sffamily X} \\ \hline
\cite{fan2023recommender} & Large Language Models & \checkmark & \text{\sffamily X} & \cite{li2023large} & Large Language Models & \checkmark & \text{\sffamily X} \\ \hline
\cite{wu2023survey} & Large Language Models & \checkmark & \text{\sffamily X} & \cite{wang2023pretrained} & Large Language Models & \checkmark & \text{\sffamily X} \\ \hline
\cite{dong2022survey} & Large Language Models & \checkmark & \text{\sffamily X} & \cite{wu2023graph} & Large Language Models & \checkmark & \text{\sffamily X} \\ \hline
\cite{liu2023pre} & Large Language Models & \checkmark & \text{\sffamily X} & \cite{do2019deep} & Aspect integration & \checkmark & \text{\sffamily X} \\ \hline
\cite{roy2022systematic} & General & \checkmark & \text{\sffamily X} & \cite{ricci2021recommender} & General & \checkmark & \text{\sffamily X} \\ \hline
\cite{karimi2018news} & News & \text{\sffamily X} & \checkmark & \cite{raza2022news} & News & \checkmark & \text{\sffamily X} \\ \hline
\cite{wu2023personalized} & News & \checkmark & \text{\sffamily X} & \cite{Himeur2022LatestPerspectives} & Privacy & \checkmark & \text{\sffamily X} \\ \hline
\cite{sarkar2023tourism} & Tourism & \checkmark & \text{\sffamily X} & \cite{evaluation-survey} & Evaluation & \checkmark & \text{\sffamily X} \\ \hline
\cite{adomavicius2005toward} & General & \checkmark & \text{\sffamily X} & \cite{ali2020graph} & General & \checkmark & \text{\sffamily X} \\ \hline
\cite{Jha2023} & Trustworthiness & \checkmark & \text{\sffamily X} & \cite{Casillo2023} & Cultural Heritage & \text{\sffamily X} & \checkmark \\ \hline
\multicolumn{8}{|c|}{\textbf{Difference:} Our survey covers the theory of RS and the application of its methods in practice.}\\ \hline
\end{longtable}
}

\textbf{Main Contributions}
\begin{enumerate}
    \item This survey provides a comprehensive review of RS, tracing their development from theoretical foundations to practical applications between 2017 and 2023. It is the first survey to specifically highlight the translation of theoretical advancements into practical solutions for industry challenges.
\item Each type of RS is thoroughly examined, including data input methods, associated challenges, relevant datasets, evaluation metrics, model accuracy, and practical applications, as presented in tables. The survey aims to offer industry professionals a set of guidelines to facilitate the deployment of these systems in real-world settings.
\item We discuss the specific challenges faced by RS in various sectors, such as e-commerce, healthcare, finance, and others. The survey emphasizes the need for scalable, real-time, and privacy-focused solutions, demonstrating how theoretical insights can address these industry-specific demands.
\end{enumerate}

\section{Background}

Recommender systems (RS) are algorithms designed to suggest items—such as books, movies, products, or content—to users based on their preferences. The primary goal of RS is to enhance user experience by personalizing content \cite{adomavicius2005toward}. At its core, an RS combines user and item profiles with a filtering mechanism to align user preferences with suitable items \cite{burke2011recommender}. User profiles gather data such as demographics and browsing history, while item profiles detail features like genres. Both explicit feedback (e.g. ratings) and implicit feedback (e.g. browsing actions) refine these recommendations.

\subsection{Historical Context} 
One of the pioneering efforts for RS is from Elen Rich in 1979 \cite{rick79} to suggest books based on user preferences categorized into ``stereotypes". Following this, Jussi Karlgren conceptualized the ``digital bookshelf" in 1990 \cite{karlgen90}, an idea later expanded by researchers at SICS, MIT, and Bellcore, with notable contributions from Pattie Maes, Will Hill, and Paul Resnick, whose GroupLens project \cite{grouplens} received the 2010 ACM Software Systems Award. Later, Adomavicius \cite{adomavicius2005toward}, Herlocker \cite{herlocker1999algorithmic}, and Beel \cite{beel} provided foundational theory on RS. 

Traditional RS methods can be categorized into collaborative filtering, content-based filtering, and hybrid approaches, aiming to improve user experience \cite{burke2011recommender}. Collaborative filtering (CF) \cite{herlocker1999algorithmic} is based on the idea that users with similar preferences will likely have similar tastes in the future. CF recommends items by finding a neighborhood of similar users or items. CF can recommend items without needing much content analysis, however, it normally faces challenges like cold starts, scalability, and sparsity \cite{wei2017collaborative} . Content-based filtering (CBF) \cite{xia2017content} recommends items based on a user past preferences and item characteristics, using techniques like Term Frequency - Inverse Document Frequency (TF-IDF), cosine similarity, and neural networks for item representation. However, it may struggle with recommending new or unseen items. Hybrid RS \cite{ccano2017hybrid} combine the strengths of both approaches, offering more accurate and personalized recommendations by integrating diverse methodologies. 

The Netflix Prize \cite{netflix}, a competition aimed at enhancing RS algorithms, significantly popularized these algorithms. While the competition focused on improving accuracy, an essential aspect of algorithmic effectiveness, it also emphasized the importance of diversity, privacy, and serendipity in boosting user satisfaction \cite{evaluation-survey}.

Machine Learning (ML) methods such as k-nearest neighbors algorithm (k-NN), deep neural networks, and Natural Language Processing (NLP) have enhanced RS over the years by providing more precise recommendations. However, key challenges and ethical issues, such as safeguarding user and data privacy, mitigating biases for fair recommendations, transparency for user trust, and keeping pace with technological advancements remain the challanges.

\subsection{Current State of Practice and Theory in Recommender Systems}
Academia focuses on the theory, methods, and algorithms in RS, while the industry emphasizes practical applications, scalability, and direct business impacts. This section explores the distinct challenges faced by these two sectors.
\paragraph*{Theoretical Research on Recommender Systems}
Theoretical research on RS is commonly initiated by academics through the development of new algorithms, models, and evaluation metrics. However, academic researchers face challenges in accessing diverse and comprehensive datasets due to privacy concerns, proprietary restrictions, and financial barriers. Additionally, data quality issues such as biases, inaccuracies, and outdated information limit the development and testing of RS in varied contexts.

The drive for high accuracy in research models can often lead to overfitting, which makes them unusable for real-world applications. Such a focus may neglect crucial aspects like diversity, novelty, and user satisfaction. Additionally, solutions from academia are frequently not easily adaptable in industry settings due to their reliance on data-intensive algorithms, complexity, and a disconnect in keeping developers updated.

\paragraph*{Practices in Recommender Systems}
The industry faces several challenges in deploying RS, particularly concerning scalability as user bases and catalog sizes expand. Adapting to constantly evolving user preferences and content availability presents ongoing difficulties. Ensuring the diversity and fairness of recommendations is crucial to avoid biases. Additionally, integrating real-time data and maintaining high performance under heavy loads are significant challenges. Balancing personalization with privacy concerns requires careful handling of user data to build trust and comply with regulations.

\textbf{Common Challanges} Both theory and practice emphasize the importance of high-quality (accurate, relevant, reliable, and representative of the intended use case or application) datasets for building RS. Academic research often relies on high-quality data for benchmarking purposes, while the industry frequently requires such data to enhance user experience and system effectiveness.

Theoretically, RS algorithms are quite advanced now, featuring layers of deep neural networks and the latest language model complexities. In practice, however, these models are not immediately applicable to real-world use cases. Industry sectors, generally running a set of standard models, require significant adaptations to implement these advanced algorithms effectively

In this survey, we examine the theoretical and practical aspects of RS, with the goal to facilitate a smooth transition from research to real-world application.

\section{Literature Review Methodology  }

To compile a comprehensive and relevant list of papers for our review, we conducted a systematic literature review, adhering to established methodology principles \cite{carrera2022conduct}. Our search query and extraction methods are detailed below.

\paragraph{Research Questions}
\begin{enumerate}
    \item How have RS algorithms evolved theoretically over the years?
    \item What strategies can be utilized to apply theoretical advancements in RS to practical applications?
\end{enumerate}

\paragraph{Databases Searched}
For selecting studies, we gathered articles published inclusively from January 2017 through April 2024. This timeframe was chosen because much of the evolution in RS is linked to deep learning algorithms, also
highlighted in the RecSys workshop in 2017 \cite{recsys2017dlrs}. We conducted our literature search across multiple academic databases and digital libraries renowned for their extensive collections of RS literature, including IEEE Xplore, ACM Digital Library, PubMed, ScienceDirect, JMLR, and Wiley.
To refine the search results, we applied specific inclusion and exclusion criteria based on the publication year, relevance to RS, the source, and the paper's focus on the technological, theoretical, and application aspects of RS. Only peer-reviewed journal articles, conference papers, and significant arXiv papers were considered. 

\paragraph{Search Query}
Our search strategy aimed to find literature across various aspects of RS, including types, algorithms, evaluation, application, user interaction, and data quality. The search query used was:

("recommender systems" OR "recommendation systems" OR "RS" OR "RecSys") AND
("content-based filtering" OR "collaborative filtering" OR "hybrid recommender systems" OR 
"context-aware recommender systems" OR "knowledge-based systems" OR "social recommender systems") AND
("matrix factorization" OR "deep learning" OR "convolutional neural networks" OR 
"recurrent neural networks" OR "reinforcement learning" OR "autoencoders" OR 
"neural collaborative filtering" OR "graph neural networks") AND
("precision" OR "recall" OR "F1 score" OR "RMSE" OR "MAE" OR "hit rate" OR 
"novelty" OR "diversity" OR "serendipity" OR "user satisfaction") AND
("e-commerce" OR "media streaming" OR "social media" OR "education" OR "healthcare" OR 
"tourism" OR "personalized news" OR "job recommenders") AND
("user interface" OR "user experience" OR "usability" OR "interaction design" OR 
"user engagement" OR "user feedback" OR "user profiling") AND
("explicit feedback" OR "implicit feedback" OR "data sparsity" OR "cold start problem" OR 
"data quality" OR "user-generated content") AND
("privacy" OR "data security" OR "ethical algorithms" OR "bias and fairness" OR 
"transparency" OR "recommendation explainability") AND
("tech industry" OR "startup case studies" OR "market analysis" OR "business models" OR 
"return on investment" OR "user retention") AND
("transformer models" OR "BERT" OR "GPT" OR "natural language understanding" OR 
"language generation" OR "sentiment analysis" OR "text embeddings") AND
("user personalization" OR "adaptive systems" OR "customization techniques" OR 
"user-adaptive content" OR "dynamic personalization")

\begin{table}[h]
\small
\centering
\caption{Inclusion and Exclusion Criteria for Literature Review}
\begin{tabular}{p{0.45\linewidth}p{0.45\linewidth}}
\toprule
\textbf{Inclusion Criteria} & \textbf{Exclusion Criteria} \\
\midrule
Articles that include at least 3-4 keywords from our search query in the title, abstract, or keywords.& Articles without relevant keywords. \\
Articles published from 2017 through January 2024. & Articles published outside this timeframe, except classical papers that need to be cited.\\
Articles that pass the initial screening based on titles and abstracts and address RQ1 or RQ2.& Articles that do not address RQ1 or RQ2.\\
Articles from published work or arXiv if it covers an important and relevant topic.& Grey literature, e.g., technical reports, or dissertations. \\
Articles in English language. & Articles not written in English. \\
\bottomrule
\end{tabular}
\label{table:criteria}
\end{table}

\begin{figure}[ht]
    \centering
    \includegraphics[width=1\linewidth, height=0.6\textheight]{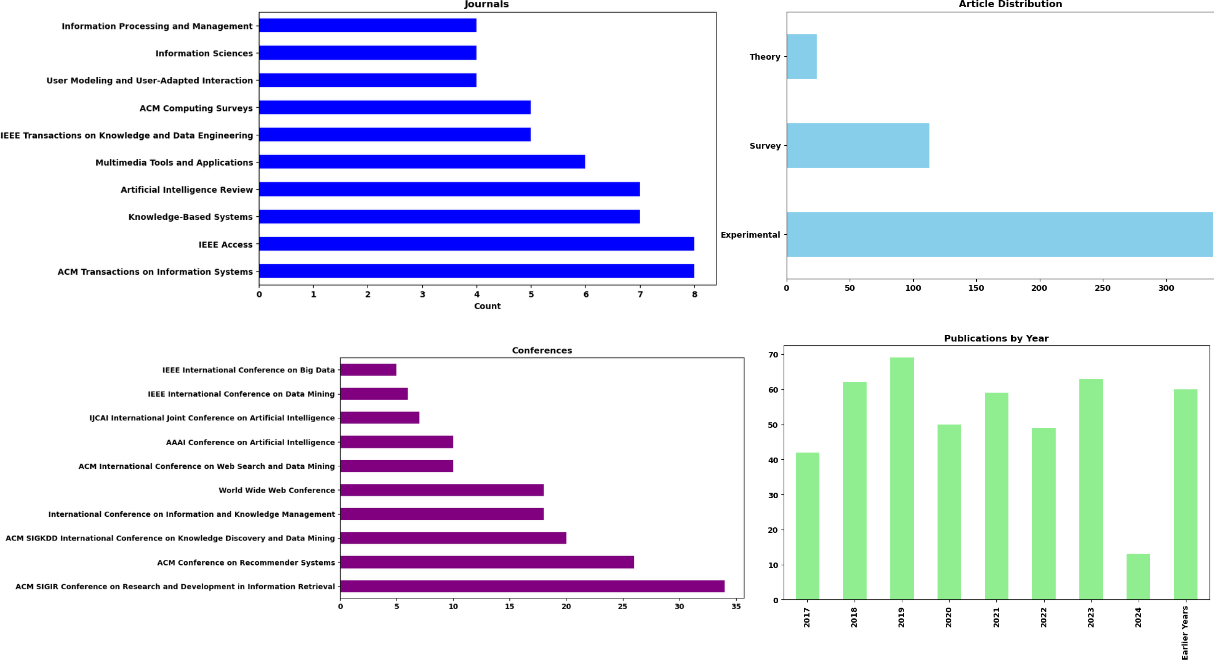}
    \caption{Overview of Publication Trends and Key Venues in RS Research.}
    \label{fig:eda}
\end{figure}

\paragraph{Quality Assessment}
We screened the articles using the inclusion and exclusion criteria detailed above in Table \ref{table:criteria}. If there was any uncertainty, the paper was briefly reviewed and then either included or excluded based on consensus from the two first co-authors. Selected papers underwent a thorough reading and were subject to a quality assessment involving a set of questions:

 A paper was considered for inclusion in our review if it received a ``yes" or ``partially" response to any of these questions:
\begin{enumerate}
    \item Does the article present a novel RS or methodology?
    \item Is the article related to a RS in an academic setting?
    \item Is the article related to a RS in an industry setting?
    \item Does the article propose a framework, tool or methodology?
    \item Has the research provided a concise statement or definition outlining its aims, goals, purposes, problems, motivations, objectives, and questions?

\end{enumerate}
The results of the literature search were finalized and categorized by database, as shown inTable~\ref{tab:num_papers}.
\begin{table}[ht]
    \centering
    \caption{Identified Papers by Database}
    \begin{tabular}{|l|c|}
        \hline
        \textbf{Publisher/Journal/Conference} & \textbf{Number of Papers} \\
        \hline
        \textbf{ACM} & 83\\
     
        \hline
        \textbf{IEEE} & 43\\
      
        \hline
        \textbf{Springer} & 32\\
      
        \hline
        \textbf{ScienceDirect}& 25\\
      
        \hline
        \textbf{arXiv} & 15\\
         \hline
        \textbf{Others} & 89\\
        \hline
    \end{tabular}
    
    \label{tab:num_papers}
\end{table}
Figure \ref{fig:eda} presents a comprehensive analysis of publications reviewed in this survey. 

\section{Challenges in Recommender Systems}

RS play a vital role in personalizing user experiences and driving business value across various domains. Despite their widespread adoption, several challenges persist in their deployment and maintenance.

\paragraph{E-commerce}
E-commerce platforms face the challenge of personalizing the shopping experience by recommending products in real-time, managing vast data, and adapting to changing consumer preferences \cite{hussien2021recommendation}. Personalization must consider factors such as time, season, location, and the user's current situation. For example, recommendations for a new parent shopping for baby products will differ significantly from those for a book lover. Introducing diversity and novelty in recommendations is crucial to keep the experience fresh and engaging.
\paragraph{Entertainment} In the entertainment industry, the challenge lies in tracking users' preferences across genres while introducing them to new content to maintain engagement. Balancing personalization and novelty is essential. Music recommendations require frequent updates due to the shorter shelf life of songs compared to movies \cite{Chen2021LearningRecommendation}. Conversely, movie recommendations are less dependent on frequent updates, as films typically have a longer shelf life. However, effective movie RS should still balance between promoting new releases and maintaining a selection of enduring favorites to satisfy a wide range of user preferences \cite{Penha2020WhatRecommendation}.

\paragraph{News}
 The news industry must deliver personalized content promptly without overwhelming users. News preferences are highly dynamic, necessitating recommendations that adapt to rapid changes in interests and current events \cite{raza2022news}. It is important to offer diverse viewpoints to prevent echo chambers, combat misinformation \cite{raza2021automatic}, and maintain user trust.

\paragraph{Tourism} Personalized booking recommendations in tourism must account for user preferences regarding destinations, travel dates, budgets, and accommodations \cite{hamid2021smart}. Integrating factors such as past travel history, seasonal trends, and real-time availability is essential. Balancing immediate needs, like dining recommendations during travel, with advance bookings for stays and major attractions enhances the overall user experience.

\paragraph{Healthcare}Healthcare RS face issues like data privacy, security, and patient consent under regulations such as 
Health Insurance Portability and Accountability Act (HIPAA). Providers, patients, and administrators all require access to relevant information tailored to their roles, necessitating role-based solutions and robust processing capabilities to handle large volumes of heterogeneous data effectively \cite{tran2020Recommender}.

\paragraph{Finance}  Financial RS need to navigate data privacy, security, and compliance with regulations like General Data Protection Regulation (GDPR) and Payment Card Industry Data Security Standard (PCI-DSS). Challenges include managing data quality, integrating diverse data sources, and providing personalized financial advice \cite{sharaf2022survey}. Ensuring the interpretability and transparency of recommendations is crucial for building trust and user confidence. Fairness of the recommendations is of utmost importance.

\paragraph{E-learning} E-learning RS face the task of addressing varied user needs, overcoming the cold start problem with new users, and handling data sparsity. Adapting to dynamic content and user behavior, ensuring contextual relevance, scalability, and employing suitable evaluation metrics to assess educational impact are fundamental challenges \cite{rahayu2022systematic}.

\paragraph{Discussion} RS across various sectors face a set of common challenges despite their industry-specific characteristics. Generally, these systems struggle with balancing personalization and user privacy, managing data scalability, and ensuring the diversity and novelty of recommendations to keep users engaged. They must also address the cold-start problem, where insufficient user data can hinder the system's ability to make accurate recommendations. Additionally, dynamic user preferences require systems to continually adapt and learn from new data, posing challenges in real-time processing and algorithmic efficiency. Lastly, ensuring fairness and avoiding bias in recommendations is crucial, as these systems often influence user decisions and can perpetuate existing disparities if not carefully managed.

In the following sections, we explore the evolution of RS and how they address these challenges.
\section{Foundational Recommender Systems}
A RS can be mathematically represented as a function \( f \) that predicts the utility of an item \( i \) for a user \( u \), denoted as \( \hat{r}_{ui} \), which estimates how much user \( u \) would prefer item \( i \).\cite{adomavicius2005toward} . This function is typically learned from historical data:
\begin{equation}
\hat{r}_{ui} = f(u, i; \Theta)
\end{equation}
where \( \Theta \) represents the parameters of the model, learned from the data. In the context of RS, the term \( \hat{r}_{ui} \)  represents a prediction of the rating or utility that a user would assign to an item 
This prediction is used for recommending items that are likely to be of interest to the user.
\begin{figure}[ht]
    \centering
    \includegraphics[width=0.75\linewidth]{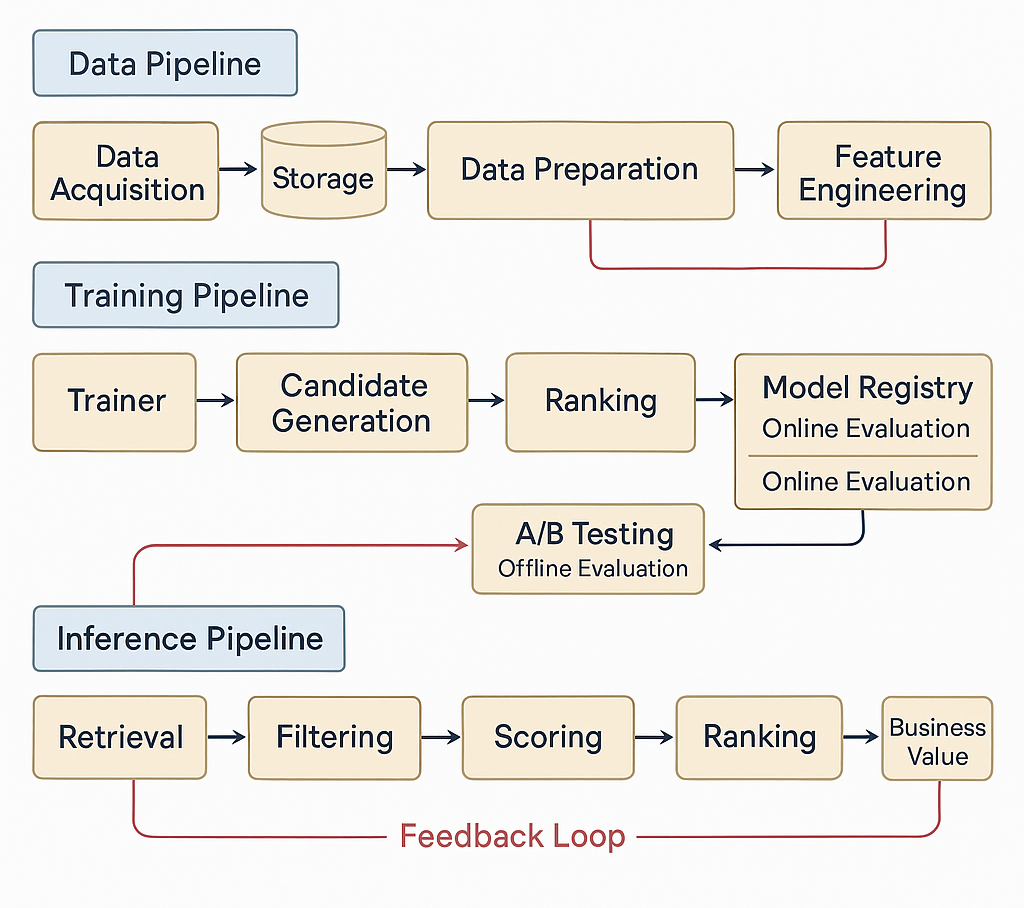}
    \caption{General Framework of RS showing an outline of a data processing pipeline. It covers stages from data acquisition, preparation, and feature engineering, through model training and validation, to deployment and monitoring. The pipeline also includes steps for candidate generation, ranking, evaluation, and a feedback loop to ensure continuous improvement and business value.}
    \label{fig:main}
\end{figure}
A general framework of RS is illustrated in Figure \ref{fig:main}. The lifecycle of a data-driven model within an RS starts with data acquisition, followed by storage and preparation. This leads to feature engineering, forming the basis of the data pipeline. The data pipeline feeds into the training pipeline, which includes model training and validation. Following training, the process involves candidate generation and ranking. This process is complemented by A/B testing, offline and/or online evaluation. The final stages include deployment and monitoring.

Below, we present an overview of foundational RS.

\subsection{Foundational Recommender Systems}
Foundational RS refers to the early models and techniques that established the core principles and methodologies in the field of recommendation engines. These systems primarily include collaborative filtering, content-based filtering, and hybrid approaches. 

\textbf{Content-Based Filtering} Content-based filtering (CBF) is a recommendation strategy that suggests items to a user based on the attributes of the items and a profile of the user's preferences, typically utilizing similarity measures to match user preferences with item attributes \cite{Pazzani2007Content-basedSystems}.  In CBF, the recommendation \( \hat{r}_{ui} \) is based on the features of items \( \phi(i) \) and a profile of the user's preferences \( \theta(u) \):
\begin{equation}
\hat{r}_{ui} = \theta(u)^\top \phi(i)
\end{equation}
\begin{itemize}
    \item \( \phi(i) \) represents the feature vector of the item.
    \item \( \theta(u) \) represents the user preference vector.
\end{itemize}

The evolution of CBF recommendation models starts from traditional methods \cite{Pazzani2007Content-basedSystems} like vector-space models, probabilistic models, and decision trees, relying on manual feature engineering and similarity calculations. Vector-space models compute item similarity through cosine similarity \cite{turney2010frequency}, probabilistic models estimate the likelihood of user preference with statistical analysis \cite{prob-cbf}, and decision trees recommend items by categorizing them based on attributes \cite{Pazzani2007Content-basedSystems}. These models laid the groundwork for personalized recommendations by leveraging explicit item features and user preferences. 

Algorithmic advancements in computer science and ML methods led to a shift to sophisticated neural networks that allow automatic feature extraction and learning complex data patterns \cite{zhang2019deep}. 
These neural network-based systems leverage deep learning to analyze user interactions and item features across different modalities, including textual (such as reviews \cite{zheng2017joint}, citations \cite{Bhagavatula2018Content-basedRecommendation}, and news \cite{Joseph2019ContentGraphs}), streaming (like music \cite{Chen2021LearningRecommendation}), and image data \cite{Deng2018LeveragingSystem}. In practice, the tunable choices are the feature-weighting/normalization scheme and the similarity metric (e.g., cosine or Pearson), whereas any similarity cutoff for top-N selection is an operational post-hoc threshold rather than a model hyperparameter~\cite{Lops2011ContentBased,Baltrunas2011ContextMatrix}.
The overall goal of these advancements is to 
generate personalized recommendations by aligning user profiles with item characteristics. 

\textit{Challenges with CBF:} In general, CBF faces challenges such as the cold-start problem, over-specialization (only suggesting items similar to those the user has already seen or liked), computational cost (which can increase quadratically or cubically with the number of users and items), and a lack of updates to user and item profiles.

\textbf{Collaborative Filtering} Collaborative filtering (CF) is a technique used by RS to predict the preferences of a user based on the preferences of similar other users \cite{wei2017collaborative}. 
CF techniques are broadly divided into two main categories: memory-based and model-based methods. Memory-based CF makes recommendations using similarities between users or items directly from user ratings. It has further two types: user-based CF predicts a user ratings based on similar users' historical ratings \cite{herlocker1999algorithmic}, while item-based CF predicts ratings based on similar items \cite{sarwar2001item}. Both methods face challenges such as scalability and sparsity. 
Model-based CF, like matrix factorization \cite{koren2009matrix}  and factorization machines \cite{rendle2012factorization}, uncover latent factors representing user preferences and item characteristics. These methods decompose the user-item interaction matrix into latent feature vectors for users and items. 

CF often starts with constructing a user-item interaction matrix \( R \) with users, items, and \( r_{ui} \) representing known interactions between users and items. One popular approach within CF is matrix factorization \cite{koren2009matrix}, where \( R \) is approximated by the product of two lower-dimensional matrices \( U \) (user features) and \( I \) (item features):

\begin{equation}
\hat{R} = U^\top I
\end{equation}

\begin{itemize}
    \item \( U \) is a \( k \times m \) matrix, with \( k \) being the number of latent factors and \( m \) the number of users.
    \item \( I \) is a \( k \times n \) matrix, with \( n \) being the number of items.
\end{itemize}

Neural extensions in CF have advanced these RS, utilizing deep learning to capture intricate user-item relationships and significantly improve recommendation accuracy. Techniques such as Neural Collaborative Filtering (NCF) \cite{he2017neuralcollab}, Sequence-Aware RS \cite{wang2022sequential}, and Graph Neural Networks (GNNs) \cite{wu2021comprehensive} have emerged as state-of-the-art approaches in CF. For MF-based CF we tune latent dimension, L2 regularization, and learning rate to balance expressiveness and overfitting~\cite{Koren2009MF}, while in neighborhood-based CF the neighborhood size is an operational setting and should be discussed with memory-based methods~\cite{Su2009SurveyCF}.

\textit{Challanges with CF} Like CBF, these methods are more accurate and robust, however they also present challenges such as computational complexity and limited interpretability, which may hinder their scalability and practical applicability in real-world scenarios.

\textbf{Hybrid Approaches} A hybrid RS combines multiple recommendation techniques, such as CBF, CF and other ML models to improve the accuracy and relevance of recommendations provided to users. The most common hybrid techniques include weighted combination, switched selection, feature combination, cascading, and feature augmentation \cite{thorat2015survey}.
 The combination can be represented as a weighted sum:
\begin{equation}
\hat{r}_{ui} = \alpha \cdot f_{CB}(u, i; \Theta_{CB}) + \beta \cdot f_{CF}(u, i; \Theta_{CF})
\end{equation}
\begin{itemize}
    \item \( f_{CB} \) and \( f_{CF} \) represent the content-based and collaborative filtering functions, respectively.
    \item \( \Theta_{CB} \) and \( \Theta_{CF} \) are the parameters for each respective model.
    \item \( \alpha \) and \( \beta \) are weights that balance the contribution of each method.
\end{itemize}
Techniques such as the Wide \& Deep Learning framework \cite{cheng2016wide}, Neural Factorization Machines (NFM) \cite{he2017neuralfact}, DeepFM \cite{guo2017deepfm}, and Deep \& Cross network \cite{wang2017deep} combine explicit feature interactions and implicit feature hierarchies, leveraging both shallow and deep learning models for enhanced recommendations. 

\begin{figure}
    \centering
    \includegraphics[width=0.75\linewidth]{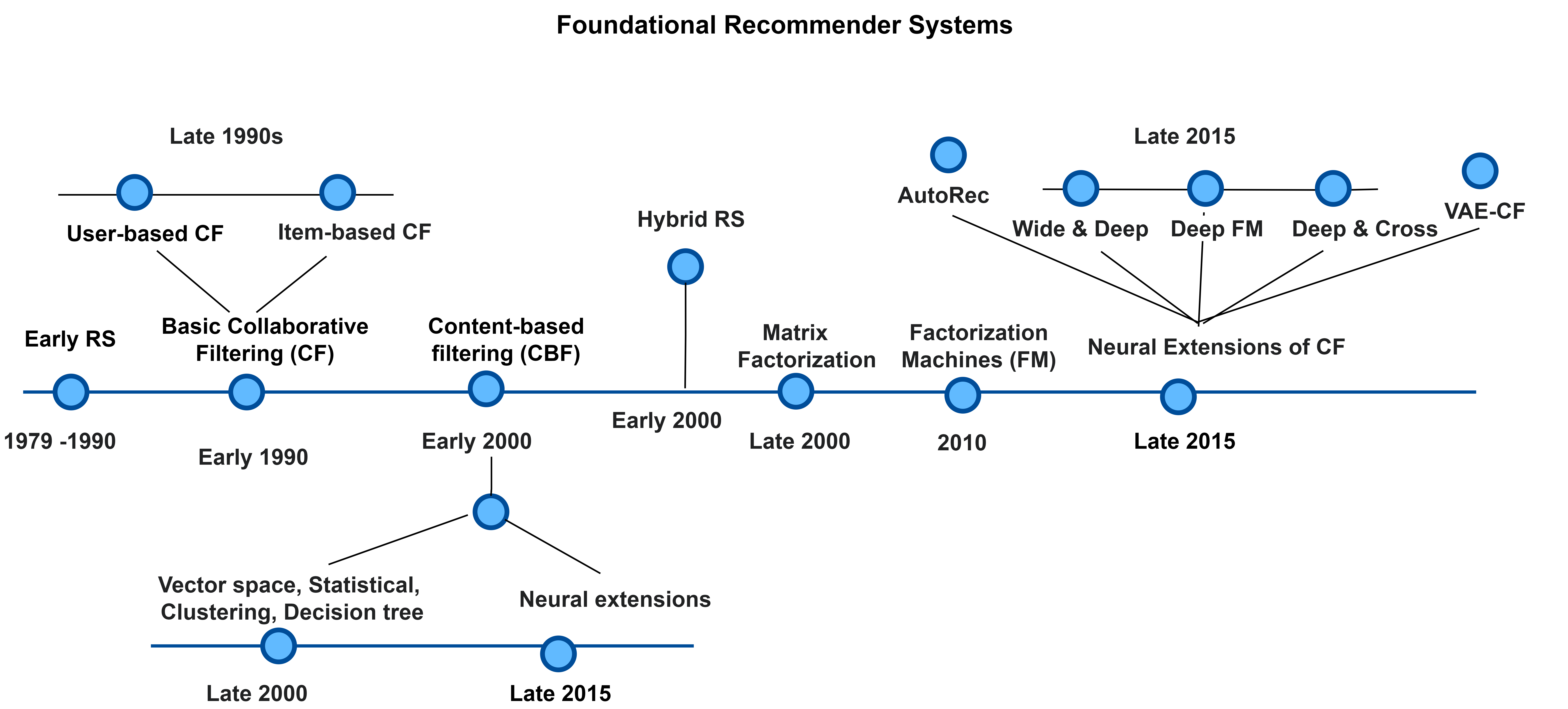}
    \caption{Timeline of Foundational Recommender Systems.}
    \label{fig:founational}
\end{figure}

A timeline illustrating the evolution in foundational RS is given in Figure \ref{fig:founational}.  Prominent publications under foundational RS are given in Table \ref{tab:frs}.
\begin{table}
    \centering
      \caption{Publications on Foundational RS.}
    \begin{tabular}{|c|>{\centering\arraybackslash}p{0.6\linewidth}|} \hline 
         Method&  Publications\\ \hline 
         CF&  \cite{grouplens,haruna2018citation, badriyah2017hybrid, fu2018novel, abdrabbah2023novel, wei2007survey, jha2023appropriate, al2019automating, burke2018balanced, chen2023bias, Raghuwanshi2019Collaborative, yu56, cke,  wang2017deep, xue2017deep, liu2017deepstyle, Xu2019E, e-learning, li2024enhancing, ma2021event, Chen2018GenerativeAU, ying2018graph,  cai2022health, wang2019heterogeneous, ding2021jel, shailaja2018machine, he2017neuralfact, an2019neural, perano2021professional, ma2018rating, cortes2017recommender, afsar2022reinforcement, lyu2021reliable, Yuan2019ARecommendation, lei2019social, geyik2018talent}\\ \hline 
         CBF&  \cite{ perez2021content,jeong2020context, reddy2019content, walek2018content, haruna2018citation,kulkarni2020context, vullam2023multi, khatter2021product, ma2018rating, Pazzani2007Content-basedSystems, Deng2018LeveragingSystem, Fan2021LighterRecommendation, geyik2018talent}\\ \hline 
         Hybrid&  \cite{ccano2017hybrid, burke2007hybrid}\\ \hline
    \end{tabular}

    \label{tab:frs}
\end{table}

\subsection{Can Foundational Recommender Systems address Practical Challenges?}

Foundational RS, which include CF, CBF, and hybrid methods, form the core of many personalized recommendation solutions. While these systems have proven effective across various industries, their ability to address practical challenges in sectors like e-commerce, entertainment, news, tourism, finance, healthcare, and e-learning is often limited. For example, CF can personalize user experiences by leveraging behavior data in a news RS \cite{raza2022news} or a music RS \cite{schedl2018current}, but struggles with the dynamic nature of user preferences and the need for real-time recommendations. In tourism, finance, healthcare, and e-learning, foundational RS can understand user preferences and behavior patterns, but issues like the cold start problem, data diversity, privacy concerns, and the need for highly personalized services require more sophisticated solutions. These solutions often blend foundational techniques with modern advancements like deep learning and specialized RS (discussed below).

\section{The Era of Deep Learning in Recommender Systems}
In recent years, deep learning has emerged as the standard in RS, as detailed in a related survey \cite{zhang2019deep}. 
In the context of this discussion, we shed some light on some popular deep neural network based RS.
\subsection{Deep Learning-based Recommender Systems}

\textbf{Multi-Layer Perceptrons} Traditional RS primarily use linear methods like matrix factorization \cite{koren2009matrix}, which struggle with capturing complex user-item interactions. In contrast, Multi-Layer Perceptrons (MLPs), a type of feedforward neural network, use deep layers to model these nonlinear interactions more accurately, improving both prediction accuracy and recommendation quality.
The evolution of MLPs is seen in RS such as Neural Collaborative Filtering (NCF) \cite{he2017neuralcollab}, Deep Factorization Machine (DFM) \cite{guo2017deepfm}, Wide \& Deep \cite{cheng2016wide}, xDeepFM \cite{lian2018xdeepfm}, Deep \& Cross Network (DCN) \cite{wang2017deep}, FMLP-Rec \cite{Zhou2022Filter-enhancedRecommendation}, a model with learnable filters for improving sequential recommendation—and FinalMLP \cite{Mao2023FinalMLP:Prediction}, which combines dual MLP architectures with feature selection for effective Click-Through Rate (CTR) prediction. MLP recommenders typically select embedding size, hidden-layer count, and dropout via validation to trade off capacity and generalization~\cite{He2017NCF,guo2017deepfm}.

\textit{Challenge with MLP:}   Despite their success, MLP models in RS face challenges like complexity, the risk of overfitting, lack of spatial invariance, issues with vanishing or exploding gradients, and explainability concerns.

\textbf{Autoencoders} are neural network architectures specifically designed for unsupervised learning and are used for an effective dimensionality reduction method. An autoencoder comprises two components: an encoder  for compressing input data into a lower-dimensional representation, and a decoder, which reconstructs the original data. Unlike traditional MLP models, autoencoders explicitly capture this encoding-decoding structure.

Notable RS include AutoRec \cite{sedhain2015autorec}, Collaborative Filtering Autoencoder, Multi-Variational Auto-Encoder (Multi-VAE) \cite{liang2018variational},  Deep Recommender (DeepRec) \cite{zhang2019deeprec}, Recommender Variational Auto-Encoder (RecVAE) \cite{shenbin2020recvae}, Item-based variational auto-encoder for fair recommendation \cite{Park2022Item-basedRecommendation}, and the Variational Bandwidth Auto-Encoder (VBAE) hybrid RS \cite{Zhu2023VariationalSystems} . These approaches address sparsity and noise challenges, making them effective for personalized recommendations. Vanilla AE recommenders tune the latent (bottleneck) size, learning rate, and L2 weight decay (or dropout), with reconstruction as the primary objective; only multi-objective variants introduce method-specific loss weights~\cite{sedhain2015autorec,Strub2016CollaborativeAE}.

\textit{Challenge with Autoencoder:}  Autoencoders are powerful for dimensionality reduction and capturing complex data structures, but one key issue is their sensitivity to noise, which can lead to poor reconstructions if the input data is noisy \cite{Berahmand123AutoencodersSurvey}. Also, the reconstruction process might not always preserve meaningful patterns essential for recommendations.

\textbf{Convolutional Neural Networks (CNNs)} can learn from visual, sequential, and multimodal data and have enhanced accuracy and personalization of recommendations. 
CNNs have been applied in RS in various settings. DeepCoNN analyzes text and visual cues to understand user preferences \cite{zheng2017joint, Sulthana2020ImprovisingLearning}. CNNs are integrated with graph structures for scalable recommendation systems \cite{ying2018graph}, employed in DKN for news recommendations \cite{wang2018dkn}, and utilized in MusicCNN for music recommendations based on audio signals \cite{pons2019musicnn}. CNN-based RS models predict next-item recommendations \cite{Yuan2019ARecommendation}, recognize user preference patterns through CoCNN \cite{chen2022cocnn}, and leverage collaborative filtering with CAGCN \cite{Wang2023Collaboration-AwareSystems}.For CNN-based recommenders, kernel size, stride, and filter count control the granularity of local pattern capture (distinct from two-tower MLP systems such as YouTube’s)~\cite{Zheng2017DeepCoNN,Kim2016CNNRec,Covington2016YouTube}. For CNN-based recommenders, kernel size, stride, and filter count control the granularity of local pattern capture~\cite{Zheng2017DeepCoNN,Kim2016CNNRec,Covington2016YouTube}.

\textit{Challenge with CNNs}: CNNs in RS face challenges such as data sparsity, scalability, privacy, and domain-specific issues \cite{Alzubaidi2021ReviewDirections}. Researchers continue to explore solutions to enhance CNN-based RS performance and usability

\textbf{Recurrent Neural Networks (RNNs)} are adept at capturing complex user-item interactions within sequential data \cite{sun64}. 
The evolution of RNN-based RS began with GRU4Rec, utilizing Gated Recurrent Units (GRUs) for session-based recommendations \cite{hidasi2016session, hidasi2018recurrent}. NARM introduced an attention mechanism to enhance accuracy \cite{li2017neural}, while SASRec used self-attention to capture long-term semantics \cite{kang2018self}. Deep attention neural networks were employed for session-based recommendations \cite{zhu2019dan}. RNNs, including LSTMs and GRUs, comprehend temporal dynamics in user behavior \cite{feng2019deep, Raza2021DeepSystem}. Integrating RNNs with CNNs via recurrent convolutional networks offers deeper insights into user preferences \cite{bach2020recurrent}, followed by CNN-RNN hybrid RS \cite{Zhou2021CNN-RNNSupport} and Reinforcement Learning-based cross-domain recommendations \cite{Guo2022ReinforcementRecommendation}. Additionally, a knowledge graph recommendation algorithm using RNN encoders has emerged \cite{Zhao2023AGRE:Encoder}. RNN sequential recommenders (e.g., GRU4Rec) typically tune hidden-state size, maximum sequence length, and dropout to balance temporal coverage and training stability~\cite{Hidasi2016GRU4Rec,Quadrana2017SequentialSurvey}.

\textit{Chalenge with RNNs}: A common challenge with RNNs is exploding and vanishing gradients \cite{batmaz2019review}. Addressing these issues often requires careful initialization, gradient clipping, or alternative architectures like LSTM networks that can mitigate gradient problems. Additionally, training is sequential, as RNNs takes the data in a sequential manner, unlike CNNs. 

\textbf{Self-supervised learning (SSL)–based recommenders.}
SSL introduces auxiliary objectives that derive supervision from unlabeled user–item interactions~\cite{Yu2024SSLSurvey}. Contrastive-learning–based models include SGL~\cite{Wu2021SGL} graph view augmentations via node and edge dropout for robustness under sparsity, SimGCL~\cite{Yu2022SimGCL} noise-perturbed embedding views without explicit graph ops, LightGCL~\cite{Cai2023LightGCL} lightweight propagation with SVD-guided signal, NCL~\cite{Lin2022NCL} neighborhood-enriched positives, HCCF~\cite{Xia2022HCCF} hypergraph views for higher-order relations, KGCL~\cite{Yang2022KGCL} adapts SSL to knowledge graphs via cross-view alignment between CF and KG signals, AdaGCL~\cite{Jiang2023AdaGCL} learns augmentation strengths with trainable generators, CL4SRec~\cite{Xie2022CL4SRec} masking/cropping/segment permutation for sequence views, TGCL~\cite{Zhang2024TGCL} time-aware transitions with temporal contrast, and IDCL~\cite{Wang2023IDCL} intent-disentangled graph contrastive learning. Generative/predictive SSL includes BERT4Rec~\cite{sun2019bert4rec} masked-item prediction, S$^3$-Rec~\cite{Zhou2020S3Rec} multi-task mutual-information objectives over items, attributes, and subsequences, and MAERec~\cite{Ye2023MAERec} masked autoencoding for recommendation. Other/hybrid SSL directions feature BUIR~\cite{Lee2021BUIR} BYOL-style bootstrap without negatives, DuoRec~\cite{Qiu2022DuoRec} dual contrastive objectives to improve sequence representations, and MMGCL~\cite{Wei2024MMGCL} multimodal contrastive learning integrating text and vision with interactions.

Key SSL hyperparameters include the temperature $\tau$ in the InfoNCE family (controls similarity sharpness), the number and sampling strategy of negatives (influences hardness and informativeness of comparisons), and augmentation probabilities (e.g., edge-drop or masking ratios) that govern view diversity. Their tuning primarily shapes the alignment–uniformity trade-off in learned embeddings~\cite{Wang2020AlignUniform,Yu2024SSLSurvey}.

\textit{Challenges with SSL:} SSL-based recommenders remain sensitive to augmentation design, negative sampling, and temperature tuning; weak or excessive perturbations can cause representation collapse or semantic drift, and optimal settings are often dataset-specific~\cite{Yu2024SSLSurvey,Wang2020AlignUniform,Qiu2022DuoRec}.

  Table \ref{tab:deepRS} shows the main publication based on deep learning RS.
\begin{table}[h]
    \centering
       \caption{Deep Learning-based RS Publications}
    \begin{tabular}{|c|>{\raggedright\arraybackslash}p{0.7\linewidth}|} \hline 
           RS Type&Publications\\ \hline 
            GNN& \cite{wu2021comprehensive,jeong2020context,guo2020deep,yin2019deeper,mansoury2021graph,ali2020graph,liu2020heterogeneous,gao2023survey,zou2020survey,guo2020survey,ekstrand2018all,liang2022aspect,bem,cai2018bibliographic,bose2019compositional,zhang2023dynamic,ma2021event,wang65,qiu2020exploiting,dai2018explore,fu2020fairness,li2019fi,kong2024gcnslim,wang2020global,liu2023gnnrec,wu2022graphconv,berg2017graph,ying2018graph,he2024graph,wu2023graph,fan2019graph,wu2022graphneural,qiu2022graph,liu2022graph,wang2019heterogeneous,Hamilton2017InductiveGraphs,zhao2019intentgc,ma67,wang2019kgat,transd,kgdef1,wu2021learning,embpropagation,he2020lightgcn,huang2021mixgcf,wei2019mmgcn,wang2023multi,mkr,mao2019multiobjective,wang2019neural,de2024personalized,rezaee2021personalized,shaikh2017recommendation,sun64,wang2020reinforced,xian2019reinforcement,lei2020reinforcement,wang2018ripplenet,wu2019session,Wang2022ARecommendation,Zhao2023AGRE:Encoder,Wang2023Collaboration-AwareSystems,Joseph2019ContentGraphs,Chen2022HybridCompletion,Gwadabe2022ImprovingInteractions,Hamilton2017InductiveGraphs,ji2021temporal,ktup,yu2023xsimgcl,Zhao2023TowardsLearning,Zhang2022TrustworthyTrends}\\ \hline 
            Sequential&\cite{kolahkaj2020hybrid,wang2021survey,yuan2020attention,chen2019behavior,sun2019bert4rec,feng2019deep,zhang2023dynamic,ludewig2021empirical,li2024enhancing,qiu2020exploiting,wang2020global,liu2023gnnrec,ksr,wang2020kerl,li2017neural,bach2020recurrent,hidasi2018recurrent,hidasi2018recurrentb,ren2019repeatnet,kang2018self,peters2017semi,quadrana2018sequence,wang2019sequential,wang2022sequential,wu2017session,jannach2017session,wu2019session,hidasi2016session,Thaipisutikul2024AnRecommendation,Thaipisutikul2024AnRecommendationb,Hou2022CORE:Space,Zhou2022Filter-enhancedRecommendation,Du2023FrequencyRecommendation,Gwadabe2022ImprovingInteractions,Zhai2023KnowledgeRecommendation,Zhai2023KnowledgeRecommendationb,Wang2023MISSRec:Recommendation,Ji2023OnlineRecommendation,Tang2018PersonalizedEmbedding,Hidasi2018RecurrentRecommendations,Guo2022ReinforcementRecommendation,de2021transformers4rec,Wu2020SSE-PT:Transformer,Hu2023AdaptiveSystems}
\\ \hline 
            KG&\cite{zou2020survey,guo2020survey,bem,wang65,fu2020fairness,ma67,wang2019kgat,transd,kgdef1,mkr,sun64,wang2020reinforced,xian2019reinforcement,wang2018ripplenet,Zhao2023AGRE:Encoder,Joseph2019ContentGraphs,Chen2022HybridCompletion,ktup}
\\ \hline
 RL&\cite{huang2021deep,wang2018reinforcement,jha2023appropriate,zhao2017deep,zhao2018deep,chen2023deep,van2016deep,arulkumaran2017deep,kokkodis2021demand,gao2019drcgr,zheng2018drn,Zhang2017DSC,Yu2024EasyRL4RecAU,Chen2018GenerativeAU,wang2020kerl,he2020learning,gottipati2023maximum,gui2019mention,Bai2019AMR,zhang2022multi,chang2021music,Wang2021ReinforcementLW,hu2017playlist,Rohde2018RecoGymAR,lei2020reinforcement,choi2018reinforcement,afsar2022reinforcement,intayoad2018reinforcement,Ie2019ReinforcementLF,hu2018reinforcement,xu2020reinforcement,sutton2018reinforcement,wiering2012reinforcement,Wang2022ARecommendation,Jiang2021DeepRepresentation,Jiang2021DeepRepresentationb,Guo2022ReinforcementRecommendation,Liaw2022REVEALScale,wang2018supervised,Zhao2023TowardsLearning}
\\\hline
 LLM&\cite{zhao2023survey,wu2023survey,kang2023llms,vats2024exploring,taylor2022galactica,ji2024genrec,sanner2023large,hou2024large,he2023large,li2023large,liu2023llmrec,valmeekam2024planning,chen2023palr,wang2023recmind,fan2023recommender,wang2023rethinking,bao2023tallrec}
\\\hline
 Multi-modals&\cite{singh2021comprehensive,wei2019mmgcn,Zhou2023MMRecSM,Wei2023MultiModalSL,chen2019personalized,Wei2024PromptMMMK,Hu2023AdaptiveSystems,Zhou2023BootstrapRecommendation,Chen2022HybridCompletion,Wang2023MISSRec:Recommendation,Ji2023OnlineRecommendation,Vuorio2018TowardMM}
\\\hline
SSL & \cite{Wu2021SGL,Yu2022SimGCL,Cai2023LightGCL,Lin2022NCL,Xia2022HCCF,Yang2022KGCL,Jiang2023AdaGCL,Xie2022CL4SRec,Zhang2024TGCL,Wang2023IDCL,sun2019bert4rec,Zhou2020S3Rec,Ye2023MAERec,Lee2021BUIR,Qiu2022DuoRec,Wei2024MMGCL,Yu2024SSLSurvey,Wang2020AlignUniform}
\\\hline
    \end{tabular}
 
    \label{tab:deepRS}
\end{table}
 
\subsection{Can Deep Learning-based Recommender Systems address Practical Challenges?}

Deep learning-based RS have effectively addressed many practical challenges faced by foundational systems. Models like NCF, DeepFM, and DeepMF have enhanced personalization by capturing complex user-item interactions in e-commerce product recommendations \cite{kumar2018recommendation}. Wide \& Deep Learning has shown improved performance in e-commerce for both product recommendations and ads \cite{zhao2021dear}. CNNs, RNNs, and their variations and hybridizations are used in content-based and sequential data recommendations, benefiting industries like news \cite{Raza2021DeepSystem} and entertainment \cite{singh2020multilingual}. Transformer models like BERT are used for movie recommendations \cite{sun2019bert4rec}. GNNs capture relationships in social networks and e-commerce, offering improved accuracy and diversity in recommendations \cite{Wang2023Collaboration-AwareSystems}. 

Advancements in deep learning have brought further changes to the theory and practice of RS, leading to the development of advanced modeling methods, which is discussed next.

\section{Advanced Modeling Techniques in Recommender Systems}

\subsection{Graph-based Recommender Systems}
Graph Neural Networks (GNNs) are specialized neural networks designed to work with graph-structured data. 
GNNs have emerged as a powerful tool in RS due to their capability to efficiently leverage complex, relational user-item interaction data, enhancing recommendation accuracy and personalization.
GNNs in RS are highlighted in the related survey articles \cite{wu2021comprehensive,gao2023survey}, showcasing their significant impact and evolution in the domain.

In RS, a graph $G = (V, E)$ represents the domain, with $V$ denoting nodes (users and items) and $E$ representing user-item interactions. Each node $v \in V$ is associated with a feature vector $\mathbf{x}_v$. GNN-based models adapt to various graph types, including homogeneous (edges link nodes of a single type), heterogeneous (nodes and edges of multiple types), and hypergraphs (edges connect more than two nodes). The core operation in GNNs, message passing, involves nodes aggregating and updating information from neighbors to refine their features, thus capturing the dynamics of user-item interactions. This process enhances recommendation accuracy and personalization by utilizing the relational data within RS. The update mechanism for a node $v$ at layer $l$ is given by \cite{gao2023survey}:

\begin{equation}
\mathbf{h}_v^{(l+1)} = \text{UPDATE}^{(l)} \left( \mathbf{h}_v^{(l)}, \text{AGGREGATE}^{(l)}\left(\left\{\mathbf{h}_u^{(l)}: u \in \mathcal{N}(v)\right\}\right) \right)
\end{equation}

Here, $\mathbf{h}_v^{(l)}$ represents node $v$'s feature vector at layer $l$, $\mathcal{N}(v)$ denotes $v$'s neighbors, and $\text{UPDATE}^{(l)}$ and $\text{AGGREGATE}^{(l)}$ are the respective update and aggregation functions. The objective of GNN-based RS is to learn a predictive function $f$ for estimating the interaction likelihood between user $u$ and item $i$, utilizing their feature vectors $\mathbf{h}_u$ and $\mathbf{h}_i$:

\begin{equation}
\hat{y}_{ui} = f(\mathbf{h}_u, \mathbf{h}_i; \Theta)
\end{equation}

In this context, $\hat{y}_{ui}$ is the interaction prediction score, with $\Theta$ indicating the model parameters. Training involves minimizing a loss function $\mathcal{L}$ that compares predicted scores $\hat{y}_{ui}$ with actual interactions $y_{ui}$:

\begin{equation}
\mathcal{L} = \sum_{(u,i) \in D} \text{loss}(\hat{y}_{ui}, y_{ui})
\end{equation}

This equation reflects the sum of losses over all observed user-item interactions in set $D$.

\textit{State-of-the-art RS models using GNNs}
GNNs have progressively transformed RS, starting from the foundational model, i.e. Graph Convolutional Matrix Completion (GCMC) \cite{berg2017graph}, which applies deep learning to user-item interaction graphs for effective link prediction. Building upon this, GraphSAGE \cite{Hamilton2017InductiveGraphs} is an inductive framework utilizing node features for dynamic environments, though it could not address the complexity of real-world interaction data. Pinterest's PinSage \cite{ying2018graph} is a scalable model for web-scale graphs, improving its predecessor model for handling billions of nodes.

The Neural Graph Collaborative Filtering (NGCF) model \cite{wang2019neural} combines collaborative signals into user and item embeddings, enhancing recommendation quality at the expense of increased complexity. Knowledge Graph Attention Network (KGAT) \cite{wang2019kgat} integrates knowledge graphs, improving recommendation diversity and explainability. The Heterogeneous Graph Attention Network (HGAT) \cite{wang2019heterogeneous} incorporates hierarchical attention into the RS, addressing the heterogeneity in relationships and node types.

Feature Interaction Graph Neural Networks (Fi-GNN) \cite{li2019fi} represented a shift towards capturing multifield feature interactions, notably in CTR prediction. Concurrently, Session-based Recommendation Graph Neural Network (SR-GNN) \cite{wu2019session} tackled session-based recommendations, enhancing accuracy by capturing item transitions. The Multi-Modal Graph Convolution Network (MMGCN) \cite{wei2019mmgcn} integrates multi-modal data into the graph-based learning, though facing scalability challenges.

The introduction of LightGCN \cite{he2020lightgcn}, with its focus on neighborhood aggregation and streamlined architecture, represents a simplification in the GNN landscape, improving efficiency without compromising performance. MixGCF \cite{huang2021mixgcf} brought forward a novel approach to negative sampling, optimizing training processes. Subsequent developments like GNNRec \cite{liu2023gnnrec} and XSimGCL \cite{yu2023xsimgcl} advanced session-based and graph contrastive learning recommendations, respectively, addressing specific challenges such as social influence integration and bias mitigation.  Ensuring trustworthiness in GNN-based RS requires enhancements in robustness, explainability, and fairness to ensure reliable recommendations \cite{Zhang2022TrustworthyTrends}.
Graph-based recommenders adjust the number of propagation layers, neighborhood-sampling size, and embedding dimension to capture higher-order connectivity without oversmoothing~\cite{Wang2019NGCF,he2020lightgcn}.

\paragraph{Practical Challenges Addressed}
GNNs can effectively address various practical challenges by modeling complex relationships in data. In e-commerce, models like LightGCN, GC-MC, NGCF, and Graph-ICF enhance personalization, scalability, and efficiency for rating, link, and item predictions. These models are capable of handling extensive product catalogs and large user bases efficiently. In social networks, GNNs such as GNN-SoR and GraphRec improve user interaction predictions, boosting content relevance and user engagement by understanding social dynamics and user relationships. In healthcare and finance, GNNs like KGAT and Fi-GNN provide secure and interpretable recommendations, ensuring data privacy and compliance with regulations. These systems have ability to address the cold-start problem by incorporating user and item features from knowledge graphs, providing accurate recommendations even with limited initial data. The practical applications and use of GNNs are further detailed in Table 6.

Table \ref{tab:GNN} presents the use of GNNs, their variants, the data being used, evaluation metrics, and applications. For detailed evaluation criteria on scalability, interpretability, computational efficiency, and reproducibility, please refer to the Appendix.
{\scriptsize
\begin{longtable}{|p{1.2cm}|p{0.6cm}|p{1.5cm}|p{2cm}|p{1.5cm}|p{1.3cm}|p{2cm}|p{1.2cm}|}
\caption{Comprehensive Overview of Graph Neural Network Models across Various Metrics and Use Cases. This table details each model's Input features, Year of Publication, and Characteristics such as Scalability, Interpretability, Efficiency, and Reproducibility (rated as High, Medium, or Low, the symbol '-' means no information available for this). It also lists the Dataset Used, Evaluation Metrics, Model Accuracy (as per evaluation metric from the previous column), Learning Task, and Application Field.}
 \\
\hline
\textbf{Model} & \textbf{Year} & \textbf{Input Data} & \textbf{Scalability, Interpretability, Efficiency, Reproducibility} & \textbf{Dataset} & \textbf{Evaluation Metrics} & \textbf{Model Accuracy} & \textbf{Application} \\ \hline
\endfirsthead

\multicolumn{8}{|c|}%
{{\bfseries \tablename\ \thetable{} -- continued from previous page}} \\
\hline
\textbf{Model} & \textbf{Year} & \textbf{Input Data} & \textbf{Scalability, Interpretability, Efficiency, Reproducibility} & \textbf{Dataset} & \textbf{Evaluation Metrics} & \textbf{Model Accuracy} & \textbf{Application} \\ \hline
\endhead

GCN \cite{he2020lightgcn}  & 2015 & MovieLens & High, Medium, High, High & MovieLens & MSE & MovieLens: 0.5 & e-commerce \\ \hline

GC-MC \cite{berg2017graph} & 2017 & MovieLens-1M, MovieLens-10M, Flixster, Douban, Yahoo Music & High, Low, -, \href{https://github.com/riannevdberg/gc-mc}{High} & MovieLens-1M, MovieLens-10M, Flixster, Douban, Yahoo Music & RMSE & MovieLens-1M: 0.832 \par MovieLens-10M: 0.777 \par Flixster: 0.941 \par Douban: 0.734 \par Yahoo Music: 20.5 & e-commerce  \\ \hline

NGCF \cite{wang2019neural} & 2019 & Gowalla, Yelp2018, Amazon-books & -, Low, Medium,  \href{https://github.com/xiangwang1223/neural_graph_collaborative_filtering}{High} & Gowalla, Yelp2018, Amazon-books & Recall, NDCG & Gowalla: 0.1569/0.1327 \par Yelp2018: 0.0579/ 0.0477 \par Amazon-books: 0.0337/0.0261 & e-commerce  \\ \hline

Graph-ICF \cite{liu2022graph} & 2022 & MovieLens-1M, Pinterest-20, Yelp & -, Low, Medium, \href{https://github.com/sunshinelium/Graph-ICF}{High} & MovieLens-1M, Pinterest-20, Yelp & HR, NDCG, MAP & MovieLens-1M: 0.7425/0.4555/0.3721 \par Pinterest-20: 0.8987/0.5830/0.4873 \par Yelp: 0.7519/0.4856/0.4033 & e-commerce \\ \hline

GNN-SoR \cite{guo2020deep} & 2020 & Epinions, Yelp, Flixster & -, -, -, Low & Epinions, Yelp, Flixster & RMSE, MAE, NDGC & Epinions: 0.880/0.791/0.792 \par Yelp: 0.820/0.871/0.687 \par Flixster: 0.863/0.859/0.594 & Social Network RecSys, e-commerce \\ \hline

GCM \cite{wu2022graphconv} & 2022 & Yelp-NC, Yelp-OH, Amazon-book & -, Low, Medium, \href{https://github.com/wujcan/GCM}{High} & Yelp-NC, Yelp-OH, Amazon-book & HR@10, NDGC@10 & Yelp-NC: 0.1046/0.0557 \par Yelp-OH: 0.2648/0.1457 \par Amazon-book: 0.0968/0.0536 & e-commerce \\ \hline

GCF-YA \cite{yin2019deeper} & 2019 & MovieLens-1M, MovieLens-10M, Taobao & -, Low, -, Low & MovieLens-1M, MovieLens-10M, Taobao & HR@10, NDGC@10 &  MovieLens-1M: 0.7818/0.4873 \par MovieLens-10M: 0.7642/0.4677 \par Taobao: 0.3662/0.2491 & e-commerce \\ \hline

DGSR \cite{zhang2023dynamic} & 2023 & Beauty, Games, CDs & -, Low, -, \href{https://github.com/CRIPAC-DIG/DGSR}{High} & Beauty, Games, CDs & NDCG@10, Hit@10 & Beauty: 52.4/35.9 \par Games: 75.57/55.7 \par CDs: 72.43/51.22 & e-commerce \\ \hline

GraphRec \cite{fan2019graph} & 2019 & Ciao, Epinions & -, Low,  -, \href{https://github.com/wenqifan03/GraphRec-WWW19}{High} & Ciao, Epinions & MAE, RMSE & Ciao: 0.7387/0.9794 \par Epinions: 0.8441/1.0878 & e-commerce\\ \hline

KGAT \cite{wang2019kgat} & 2019 & Amazon-book, Last-FM, Yelp2018 & -, High, High,  \href{https://github.com/xiangwang1223/knowledge_graph_attention_network}{High} &  Amazon-book, Last-FM, Yelp2018 & Recall@20
NDCG@20 & Amazon-book: 0.1489/0.1006 \par  Last-FM:0.0870/0.1325 \par Yelp2018: 0.0712/0.0867 & e-commerce \\ \hline
\label{tab:GNN}
\end{longtable}
}

\subsection{Sequential and Session-based Recommender Systems}
Traditional models like Markov chains \cite{salakhutdinov2008bayesian}, pattern/rule mining \cite{kolahkaj2020hybrid}, and latent factorization techniques \cite{pena2020combining} have been long used in analyzing sequential data and user-item relationships by examining transitions, patterns, and latent connections. However, they often struggle with dynamically predicting user preferences, typically due to a narrow focus on immediate past users' interactions or statistical correlations. This limitation is overcome by sequential RS \cite{wang2022sequential}, which exploit the temporal order and context of user interactions. The evolution of sequential RS has transitioned from Markov Chains and session-based KNN to sophisticated deep learning approaches, including RNNs, LSTMs, attention mechanisms, and transformer architectures. 

Sequential recommendation is commonly viewed as a next-item or next-basket prediction challenge \cite{quadrana2018sequence}. Both the sequential and session-based RS leverage user action sequences to anticipate users’ future preferences \cite{wang2022sequential}. Specifically, sequential RS consider the interaction histories of the users to predict future behaviour or users' preferences. In contrast, session-based RS, detailed in survey \cite{wang2021survey}, focus on short-term user activity for real-time recommendations. These approaches collectively enhance personalization and relevance across diverse platforms. 

A sequential RS model can be defined as:
\[
i_{\text{next}} = f(\text{history}(u)),
\]
where $i_{\text{next}}$ is the next recommended item, $\text{history}(u)$ is the user $u$'s interaction sequence, and $f$ models sequential behavior to predict future interactions.

A session-based RS model can be defined as:
\[
i_{\text{session-next}} = g(s_{\text{current}}),
\]
with $i_{\text{session-next}}$ as the imminent session recommendation, $s_{\text{current}}$ representing the ongoing session interactions, and $g$ predicting the next item considering the session's context.

The evolution of sequential and session-based RS has seen significant advancements with various models. For example, Translation-based RS (TransRec) \cite{He2017Translation-basedRecommendation}, integrates third-order interactions to enhance sequential predictions. The research has progressed to using RNNs with GRU4Rec \cite{hidasi2016session} and its enhancement, GRU4Rec+ \cite{hidasi2018recurrent}, improving session-based recommendations through refined loss functions and sampling strategies.

Subsequently, CNNs are applied in models like Convolutional Sequence Embedding Recommendation Model  (Caser) \cite{Tang2018PersonalizedEmbedding} and NextItNet \cite{Yuan2019ARecommendation}, targeting effective session-based recommendations. The introduction of self-attention mechanisms in Self-Attention based RS (SASRec) \cite{kang2018self} for sequential model, and the exploration of item transitions with Session-based Recommendations with Graph Neural Networks (SR-GNN) \cite{wu2019session}, showed further progress.

Recent developments have seen the application of the Transformers architecture, with models like Bert for RS (BERT4Rec) \cite{sun2019bert4rec} using bidirectional self-attention for deep sequence analysis, and Transformers4Rec \cite{de2021transformers4rec} adapting NLP transformers for recommendation contexts.

GNNs have been employed for modeling session-based interactions in GRASER \cite{Gwadabe2022ImprovingInteractions}, and LightSANs \cite{Fan2021LighterRecommendation} improved traditional Self-Attention Networks (SANs) by reducing complexity and refining sequence modeling with low-rank decomposed self-attention.

Frequency Enhanced Hybrid Attention Network (FEARec) \cite{Du2023FrequencyRecommendation} and Knowledge Prompt-tuning for Sequential Recommendation (KP4SR) \cite{Zhai2023KnowledgeRecommendation} advance sequential recommendation by leveraging hybrid attention mechanisms and integrating external knowledge bases, respectively, for better model performance.
Transformer-style sequential models (e.g., SASRec, BERT4Rec) tune hidden size, number of self-attention heads, number of layers (stack depth), and maximum sequence length to set the temporal range and computational cost~\cite{Kang2018SASRec,sun2019bert4rec}.

\paragraph{Practical Challenges Addressed}

Sequential and session-based RS effectively tackle practical challenges by capturing temporal dynamics and sequential patterns in user behavior. Models like TransRec, GRU4Rec, and GRU4Rec+ use recurrent neural networks to ensure scalability and computational efficiency, making them ideal for e-commerce and video streaming. Caser and NextItNet enhance these capabilities with convolutional layers, improving accuracy. SASRec and SR-GNN apply self-attention mechanisms and graph neural networks to capture complex user-item interactions in e-commerce and video games. BERT4Rec and Transformers4Rec leverage transformer architectures to model long-range dependencies, achieving high accuracy across datasets like Amazon Beauty, Steam, and MovieLens.

These models also have the ability to address data sparsity and cold start issues by considering both short-term session-based and long-term sequential preferences. They adapt to rapidly changing user interests and provide real-time recommendations, making them effective in industries like entertainment and news. For instance, in the news industry, they offer timely and relevant articles . In e-commerce, they track user interactions within a session to provide context-aware product suggestions, enhancing the shopping experience and increasing the likelihood of immediate purchases. The practical applications and use of these systems are further detailed in Table \ref{tab:seq}.


{\scriptsize
\begin{longtable}{|p{1.2cm}|p{0.6cm}|p{1.5cm}|p{2cm}|p{1.5cm}|p{1.3cm}|p{2cm}|p{1.2cm}|}

\caption{Sequential Models. This table provides a detailed overview of various sequential models in recommendation systems, showcasing their combined characteristics of Scalability, Interpretability, Computational Efficiency, and Reproducibility (rated as High, Medium, or Low). Additionally, the table includes information on datasets used, evaluation metrics, model accuracy, publication year, and application fields.} \\
\hline

\textbf{Model} & \textbf{Year} & \textbf{Input Data} & \textbf{Scalability, Interpretability, Efficiency, Reproducibility} & \textbf{Dataset} & \textbf{Evaluation Metrics} & \textbf{Model Accuracy} & \textbf{Application} \\ \hline
\endfirsthead

\multicolumn{8}{|c|}%
{{\bfseries \tablename\ \thetable{} -- continued from previous page}} \\
\hline
\textbf{Model} & \textbf{Year} & \textbf{Input Data} & \textbf{Scalability, Interpretability, Efficiency, Reproducibility} & \textbf{Dataset} & \textbf{Evaluation Metrics} & \textbf{Model Accuracy} & \textbf{Application} \\ \hline
\endhead

TransRec \cite{He2017Translation-basedRecommendation} & 2017 & User-item interaction; Sequential behavior & High, Medium, High, Yes \footnote{https://sites.google.com/a/eng.ucsd.edu/ruining-he/} & Epinions; Automotive; Google Local; Office Products; Toys and Games; Clothing, Shoes, and Jewelry; Cell Phones and Accessories; Video Games; Electronics; Foursquare; Flixter & AUC; Hit@50 & Epinions: 0.6133, 4.63\%; Automotive: 0.6868, 5.37\%; Google Local: 0.8691, 6.84\%; Office Products: 0.7302, 6.51\%; Toys and Games: 0.7590, 5.44\%; Clothing, Shoes, and Jewelry: 0.7243, 2.12\%; Cell Phones and Accessories: 0.8104, 9.54\%; Video Games: 0.8815, 16.44\%; Electronics: 0.8484, 5.19\%; Foursquare: 0.9651, 67.09\%; Flixter: 0.9750, 35.02\% & E-commerce, Video Streaming, Social Media \\ \hline

GRU4Rec \cite{hidasi2016session} & 2016 & User-item graphs; node features & High , Medium , High , Yes \footnote{https://github.com/hidasib/GRU4Rec} & RecSys Challenge 2015 \footnote{https://recsys.acm.org/recsys15/challenge/}; Youtube-like OTT video service platform. & Recall@20; MRR@20 & Item-KNN for: RSC15: 0.5065, 0.2048; VIDEO: 0.5508, 0.3381& E-commerce; Video Streaming \\ \hline

GRU4Rec+ \cite{hidasi2018recurrent} & 2018 & Session-based; RNN; GRU & High, Medium, High, Yes \footnote{https://github.com/hidasib/GRU4Rec} & RSC15; VIDEO; VIDXL; CLASS & Recall@20; MRR@20 & RSC15: 0.7208, 0.3137; VIDEO: 0.6400, 0.3079; VIDXL: 0.8028, 0.5031; CLASS: 0.3137, 0.1167 & E-commerce; Video Streaming; Classifieds \\ \hline

Caser \cite{Tang2018PersonalizedEmbedding} & 2018 & Sequential; CNN & High, Medium, High, Yes \footnote{https://github.com/graytowne/caser} & MovieLens; Gowalla; Foursquare; Tmall & Precision@N; Recall@N; MAP & MovieLens: 0.2502, 0.0632, 0.1507; Gowalla: 0.1961, 0.0845, 0.0928; Foursquare: 0.1351, 0.1035, 0.0909; Tmall: 0.0312, 0.0366, 0.0310 & Various domains \\ \hline

NextItNet \cite{Yuan2019ARecommendation} & 2019 & Sequential; CNN; Dilated convolution & High, Medium, High, Yes \footnote{https://github.com/fajieyuan/NextItNet} & Yoochoose-buys; Last.fm & MRR@20; HR@20; NDCG@20 & Yoochoose-buys: 0.1901, 0.4645, 0.2519; Last.fm: 0.3223, 0.4626, 0.3542 & E-commerce; Music \\ \hline

SASRec \cite{kang2018self} & 2018 & User-item graphs; node features & High, Medium , High , Yes reproducibility\footnote{https://github.com/kang205/SASRec} & Amazon - Beauty; Amazon - Games; Steam; MovieLens-1M & Hit@10; NDCG@10 & Amazon - Beauty: 0.4854, 0.3219; Amazon - Games: 0.7410, 0.5360; Steam: 0.8729, 0.6306; ML-1M: 0.8245, 0.5905 & E-Commerce; Video Games; Movies\\ \hline


SR-GNN \cite{wu2019session} & 2019 & User-item graphs; node features & High, Medium , High , Yes \footnote{https://github.com/CRIPAC-DIG/SR-GNN} & YOOCHOOSE 1/64: ; YOOCHOOSE 1/4; DIGINETICA & P@20; MRR@20 & YOOCHOOSE 1/64: 0.7057, 0.3094; YOOCHOOSE 1/4: 0.7136, 0.3189; DIGINETICA: 0.5073, 0.1759 & E-Commerce \\ \hline

BERT4Rec \cite{sun2019bert4rec} & 2019 & User-item graphs; node features & High , Medium , High , No  & Amazon Beauty; Steam; MovieLens-1m; MovieLens-20m & HR@10; NDCG@10; MRR\footnote{HR@1, HR@5, NDCG@5 metrics dropped for simplicity.} & Amazon Beauty: 0.1599, 0.1862, 0.1701; Steam:0.4013, 0.2261, 0.1949; ML-1m: 0.6970, 0.4818, 0.4254; ML-20m: 0.7473, 0.5340, 0.4785 & E-commerce; Movies \\ \hline


Transformers 4Rec \cite{de2021transformers4rec} & 2021 & User-item graphs; node features & High , Medium , High  Yes \footnote{https://paperswithcode.com/paper/behavior-sequence-transformer-for-e-commerce} & REES46; YOOCHOOSE; G1; ADRESSA& NGCG@20; HR@20 & REES46: 0.2542, 0.4858; YOOCHOOSE: 0.3776, 0.6384; G1: 0.3675, 0.6721; ADRESSA: 0.3912, 0.7488 & E-commerce, News\\ \hline

GRASER \cite{Gwadabe2022ImprovingInteractions} & 2022 & Session-based; Graph Neural Networks; Non-sequential interactions & High, Medium, High, Yes \footnote{https://github.com/tgdabe/GRASER} & Yoochoose; Diginetica & MRR@20; P@20 & Yoochoose: 0.3497, 71.37; Diginetica: 0.2045, 53.45 & E-commerce \\ \hline

LightSANs \cite{Fan2021LighterRecommendation} & 2021 & Sequential; Low-rank decomposed self-attention & High, Medium, High, Yes \footnote{https://github.com/RUCAIBox/LightSANs} & Yelp; Books; ML-1M & HIT@10; NDCG@10 & Yelp: 0.5480, 0.2890; Books: 0.8760, 0.4250; ML-1M: 0.2284, 0.1145 & E-commerce; Books; Movies \\ \hline

FEARec \cite{Du2023FrequencyRecommendation} & 2023 & Sequential; Frequency-based self-attention & High, Medium, High, Yes \footnote{https://github.com/sudaada/FEARec} & Beauty; Clothing; Sports; ML-1M & HR@5; HR@10; NDCG@5; NDCG@10 & Beauty: 0.0597, 0.0884, 0.0366, 0.0459; Clothing: 0.0214, 0.0323, 0.0121, 0.0156; Sports: 0.0353, 0.0547, 0.0216, 0.0272; ML-1M: 0.2212, 0.3123, 0.1523, 0.1861 & E-commerce; Movies \\ \hline

KP4SR \cite{Zhai2023KnowledgeRecommendation} & 2023 & Sequential; Knowledge graph; Prompt-tuning & High, Medium, High, Yes \footnote{https://github.com/zhaijianyang/KP4SR} & Books; Music; Movies & NDCG@5; HR@5 & Books: 0.0609, 0.0824; Music: 0.0906, 0.1108; Movies: 0.0755, 0.1058 & E-commerce; Music; Movies \\ \hline

\label{tab:seq}

\end{longtable}
}

\subsection{Knowledge-based Recommender Systems}

Knowledge Bases (KB), particularly Knowledge Graphs (KG), have been extensively used in the literature, for enhancing personalized recommendations by leveraging user/item information~\cite{zou2020survey}. A KG is a directed graph $G=(V,E)$, where $V$ and $E$ represent entities and relations between them, respectively, with $E \subseteq V\times V$. It includes entity type function $\Phi : V \rightarrow A$ and relation type function $\Psi: E \rightarrow R$, mapping entities to types $A$ and relations to types $R$. KGs are depicted as sets of triples $\langle e_h,r,e_t\rangle$, signifying a relation $r$ from $e_h$ to $e_t$. This relational information helps RS understand user preferences and item relations, employing various methods to integrate KGs for improved recommendations.
KG-based RS can be observed through three primary approaches: Embedding-based, Path-based, and Propagation-based approaches, each advancing the way RS leverage the rich relational data within KGs, as classified by \cite{zou2020survey}. 

Embedding-based approaches focus on learning and applying embeddings to represent KG entities (nodes) and relations (edges), enhancing user and item representations. They typically start with initial embedding generation using models like TransE \cite{bordes2013translating}, TransD \cite{transd}, and node2vec \cite{node2vec}, followed by their application in RS through attention mechanisms in KSR \cite{ksr} or generative models like BEM \cite{bem} and KTGAN \cite{ktgan}.

Joint Learning Methods optimize both KG embeddings and recommendation components simultaneously using a unified loss function. Examples include CKE \cite{cke}, which integrates auto-encoders for item representations, and SHINE \cite{shine}, which acquires user embeddings from heterogeneous graphs. Multi-Task Methods such as KTUP \cite{ktup} and MKR \cite{mkr} address KG-enhanced recommendation and KG completion concurrently, improving both entity/relation representations and recommendations.

Path-based approaches utilize KG connectivity patterns. Meta-Structure-based Methods like KGCN \cite{yu56} maintain entity proximity in the latent space using graph convolution. Path-Embedding-based Methods, such as MCRec \cite{hu1} and RKGE \cite{sun64}, derive preference scores from path embeddings, incorporating meta-path information and RNN-based path semantics.

Propagation-based approaches influence embeddings through multi-hop neighbor interactions within the KG. Item KG-based methods like Ripplenet \cite{wang2018ripplenet} aggregate item-related embeddings to derive user interests, whereas User-Item KG-based methods such as KGAT \cite{wang2019kgat} and Intentgc \cite{zhao2019intentgc} refine both user and item embeddings by propagating embeddings across a user-item graph, enhancing recommendation accuracy.
For knowledge-graph–based recommenders (as opposed to symbolic rule systems), key hyperparameters include embedding dimension, number of message-passing layers, and regularization strength~\cite{wang2019kgat,Wang2019KGCN}.

\paragraph{Practical Challenges Addressed}

KGs have become increasingly instrumental across various industries, leveraging complex and rich datasets to build RS. For instance, in e-commerce, methods like TransE \cite{bordes2013translating} and Node2Vec \cite{node2vec} have been used to accurately suggest products by understanding the underlying connections between items and user preferences. Similarly, in the movie recommendation space, models like KSR \cite{ksr} and KTUP \cite{ktup} utilize user-item interactions and entity graphs to provide personalized movie suggestions. Social network platforms benefit as well, with systems like SHINE \cite{shine} analyzing sentiment and social networks to enhance user engagement. Overall, these systems enable more contextually aware, personalized, and efficient recommendation systems, significantly improving user experience across these sectors. More details in Table \ref{tab:kg}.

{\scriptsize
\begin{longtable}{|p{1.2cm}|p{0.6cm}|p{1.5cm}|p{2cm}|p{1.5cm}|p{1.3cm}|p{2cm}|p{1.2cm}|}

\caption{Comprehensive Overview of Knowledge Graph Based Recommender System Models across Various Metrics and Use Cases. This table details each model's Input features, Year of Publication, and Characteristics such as Scalability, Interpretability, Efficiency, and Reproducibility (rated as High, Medium, or Low). It also lists the Dataset Used, Evaluation Metrics, Model Accuracy, Learning Task, and Application Field.}
\\
\hline
\textbf{Model} & \textbf{Year} & \textbf{Input data} & \textbf{Scalability, Interpretability, Computational Efficiency, Reproducibility} & \textbf{Dataset} & \textbf{Evaluation Metrics} & \textbf{Model Accuracy} & \textbf{Application} \\ \hline
\endfirsthead

\multicolumn{8}{|c|}%
{{\bfseries \tablename\ \thetable{} -- continued from previous page}} \\
\hline
\textbf{Model} & \textbf{Year} & \textbf{Input data} & \textbf{Scalability, Interpretability, Computational Efficiency, Reproducibility} & \textbf{Dataset} & \textbf{Evaluation Metrics} & \textbf{Model Accuracy} & \textbf{Application} \\ \hline
\endhead

TransE\cite{bordes2013translating} & 2013 & Item-item graph, Multi-relational relationships & High, Medium, -, \href{https://github.com/pyg-team/pytorch_geometric/blob/master/torch_geometric/nn/kge/transe.py}{High} & Wordnet, Freebase15k, Freebase1M & Mean Rank, Hits@10 (Raw/filtered) & Wordnet: 263/251, Freebase15k: 75.4/89.2, Freebase1M: 243/125, 34.9/47.1, 14615/34.0 & Social network analysis \\ \hline

Hete-MF\cite{yu56} & 2013 & User-item interaction, Entity-relation graph & -, -, High, Low & IMDb-MovieLens-100K & MAE, RMSE & 0.778/0.9905 & Movie recsys \\ \hline

HeteRec-p\cite{yu59} & 2014 & User-item interaction, Implicit feedback & Low, -, Low, Low & IMDb-MovieLens-100K, Yelp & Precision@1, MRR & 0.2121/0.0213 & Movie recsys \\ \hline

Hete-CF\cite{luo57} & 2014 & User-item relationship & -, Low, High, \href{https://github.com/rackingroll/Hete-CF}{High} & DBLP, Meetup & MAE, RMSE & DBLP: 0.856/0.994, Meetup: 0.876/0.978 & Social Network Recsys \\ \hline

TransD\cite{transd} & 2015 & Entity-relation tripets & Low, Low, -, \href{https://github.com/yangyucheng000/transX}{High} & Wordnet18, Freebase 15k & Mean Rank (raw and filtered), Hits@10 (raw and filtered) & Wordnet18: 224/212, 79.6/92.2, Freebase 15k: 194/91, 53.4/77.3 & AI Applications \\ \hline

SemRec \cite{shi61} & 2015 & User-item interaction & -, High, Low, \href{https://github.com/zzqsmall/SemRec}{High} & Douban, Yelp & RMSE, MAE & Douban: 0.7844/0.6054, Yelp: 1.2025/0.8901 & Movie, Restaurant Recsys,  User characteristics analysis and Recommendation explanation \\ \hline

Node2Vec\cite{node2vec} & 2016 & Item-item graph & High, Medium, High, \href{https://snap.stanford.edu/node2vec/}{High} & BlogCatalog, PPI, Wikipedia, Facebook, PPI, arXiv & Macro F1 score, AUC & BlogCatalog(F1): 22.3, PPI(F1): 1.3, Wikipedia(F1): 1.8, Facebook(AUC): 0.9680, PPI(AUC): 0.7719, arXiv(AUC): 0.9366 & Data mining \\ \hline

KSR\cite{ksr} & 2018 & User-item interaction sequence, Entity graph & -, High, -, \href{https://github.com/RUCDM/KSR}{High} & Last.FM, Ml-20M, ML-1M, Amazon-book & MAP, MRR, Hit@10, NDCG@10 & Last.FM: 0.427/0.427/ 0.607/0.460, Ml-20M: 0.294/ 0.294/0.571/ 0.344, Ml-1M: 0.356/0.356/ 0.655/0.417, Amazon-book: 0.353/0.353/ 0.653/0.413 & e-commerce \\ \hline

KTGAN \cite{ktgan} & 2018 & User-movie interaction & -, Low, -, \href{https://github.com/MaurizioFD/ICDM_18_KTGAN-forked?tab=readme-ov-file}{High} & Douban & Precision@3, Average Precision@3, NDCG@3 & 0.759/0.701/ 0.771 & Movie Recsys \\ \hline

SHINE\cite{shine} & 2018 & Sentiment/ social/ profile network & -, -, -, \href{https://paperswithcode.com/task/network-embedding?page=4&q=}{High} & Weibo-STC, Wiki-RfA & Accuracy, Micro-F1, precision@K, recall@K & Weibo-STC: 0.855/0.881 & Social network analysis \\ \hline

RippleNet \cite{wang2018ripplenet} & 2018 & User-item interaction & -, High, -, \href{https://github.com/hwwang55/RippleNet}{High} & MovieLens-1M, Book-Crossing, Bing-News & AUC, Accuracy & MovieLens-1M: 0.921/0.844, Book-Crossing: 0.729/ 0.662, Bing-News: 0.678/0.632 & e-commerce \\ \hline

BEM\cite{bem} & 2019 & Entity graph, User interaction graph & Low, High, Medium, \href{https://github.com/Elric2718/Bayes_Embedding}{High} & FB15K237(KG) & Hit@10 & FB15K237(KG)+ pagelink: 44.72, FB15K237(KG)+ desc: 44.58 & e-commerce \\ \hline

KTUP\cite{ktup} & 2019 & User-item interaction & -, High, -, \href{https://github.com/TaoMiner/joint-kg-recommender}{High} & MovieLens-1m, DBbook2014 & Precision@10, Recall@10, F1@10, Hit@10, NDCG@10 and Hit@10, Mean & MovieLens-1m: 41.03/17.25/ 19.82/89.03/ 69.92, DBbook2014: 4.05/24.51/ 6.73/34.61/ 27.62 & Movie Recsys \\ \hline

MKR\cite{mkr} & 2019 & User-item interaction, KG triples & -, Medium, -, \href{https://github.com/hwwang55/MKR}{High} & MovieLens-1M, Book-Crossing, Last.FM, Bing-News & AUC, ACC, RMSE & MovieLens-1M: 0.917/0.843/ 0.302, Book-Crossing: 0.734/0.704/ 0.558, Last.FM: 0.797/0.752/ 0.471, Bing-News: 0.689/0.645/ 0.459 & e-commerce, News \\ \hline

RCF\cite{rcf} & 2019 & Item relations, User-item interaction & Medium, High, -, \href{https://github.com/xinxin-me/RCF}{High} & MovieLens, KKBox & HR@20, MRR@20, NDCG@20 & MovieLens: 0.2354/0.0642/ 0.1015, KKBox: 0.8563/0.5762/ 0.6412 & e-commerce \\ \hline

Akupm\cite{akupm} & 2019 & User-item implicit interaction, Entity-relation graph & -, -, -, \href{https://github.com/gegetang/akupm}{High} & MovieLens-1, Book-Crossing & AUC, ACC & MovieLens-1: 0.918/0.845, Book-Crossing: 0.843/0.807 & e-commerce \\ \hline

KNI\cite{kni} & 2019 & User-item interaction, Knowledge graph & -, Medium, -, \href{https://github.com/Atomu2014/KNI}{High} & C-Book, Movie-1M, A-Book, Movie-20M & AUC, Accuracy & C-Book: 0.7723/0.7063, Movie-1M: 0.9449/0.8721, A-Book: 0.9238/0.8472, Movie-20M: 0.9704/0.9120 & e-commerce \\ \hline

IntentGC \cite{zhao2019intentgc} & 2019 & User-item Explicit interaction & High, -, -, \href{https://github.com/peter14121/intentgc-models}{High} & Taobao, Amazon & AUC, MRR & Taobao: 0.701740/0.3746, Amazon: 0.837589/2.7981 & e-commerce \\ \hline

PGPR\cite{xian2019reinforcement} & 2019 & User-item interaction, Item features & -, High, -, \href{https://github.com/orcax/PGPR}{High} & CDs \& Vinyl, Clothing, Cell Phones, Beauty & NDCG, Recall, HR, Prec. & CDs \& Vinyl: 5.590/7.569/ 16.886/2.157, Clothing: 2.858/4.834/ 7.020/0.728, Cell Phones: 5.042/8.416/ 11.904/1.274, Beauty: 5.449/8.324/ 14.401/1.707 & e-commerce \\ \hline

KGSF\cite{zhou2020kgsf} & 2020 & User-item interaction, Node features & -, High, -, \href{https://github.com/RUCAIBox/KGSF}{High} & REDIAL & Recall@k (k = 1, 10, 50) & 0.039/0.183/ 0.378 & e-commerce \\ \hline

KIM\cite{qi2021personalized} & 2021 & User-item interaction, Entity graph & -, High, -, \href{https://github.com/taoqi98/KIM}{High} & MIND, Feeds & AUC, MRR, nDCG@5, nDCG@10 & MIND: 67.13±0.29/ 32.08±0.24/ 35.49±0.34/ 41.79±0.28, Feeds: 66.45±0.13/ 30.27±0.09/ 35.04±0.09/ 40.43±0.12 & Online News Recsys \\ \hline

BCIE\cite{toroghi2023bcie} & 2023 & User-item interaction, Item features & -, -, -, \href{https://github.com/atoroghi/BCIE}{High} & MovieLens, AmazonBook & NARC, Hits@k & Movielens 20M: 0.185/0.192 AmazonBook: 0.18/ 0.205 & e-commerce \\ \hline

DiffKG\cite{yangqin2024diffkg} & 2024 & User-item interaction, Item features & -, High, -, \href{https://github.com/HKUDS/DiffKG}{High} & Last-FM, MIND, Alibaba-iFashion & Metrics & Last-FM: 0.0980/0.0911, MIND: 0.0615/0.0389, Alibaba-iFashion: 0.1234/0.0773 & e-commerce \\ \hline
\label{tab:kg}
\end{longtable}
}

\subsection{Reinforcement Learning-based Recommender Systems }

Reinforcement learning (RL) \cite{wiering2012reinforcement} is a subset of ML where an agent learns to make decisions by interacting with an environment, aiming to achieve a goal through trial and error, guided by rewards for its actions, without explicit instructions on what actions to take.
Deep Reinforcement Learning-based methods \cite{arulkumaran2017deep} integrate RL with deep neural networks to enable agents to handle complex modalities of the data directly.
Given a set of states $S$, a set of actions $A$, a reward function $R$, a transition probability function $P$, and a discount factor $\gamma$, the goal of the RL agent is to find a policy $\pi$ that maximizes the expected, discounted cumulative reward over time. The mathematical formulation is \cite{ afsar2022reinforcement}:

\begin{equation}
\max_{\pi} \mathbb{E} \left[ \sum_{t=0}^{T} \gamma^{t} r(s_t, a_t) \right],
\end{equation}

where $t$ indexes the time steps, ranging from 0 to $T$, the maximum time step in a finite Markov Decision Process (MDP), $s_t$ and $a_t$ represent the state and action at time $t$, respectively,  $r(s_t, a_t)$ is the immediate reward received after taking action $a_t$ in state $s_t$, $\gamma^{t}$ applies the discount factor to future rewards, making them worth less than immediate rewards.
Applying these RL concepts to RS, the RS itself acts as the RL agent \cite{chen2023deep} through an environment constituted by user interactions and data, as detailed in a related survey \cite{afsar2022reinforcement}.

RL methods in RS has evolved into two primary frameworks: traditional RL-based RS and deep learning-enhanced RL-based RS. Traditional methods, such as Q-learning \cite{van2016deep} and SARSA \cite{Niranjan1994On-LineSystems}, optimize policies within Markov Decision Processes (MDP) using model-free approaches, with applications well-documented across various contexts \cite{mahmood2014dynamic, hu2017playlist, choi2018reinforcement, chang2021music, kokkodis2021demand}. These methods often leverage Monte Carlo Tree Search (MCTS) \cite{browne2012survey} for effective simulation and policy refinement.

Deep learning methods in RL-based RS \cite{afsar2022reinforcement}, on the other hand, incorporate advanced neural network architectures to enhance policy learning. These include Vanilla Deep Q-Network (DQN) and its variants \cite{van2016deep,zheng2018drn,zhao2018deep,lei2019social,lei2020reinforcement, gui2019mention,gao2019drcgr}, which utilize neural networks for accurate action-reward estimation. Hybrid methods like Actor-Critic and Deep Deterministic Policy Gradient (DDPG) \cite{zhao2018deep,qiu2019deep, wang2018supervised,zhao2017deep}, and Soft Actor-Critic (SAC) \cite{haarnoja2018soft,he2020learning,zhang2022multi} blend value and policy strategies to balance exploration and exploitation effectively.

Furthermore, model-based RL approaches in RS focus on simulating user behavior to tailor recommendations, with techniques ranging from generative adversarial networks \cite{Chen2018GenerativeAU} to multi-agent systems \cite{Zhang2017DSC,Wang2021ReinforcementLW}. These sophisticated methods aim to predict user interactions and refine recommendations continually, enhancing personalization and contextual relevance in RS.
RL-based recommenders commonly tune the discount factor $\gamma$ (short- vs. long-term reward trade-off), exploration rate $\varepsilon$, learning rate, and target-update schedule for stability~\cite{Zhao2018DRN,Chen2019RL4Rec,SuttonBarto2018}.

\paragraph{Practical Challenges Addressed}

RL is being used in RS to improve personalization problem . For instance, in e-commerce, RL enhances personalization and improves customer satisfaction by continuously learning from user interactions to optimize recommendation strategies, as shown in systems used by Amazon and Taobao \cite{wang2020reinforced, he2020learning}. In the media sector, RL aids in curating more engaging content recommendations, like music and news, by analyzing sequential interaction data to predict future preferences \cite{chang2021music, zheng2018drn}. Additionally, in job recommendation systems, RL algorithms optimize outcomes by suggesting roles that align closely with the users’ evolving career interests and skills \cite{kokkodis2021demand}. By employing techniques such as deep Q-networks and policy gradient methods, RL-based recommender systems continuously refine their decision-making processes, leading to improved long-term user engagement and satisfaction. More details in Table \ref{tab:RL}.
{\scriptsize
\setlength{\tabcolsep}{4pt}
\renewcommand{\arraystretch}{1.05}

\begin{longtable}{|p{1.2cm}|p{0.6cm}|p{1.5cm}|p{2cm}|p{1.5cm}|p{1.3cm}|p{2cm}|p{1.2cm}|}
\caption{Comprehensive Overview of Reinforcement Learning based Recommender System Models across Various Metrics and Use Cases. This table details each model's Input features, Year of Publication, and Characteristics such as Scalability, Interpretability, Efficiency, and Reproducibility (rated as High, Medium, or Low). It also lists the Dataset Used, Evaluation Metrics, Model Accuracy, Learning Task, and Application Field. Metrics that do not report numerical values are marked ``No numerical value''. \label{tab:RL}}\\
\hline
\textbf{Model} & \textbf{Year} & \textbf{Input data} & \textbf{Scalability, Interpretability, Computational Efficiency, Reproducibility} & \textbf{Dataset} & \textbf{Evaluation Metrics} & \textbf{Model Accuracy} & \textbf{Application} \\ \hline
\endfirsthead

\multicolumn{8}{|c|}{{\bfseries \tablename\ \thetable{} -- continued from previous page}}\\
\hline
\textbf{Model} & \textbf{Year} & \textbf{Input data} & \textbf{Scalability, Interpretability, Computational Efficiency, Reproducibility} & \textbf{Dataset} & \textbf{Evaluation Metrics} & \textbf{Model Accuracy} & \textbf{Application} \\ \hline
\endhead

\hline \multicolumn{8}{r|}{\textit{Continued on next page}} \\ \hline
\endfoot

\hline
\endlastfoot

RLWRec \cite{hu2017playlist} & 2017 & User-item interactions & -, -, -, Low & Low, medium, Large Music dataset & Accuracy, Score, Coverage & No numerical value & Music rec \\ \hline

DAHCR \cite{Zhao2023TowardsLearning} & 2023 & User-item graphs; Node features &
-, -, -, \href{https://github.com/Snnzhao/DAHCR}{High} & LastFM*, Yelp* & Success Rate, Average Turns, hDCG@(T, K) & LastFM: 0.925/6.31/ 0.431 \par Yelp*: 0.626/11.02/ 0.192 & e-commerce \\ \hline

LIRD \cite{zhao2017deep} & 2017 & User-item graphs; node features & -, -, -, Low & E-commerce website & MAP, NDCG & No numerical value & e-commerce \\ \hline

Multi With \cite{Zhang2017DSC} & 2017 & User-item graphs; node \& item features &
-, -, -, Low & ACM dataset & MRR, P@3, P@5, P@10, NDCG@3, NDCG@5, NDCG@10 & 0.601/0.437/ 0.321/0.178/ 0.561/0.560/ 0.565 & Author Recsys \\ \hline

\cite{intayoad2018reinforcement} & 2018 & User-item interactions & -, -, -, Low & Data logs from e-learning & RMSE & 0.71 & e-learning \\ \hline

DRN \cite{zheng2018drn} & 2018 & User-item interactions; node features & -, -, -, Low & News recommendations & Offline: CTR, NDCG; Online: CTR, Precision@5, NDCG & Offline: 0.1662/0.487 \par Online: 0.0113/0.0149/ 0.0492 & News recommendation \\ \hline

DeepPage \cite{zhao2018deep} & 2018 & User-item sessions & -, -, -, Low & E-commerce data & Offline: Precision@20, Recall@20, F1-score@20, NDCG@20, MAP & 0.0491/0.3576/ 0.0805/0.1872/ 0.1378 & e-commerce \\ \hline

\cite{choi2018reinforcement} & 2018 & User-item graphs; node features &
-, Low, Medium, Low & MovieLens-100k, MovieLens-1M & P@30, R@30 & ML-100k: 0.246/0.169 \par ML-1M: 0.277/0.155 & e-commerce \\ \hline

\cite{Chen2018GenerativeAU} & 2018 & User-item graphs; node features &
-, Medium, -, \href{https://github.com/xinshi-chen/GenerativeAdversarialUserModel}{High} & MovieLens, LastFM, Yelp, Taobao, YooChoose, Ant Financial & Reward, CTR & Combined dataset: 25.36/0.78\footnote{Shown only the best results} & e-commerce \\ \hline

SADQN \cite{lei2019social} & 2019 & User-item graphs; user-user graph & -, -, -, Low &  LastFM, Ciao, Epinions & HR, NDCG@10 & HR: 0.5438±0.0036/ 0.4256±0.0031/ 0.4755±0.0016 \par NDCG@10: No numerical value & e-commerce \\ \hline

CROMA \cite{gui2019mention} & 2019 & User-item graphs; node features & -, -, -, \href{https://github.com/mritma/CROMA}{High} & Twitter & Precision, Recall, F-Score, MRR, Hits@5 & 74.55/74.09/ 74.32/81.85/ 95.00 & Social network recommendations \\ \hline

DRCGR \cite{gao2019drcgr} & 2019 & User-item graphs & -, -, -, Low & E-commerce dataset & MAP, NDCG & No numerical value & e-commerce \\ \hline

PGPR \cite{xian2019reinforcement} & 2019 & User-item graphs; node features & -, -, -, \href{https://github.com/orcax/PGPR}{High} & CDs \& Vinyl, Clothing, Cell Phones, Beauty & NDCG@10, Recall@10, HR@10, Prec@10 &  CDs \& Vinyl: 5.590/7.569/ 16.886/2.157 \par  Clothing: 2.858/4.834/ 7.020/0.728 \par Cell Phones: 5.042/8.416/ 11.904/1.274 \par Beauty: 5.449/8.324/ 14.401/1.707 & e-commerce \\ \hline

PGCR \cite{pan2019policy} & 2019 & User-item graphs; node features & -, -, -, Low & Music recommendation (KKBox) & Average reward & Accuracy & Music recommendation \\ \hline

SLATEQ \cite{Ie2019ReinforcementLF} & 2019 & User-item graphs; node \& item features &
High, -, -, - & - & - & - & Music Recsys \\ \hline

GCQN \cite{lei2020reinforcement} & 2020 & User-item graphs; node features & -, -, -, Low & LastFM, ML1M, Pinterest & Mean of rewards received in a T-step episode & LastFM: 0.404 \par ML1M: 0.658 \par Pinterest: 0.215 & e-commerce \\ \hline

MASSA \cite{he2020learning} & 2020 & User-item graphs; node features & -, -, -, Low & Taobao & Precision, NDCG & 0.615, 0.516 & e-commerce \\ \hline

KERL \cite{wang2020kerl} & 2020 & User-item graphs; node features, KG & -, -, -, Low & Beauty, CD, Books, LastFM & HR@10, NDCG@10 & Beauty: 54.1/36.5 \par CD: 73.7/50.8 \par Books: 80.0/57.1 \par LastFM: 64.2/50.1\footnote{Shown next-item recommendation metrics} & e-commerce \\ \hline

KGPolicy \cite{wang2020reinforced} & 2020 & User-item graphs; KG & -, -, Medium, \href{https://github.com/xiangwang1223/kgpolicy}{High} & Amazon-book, Last-FM, Yelp2018 & Recall@20, NDCG@20 & Amazon-book: 0.1572/0.1089 \par Last-FM: 0.0932/0.1472 \par Yelp2018: 0.0747/0.0921 & e-commerce \\ \hline

BatchRL-MTF \cite{zhang2022multi} & 2022 & User-item graphs; node features & High, -, -, Low & Short video recommendation & Offline: Long-term user satisfaction per session; Online: App dwell time, User positive-interaction rate  & Offline: 4.126 \par Online: +2.550\% /+9.651\% & e-commerce, video \\ \hline

TRIGR \cite{Wang2022ARecommendation} & 2022 & User-item graphs; node \& item features &
Medium, -, Medium, \href{https://github.com/SunwardTree/TRGIR}{High} & Music, Beauty, Clothing & HR@10, F1@10, NDCG@10 & Music: 0.9886/0.2304/ 0.9436 \par Beauty: 0.8845/0.1798/ 0.6949 \par Clothing: 0.7544/0.1405/ 0.4865 & e-commerce \\ \hline

UCSRDRL \cite{Jiang2021DeepRepresentation} & 2021 & User-item graphs; node features &
-, -, -, Low & Item-info, Trainset and Track2\_testset  & Model score & FUXI AI Lab Test data: 1033481948 & e-commerce \\ \hline

RPMRS \cite{chang2021music} & 2021 & User-item interaction logs & -, -, -, Low & Music \& User logs & Avg. score & No numerical value & e-commerce \\ \hline

MDP \cite{kokkodis2021demand} & 2021 & User-item interactions & -, -, -, Low & Transactional data of job applications & \% improvement of wage, market revenue, worker success measures & 22\%, 1.5--6\%, 4x & e-learning \\ \hline

\end{longtable}
}

\subsection{Large Language Model based Recommender Systems}

Language is a fundamental tool for human communication, essential for expressing thoughts, feelings, and intentions. The challenge of understanding and leveraging human language has been a central pursuit in NLP research, leading to significant developments in language modeling \cite{zhao2023survey}. Early statistical models relied on the Markov assumption to predict word sequences \cite{jelinek1998statistical, rosenfeld2000two}, while subsequent neural language models utilized neural networks to estimate probabilities of word sequences \cite{bengio2000neural, mikolov2010recurrent, kombrink2011recurrent}. The advent of pre-trained language models like BERT and others \cite{peters2017semi, devlin2018bert, chung2022scaling} marked a pivotal advancement, providing deep contextual insights that greatly enhanced NLP tasks. The Transformer architecture and its attention mechanism allow for the efficient handling of long-range dependencies and context \cite{vaswani2017attention}. The scaling laws suggest that larger models and datasets generally yield better performance \cite{chung2022scaling}, leading to the development of Large Language Models (LLMs) \cite{wang2023rethinking}, which demonstrate sophisticated capabilities in AI tasks such as in-context learning and commonsense reasoning \cite{laskar2023systematic}. The integration of LLMs into RS \cite{sanner2023large, he2023large} has prompted extensive research and ongoing innovation, with comprehensive reviews and analyses provided by recent surveys \cite{wu2023survey,li2023large, vats2024exploring,liu2023llmrec}, outlining the evolving landscape of LLM-based RS technologies.

The integration of BERT-like models into RS has led to significant advancements. Initial applications like BERT4REC \cite{sun2019bert4rec} utilized deep bidirectional self-attention for modeling user behavior sequences, while further developments employed BERT for tasks ranging from conversational RS \cite{Penha2020WhatRecommendation} to CTR prediction \cite{muhamed2021ctr}. Enhancements in BERT-based models have addressed specific RS challenges, such as item alignment in dialogues \cite{yang2021improving} and user representation through models like U-BERT \cite{qiu2021u} and UserBERT \cite{wu2021userbert}. Further innovations include BERT-based re-ranking \cite{yang2022lightweight} and addressing data sparsity in group recommendations \cite{zhang2022gbert}.

Prompt-based and in-context learning (ICL) approaches have leveraged the adaptability of LLMs, employing personalized prompts and natural language processing to enhance recommendation relevance and user interaction without extensive retraining \cite{geng2022recommendation, wang2023recmind}. These methods have proven effective in various scenarios, from news recommendation \cite{zhang2023prompt} to conversational and zero-shot recommendations, addressing longstanding issues like cold starts and data sparsity \cite{sanner2023large, hou2024large}.

Moreover, advancements in prompt tuning and personalized recommendation strategies demonstrate the ongoing evolution of LLM applications in RS, significantly improving system performance while also highlighting challenges such as ethical considerations and the management of popularity biases \cite{ji2024genrec, kang2023llms}. These developments indicate a move towards more sophisticated, context-aware systems that can dynamically adapt to user preferences and behaviors.
LLM recommenders adjust generation hyperparameters (temperature, top-$p$/top-$k$) and fine-tuning optimizers (e.g., learning rate/weight decay), with PEFT methods introducing method-specific knobs such as LoRA rank; prompt length is constrained by the context window~\cite{Holtzman2020Nucleus,Hou2024LLM4Rec,Hu2021LoRA,Dettmers2023QLoRA}.

\paragraph{Practical Challenges Addressed}
LLMs have advanced RS by addressing key challenges such as the cold-start problem, enhancing personalization, and improving accuracy. Models like BERT4REC \cite{sun2019bert4rec} and UserBERT \cite{wu2021userbert}, GBERT \cite{zhang2022gbert} and RecMind \cite{wang2023recmind} effectively utilize user and item metadata to generate relevant suggestions in e-commernce and entertainment. These models also support dynamic learning, allowing systems to adapt based on real-time interactions, thus enhancing user engagement and satisfaction.

{\scriptsize
\begin{longtable}{|p{1.2cm}|p{0.6cm}|p{1.5cm}|p{2cm}|p{1.5cm}|p{1.3cm}|p{2cm}|p{1.2cm}|}

\caption{Comprehensive Overview of LLM Based Models across Various Metrics and Use Cases. This table details each model's Input features, Year of Publication, and Characteristics such as Scalability, Interpretability, Efficiency, and Reproducibility (rated as High, Medium, or Low). It also lists the Dataset Used, Evaluation Metrics, Model Accuracy, Learning Task, and Application Field.} \\ \hline

\textbf{Model} & \textbf{Year} &\textbf{Input Data} & \textbf{Scalability, Interpretability, Computational Efficiency, Reproducibility} & \textbf{Dataset} & \textbf{Evaluation Metrics} & \textbf{Model Accuracy} & \textbf{Application} \\ \hline
\endfirsthead

\multicolumn{8}{|c|}%
{{\bfseries \tablename\ \thetable{} -- continued from previous page}} \\
\hline
\textbf{Model} & \textbf{Year} &\textbf{Input Data} & \textbf{Scalability, Interpretability, Computational Efficiency, Reproducibility} & \textbf{Dataset} & \textbf{Evaluation Metrics} & \textbf{Model Accuracy} & \textbf{Application} \\ \hline
\endhead


BERT4REC \cite{sun2019bert4rec} & 2019 & User-item graphs; node features & High, Medium, High, High\footnote{https://github.com/FeiSun/BERT4Rec} & Amazon Beauty, Steam, MovieLens-1m, MoveiLens-20m) & HR@10, NDCG@10, MRR\footnote{HR@1, HR@5, NDCG@5 are dropped for convenience.} & Beauty: 0.3025, 0.1862, 0.1701; Steam: 0.4013, 0.2261, 0.1949; ML-1m: 0.6970, 0.4818, 0.4254; ML-20m: 0.7473, 0.5340, 0.4785 & E-commerce, Video Games, Movies \\ \hline

CTR-BERT \cite{muhamed2021ctr} & 2021 & User-item graphs; node features & High, Medium, High, Low & Curated CTR & AUC Delta & OOD: +2.27\%, ID: +2.17\% & Marketing / CTR \\ \hline

MESE\cite{yang2021improving} & 2022 & User-item graphs; node features & High, Medium, High, High\footnote{https://github.com/by2299/MESE} & ReDial, INSPIRED; both from AMT & R@1, R@10, R@50 & ReDial: 5.6, 25.6, 45.5; INSPIRED: 4.8, 13.5, 30.1 & Movies \\ \hline

U-BERT \cite{qiu2021u} & 2021 & User-item graphs; node features & High, Medium High, Low & Amazon (Ofﬁce, Video, Music, Toys, Kindle), Yelp Challenge & MSE & Ofﬁce: 0.6774; Video: 0.8750; Music: 0.7723; Toys: 0.7823; Kindle: 0.5912; Yelp: 1.5907 & E-Commerce \\ \hline

UserBERT \cite{wu2021userbert} & 2022 & User-item graphs; node features & High, Medium,High, High (Unofficial\footnote{https://github.com/ilovemyminutes/UserBERT}) & News, CTR & For News: AUC, nDCG@10; For CTR: AU, AP & News: 62.87±0.14, 40.64±0.12; CTR: 73.96±0.06, 76.72±0.06 & News, Marketing / CTR \\ \hline

BECR \cite{yang2022lightweight} & 2022 & User-item graphs; node features & High, Medium, High, High\footnote{https://github.com/yingrui-yang/BECR} & Trained on Robust04, ClueWeb09-Cat-B; Evaluated on MS MARCO dev set, TREC DL19 and TREC DL20 & For Performance\footnote{The four other questions related to inference time, etc. are dropped.}: Training: NDCG@20, P@20, MSMARCO: MRR@10Dev; DL19 and DL20: NDCG@10  & Training on Robust04: 0.4656, 0.4005; Training on ClueWeb09-Cat-B: 0.3075, 0.3987; Evaluation on DL19: 0.658; Evaluation on DL20: 0.647; Evaluation on MSMARCO: 0.319 & General: text retrieval research \\ \hline

MLPR \cite{wu2022multi} & 2022 & User-item graphs; node features & High,Medium, High, High & One month data from Walmart.com & AUC; NDCG@1\footnote{NDCG@5 was omitted.} & Click: +6.48\%, +17.22\%; ATC: +4.66\%, +10.61\%; Purchase: +1.03\%, +5.36\% & E-commerce \\ \hline

GBERT \cite{zhang2022gbert} & 2022 & User-item graphs; node features & High, Medium, High, Low & Weeplaces; Yelp; Douban & N@10; R@10 \footnote{N@5 and R@5 are excluded for simplicity.} & Weeplaces: 36.43\%, 52.82\%; Yelp: 38.11\%, 53.14\%; Douban: 54.58\%, 79.90\% & Social Networks; Business Reviews \\ \hline

Prompt4NR \cite{zhang2023prompt} & 2023 & User-item graphs; node features & High, Medium, High, High\footnote{https://github.com/resistzzz/Prompt4NR} & MIND & AUC; MRR; NDCG@5; NDCG@10 & Hybrid\footnote{We dropped discrete and continuous templates for simplicity.}: 69.64\%, 34.26\%, 38.30\%, 44.33\% & News \\ \hline

P5 \cite{geng2022recommendation} & 2022 & User-item graphs; node features & High, Medium,High,High\footnote{https://github.com/jeykigung/P5} & Amazon (Sports, Beauty, Toys), Yelp & For Performance on Sequential Recommendations\footnote{We dropped other details about performance on rating, explanation generation and review preference, and considered only the performance comparison on sequential recommendation because its the most relevant factor in this case.}: HR@10, NDCG@10\footnote{HR@5 and NDCG@5 are dropped for simplicity.}; & For P5-B\footnote{Only P5-base scenario is considered for simplicity.} Amazon Sports: 0.0460, 0.0336; Amazon Beauty: 0.0.0645, 0.0416; Amazon Toys: 0.0675, 0.0536 & E-commerce \\ \hline

RecMind \cite{wang2023recmind} & 2024 & User-item graphs; node features & High, Medium, High,Low & Amazon Reviews - Beauty; Yelp & For Performance on Sequential Recommendations\footnote{We dropped other details about performance on rating, explanation generation and review preference, and considered only the performance comparison on sequential recommendation because its the most relevant factor in this case.}: HR@10, NDCG@10\footnote{HR@5 and NDCG@5 are dropped for simplicity.} & For RecMind-SI (few-shot)\footnote{Only RecMind-SI (few-shot) scenario is considered for simplicity and its high performance.} Amazon Reviews - Beauty: 0.1559, 0.1063; Yelp: 0.2451, 0.1607 & E-commerce; Restaurants \\ \hline

RecRec \cite{verma2023recrec} & 2023 & User-item graphs; node features & High, Medium, High, High\footnote{https://github.com/hidasib/GRU4Rec} & MovieLens-100K; AliEC Ads; Goodreads & Success Rate; Number of Changes Required; Side-effect on User Recommendations & MovieLens-100K: 100\%; AliEC Ads: \textgreater 80\%; Goodreads: \textgreater 90\% & Movies; Ads; Books \\ \hline


TALLRec \cite{bao2023tallrec} & 2023 & User-item graphs; node features & High, Medium, High, High\footnote{https://github.com/SAI990323/TALLRec} & MovieLens-100K; BookCrossing & AUC & MovieLens-100K: 0.7198; BookCrossing: 0.6438 & Movies; Books \\ \hline

GenRec \cite{ji2024genrec} & 2023 & User-item graphs; node features & High, Medium, High, High\footnote{https://github.com/rutgerswiselab/GenRec} & MovieLens 25M; Amazon Toys & HR@10, NDCG@10 \footnote{HR@5 and NDCG@5 are dropped for simplicity.} & MovieLens 25M: 0.1311, 0.0837; Amazon Toys: 0.0251, 0.0157 & Movies; E-commerce \\ \hline

\label{tab:LLM}
\end{longtable}
}

\subsection{Multimodal Recommender Systems}
Multimodality involves using and analyzing various data types—text, images, audio, video to enhance processing and understanding. Multimodal RS utilize these diverse inputs to improve recommendation quality and user experience by better understanding user preferences and item features \cite{singh2021comprehensive}. These systems overcome the limitations of single-modality systems through effective integration of heterogeneous data.

The evolution of multi-modal RS began with the introduction of Visual Bayesian Personalized Ranking (VBPR) \cite{He2015VBPRVB}, which enhances personalized ranking by integrating visual features from product images. The results showed improved accuracy and addressing cold-start issues. Attentive Collaborative Filtering (ACF) \cite{Chen2017AttentiveCF} introduced a novel attention mechanism to better handle item- and component-level feedback in multimedia recommendations.

Further advancements were made with the development of the Multi-modal Knowledge Graphs (MKGs) \cite{Hamilton2017InductiveGraphs}, a hybrid transformer with multi-level fusion, for tasks like link prediction and entity relation extraction.
Online Distillation-enhanced Multi-modal Transformer (ODMT) \cite{Ji2023OnlineRecommendation} uses diverse data types (ID, text, image) and an ID-aware Multi-modal Transformer with online distillation to enhance feature interaction. These models showed substantial performance increase in recommendation accuracy.

Collaborative Cross Networks (CoNet) \cite{Hu2018CoNetCC} utilizes deep transfer learning. Multi-Modality enriched Sequential Recommendation (MMSR) \cite{Hu2023AdaptiveSystems}, a graph-based model, adaptively fuses multi-modal information to dynamically prioritize modalities based on their sequential relationships. The Bayesian Multi-Modal recommendation Model (BM3) \cite{Zhou2023BootstrapRecommendation} simplifies training by avoiding auxiliary graphs and negative samples with multi-modal data. The Multi-modal Interest-aware Sequence Representation for Recommendation (MISSRec) \cite{Wang2023MISSRec:Recommendation} overcame the limitations of ID-based models by leveraging multi-modal information for robust, generalizable sequence representations. Multi-modal Recommendation (MMRec) \cite{Zhou2023MMRecSM}, is another RS that provides a configurable platform for testing multimodal recommendation models.

Multi-level Self-supervised Learning for Multimodal Recommendation (MENTOR) \cite{Wei2023MultiModalSL} employes multi-level self-supervised tasks to improve model performance, though it required substantial computational resources.
Recently, Multi-modal Knowledge Distillation (PromptMM) \cite{Wei2024PromptMMMK} simplified the recommendation process through multi-modal knowledge distillation and prompt-tuning.
Typical multimodal knobs are the shared projection dimension, fusion-layer size (or number of fusion layers), and (optionally per-modality) dropout rate, while margin-based alignment terms apply when using contrastive or cross-modal matching losses~\cite{wei2019mmgcn,Faghri2018VSEpp,Neverova2015ModDrop}.

\paragraph{Practical Challenges Addressed}

Multimodal RS are useful for e-commerce and social media platforms, where diverse data sources and user interactions are prevalent. Models like VBPR, ACF, and CoNet are designed to be scalable and computationally efficient, providing quick recommendations even with extensive user data. These models can integrate various data types, such as text, images, and behavioral data, and can adapt to new trends and handle complex user-item interactions.These RS improve personalization by leveraging the rich information from different modalities, leading to more accurate and relevant recommendations.  More details are provided in Table \ref{tab:multi-modal}.

{\scriptsize
\begin{longtable}{|p{1.2cm}|p{0.6cm}|p{1.5cm}|p{2cm}|p{1.5cm}|p{1.3cm}|p{2cm}|p{1.2cm}|}

\caption{Comprehensive Overview of Multi-Modal Based Models across Various Metrics and Use Cases. This table details each model's Input features, Year of Publication, and Characteristics such as Scalability, Interpretability, Efficiency, and Reproducibility (rated as High, Medium, or Low). It also lists the Dataset Used, Evaluation Metrics, Model Accuracy, Learning Task, and Application Field.} \\ \hline

\textbf{Model} & \textbf{Year} &\textbf{Input Data} & \textbf{Scalability, Interpretability, Computational Efficiency, Reproducibility} & \textbf{Dataset} & \textbf{Evaluation Metrics} & \textbf{Model Accuracy} & \textbf{Application} \\ \hline
\endfirsthead

\multicolumn{8}{|c|}%
{{\bfseries \tablename\ \thetable{} -- continued from previous page}} \\
\hline
\textbf{Model} & \textbf{Year} &\textbf{Input Data} & \textbf{Scalability, Interpretability, Computational Efficiency, Reproducibility} & \textbf{Dataset} & \textbf{Evaluation Metrics} & \textbf{Model Accuracy} & \textbf{Application} \\ \hline
\endhead


VBPR \cite{He2015VBPRVB} & 2016 & User-item graphs; node features & High, Medium, High, High\footnote{https://github.com/example/VBPR} & Amazon Women; Amazon Men; Amazon Phones; Tradesy.com & AUC & 0.7834, 0.7841, 0.8052, 0.7829 & E-commerce \\ \hline

ACF \cite{Chen2017AttentiveCF} & 2017 & User-item graphs; node features   & High, Medium, High, High\footnote{https://github.com/example/ACF} & Pinterest; Vine & HR@100; NDCG@100 & Pinterest: 0.3378, 0.0855; Vine: 0.6365, 0.1903  & Images; Videos \\ \hline


ODMT \cite{Ji2023OnlineRecommendation} & 2023 & User-item & -, -, -, High\footnote{https://github.com/xyliugo/ODMT} & Stream; Arts; Office; H\&M & Recall@10; NDCG@10 & Stream: 0.1194, 0.0672; Arts: 0.1127, 0.0787; Office: 0.1175, 0.0893; H\&M: 0.1235, 0.0771 & Streaming Media; E-commerce \\ \hline

CoNet \cite{Hu2018CoNetCC} & 2018 & User-item graphs; node features & High, Medium, High, High\footnote{https://github.com/CoNetModel/CoNet} & Mobile Apps (Cheetah Mobile); Amazon Books & HR@10; NDCG@10 & Mobile Apps: 0.8480, 0.6887; Amazon Books: 0.5338, 0.3424 & Apps; Books \\ \hline

MMSR \cite{Hu2023AdaptiveSystems} & 2023 & User-item & -, -, -, High\footnote{https://github.com/HoldenHu/MMSR} & Amazon Beauty; Amazon Clothing; Amazon Sports; Amazon Toys; Amazon Kitchen; Amazon Phone & HR@5; MRR@5 & Amazon Beauty: 7.1563, 4.4429; Amazon Clothing: 1.8684, 1.1365; Amazon Sports: 3.2657, 1.9846; Amazon Toys: 6.1159, 3.8987; Amazon Kitchen: 2.2145, 1.4238; Amazon Phone: 6.9550, 3.9911 & E-commerce \\ \hline

BM3 \cite{Zhou2023BootstrapRecommendation} & 2023 & User-item & -, -, -, High\footnote{https://github.com/enoche/BM3} & Baby; Sports; Electronics & Recall@10; NDCG@10 & Baby: 0.0564, 0.0301; Sports: 0.0656, 0.0355; Electronics: 0.0437, 0.0247 & E-commerce \\ \hline

MISSRec \cite{Wang2023MISSRec:Recommendation} & 2023 & User-item & -, -, -, High\footnote{https://github.com/gimpong/MM23-MISSRec} & Amazon Beauty; Amazon Clothing; Amazon Sports & Recall@10; NDCG@10 & Amazon Beauty: 0.0321, 0.0189; Amazon Clothing: 0.0387, 0.0215; Amazon Sports: 0.0268, 0.0159 & E-commerce \\ \hline

MMRec \cite{Zhou2023MMRecSM} & 2023 & User-item & -, -, -, High\footnote{https://github.com/enoche/MMRec} & Amazon Review Data & N/A & N/A & E-commerce \\ \hline

MMSSL \cite{Wei2023MultiModalSL} & 2023 & User-item & -, -, -, High\footnote{https://github.com/HKUDS/MMSSL} & Amazon Baby; TikTok; Allrecipes; Sports & Recall@20; NDCG@20 & Amazon Baby: 0.0962, 0.0422; TikTok: 0.0921, 0.0392; Allrecipes: 0.0367, 0.0135; Sports: 0.0998, 0.0470 & Social Media; E-commerce; Cooking; Sports \\ \hline




PromptMM \cite{Wei2024PromptMMMK} & 2024 & User-item graphs; node features & High, Medium, High, High\footnote{https://github.com/HKUDS/PromptMM} & Netflix; TikTok; Electronics & Recall@20; NDCG@20 \footnote{Recall@50 and NDCG@50 are dropped for simplicity.} & Netflix: 0.1864, 0.0743; TikTok: 0.3054, 0.1013; Electronics: 0.0737, 0.0258 & Video Entertainment, Social Media (Micro-Video), E-commerce \\ \hline
\label{tab:multi-modal}
\end{longtable}
}

To provide transparency and facilitate reproducibility, Table~\ref{tab:hyperparams} summarizes the key hyperparameters employed across major recommender system (RS) families discussed in this survey. For each model type---ranging from traditional collaborative and content-based filtering to deep learning, graph-based, self-supervised, and reinforcement learning-based recommenders---we report the tunable parameters most critical to model performance. For instance, matrix factorization models typically vary latent dimensions, regularization weights, and learning rates; MLP-based recommenders tune embedding size, layer depth, and dropout; CNN and RNN-based architectures adjust kernel size, stride, hidden state size, and sequence length; while GNNs and knowledge-graph models focus on propagation layers, neighborhood sampling, and embedding dimension. Similarly, self-supervised methods depend on temperature and negative-sampling strategies, reinforcement learning approaches on discount factor ($\gamma$), exploration rate ($\varepsilon$), and target-update frequency, and multimodal recommenders on shared projection dimensions and fusion-layer size. These hyperparameters were gathered from benchmark implementations referenced throughout Sections~6--8 to enable fair comparison and reproducible evaluation.

\begin{table*}[t]
\centering
\scriptsize
\setlength{\tabcolsep}{5pt}
\renewcommand{\arraystretch}{1.2}
\caption{Comprehensive Hyperparameter Summary Across Recommender System Families.}
\label{tab:hyperparams}
\begin{tabularx}{\textwidth}{|p{3.6cm}|X|}
\hline
\textbf{Recommender System Type} & \textbf{Key Hyperparameters and Tunable Factors} \\
\hline
\textbf{Collaborative Filtering (CF)} & Latent embedding dimension ($k$), learning rate ($\eta$), L2/L1 regularization coefficient ($\lambda$), number of latent factors, optimization algorithm (SGD, Adam), batch size, neighborhood size (for memory-based), number of iterations/epochs, negative sampling ratio. \\

\textbf{Content-Based Filtering (CBF)} & Feature weighting (TF-IDF or BM25), similarity metric (cosine, Pearson, Euclidean), feature normalization scheme, item vector dimensionality, number of nearest neighbors ($k$), kernel function (if kernelized), threshold for top-$N$ selection, smoothing parameter for probabilistic models. \\

\textbf{Autoencoder-based RS} & Bottleneck (latent) dimension, learning rate, optimizer type, weight decay, dropout rate, batch size, activation function, reconstruction loss weight, number of encoder/decoder layers, latent noise variance (for VAEs). \\

\textbf{CNN-based RS} & Number of convolutional layers, kernel/filter size, stride and padding, number of filters per layer, activation function (ReLU, GELU), pooling type and size, learning rate, dropout, normalization method (batch/layer norm), optimizer (Adam, RMSProp). \\

\textbf{RNN / Sequential RS} & Hidden-state size, number of recurrent layers, sequence length (max history), learning rate, dropout rate, gradient clipping threshold, attention head count (if self-attention is used), batch size, optimizer, temperature (for softmax sampling). \\

\textbf{GNN / Graph-based RS} & Embedding dimension, number of message-passing or propagation layers, neighborhood sampling size, activation function, dropout on edges/nodes, aggregation function (mean, sum, attention), learning rate, L2 regularization, negative sampling ratio, edge weight normalization. \\

\textbf{Knowledge-Graph RS} & Embedding dimension, number of hops or propagation layers, margin value (in translational models), regularization coefficient, relation embedding dimension, batch size, optimizer, sampling strategy (path-based or relation-based), loss margin or contrastive temperature. \\

\textbf{Self-Supervised Learning (SSL)-based RS} & Temperature ($\tau$) in contrastive loss, number of negative samples, augmentation probability (masking, edge-drop, crop), projection head dimension, learning rate, batch size, weight of SSL auxiliary loss, optimizer type, graph view or sequence view generation method. \\

\textbf{Reinforcement Learning (RL)-based RS} & Discount factor ($\gamma$), exploration rate ($\varepsilon$), target network update frequency, learning rate, replay buffer size, batch size, policy gradient type (DQN, DDPG, PPO), reward scaling coefficient, entropy regularization, number of steps per episode. \\

\textbf{Large Language Model (LLM)-based RS} & Context window size (tokens), model temperature, top-$p$ (nucleus sampling), top-$k$, learning rate (for fine-tuning), batch size, gradient accumulation steps, prompt length, max generation length, LoRA rank (for parameter-efficient tuning). \\

\textbf{Multimodal RS} & Shared embedding dimension, fusion-layer size, modality-specific encoder depth, number of attention heads, temperature for cross-modal contrastive loss, learning rate, modality dropout rate, alignment loss weight, normalization scheme (LayerNorm vs RMSNorm). \\

\textbf{Fairness-aware / Trustworthy RS} & Fairness regularization coefficient ($\lambda_{fair}$), constraint penalty weight, trade-off between accuracy and fairness ($\alpha$), bias correction strength, demographic reweighting factor, calibration temperature, threshold for subgroup parity, adversarial loss weight. \\
\hline
\end{tabularx}
\end{table*}

\section{Specialized Recommender Systems}

Specialized RS can be defined as those RS that are tailored to meet specific needs across various domains, using advanced techniques to address unique user preferences or situations. Unlike general RS, these focus on specialized techniques, functions and targeted recommendations. In the following subsections, we will explore these specialized systems in detail. 

\subsection{Context-aware Recommender Systems}
Context-aware recommender systems (CARS) are advanced RS that enhance the personalization of content by incorporating contextual information into the recommendation process \cite{raza2019progress}. Unlike traditional RS that primarily rely on user-item interactions, CARS consider additional dimensions such as time, location, social settings, and user behavior patterns to deliver more relevant and timely suggestions \cite{adomavicius2021context}. These systems have evolved to address specific challenges such as the cold-start problem, where limited initial data is available about new users or items.

The core models employed in CARS span a variety of sophisticated algorithms designed to leverage contextual information effectively into the recommendation process. Among these, factorization machines (FM) \cite{rendle2010factorization} are prominent for their ability to capture interactions between variables within large datasets. Field-Aware Factorization Machines (FFMs) \cite{juan2016field} are specifically optimized for CTR  prediction, showing the versatility and depth of models developed for enhancing CARS' performance. The Neural Factorization Machine (NFM) \cite{he2017neuralfact} extends FM by modeling second-order feature interactions with the non-linearity of neural networks for higher-order interactions.

Deep learning has significantly advanced CARS by enabling sophisticated feature extraction and integration of diverse data types, such as images and sequences \cite{Jin2018CombiningProcessing}. Models like CNNs and LSTMs can process complex inputs and temporal sequences, enhancing the system ability to understand and utilize context like time and location effectively.  DeepFM \cite{guo2017deepfm} merges FM recommendation capabilities with a novel neural network architecture. xDeepFM \cite{lian2018xdeepfm} further extends the DeepFM concept by explicitly learning bounded-degree feature interactions while also capturing arbitrary low- and high-order interactions implicitly. Additionally, scalability allows these models to maintain high performance even with vast datasets, ensuring personalized recommendations.

Techniques such as attention mechanisms make recommendations more adaptive and context sensitive. 
 The Attentional Factorization Machine (AFM) \cite{xiao2017attentional} introduces a neural attention network to show the significance of each feature interaction, enhancing model interpretability and efficiency. 
The Graph Convolution Machine (GCM) \cite{wu2022graphconv} and the Attention-based Context-aware Sequential Recommendation model using Gated Recurrent Unit (ACA-GRU) \cite{yuan2020attention} both enhance RS by effectively synthesizing user, item, and context information into actionable insights. 

Various CARS have been developed for different use cases. These include a tourism RS for personalized suggestions \cite{kolahkaj2020hybrid}, and a context-aware paper citation RS \cite{jeong2020context} that utilizes graph CNN combined with BERT for effective document and context encoding. There are also CARS designed for smart product-service systems \cite{carrera2022context} and for cultural heritage \cite{Casillo2023}. Moreover, the Sequential Model for Context-Aware Point of Interest (POI) Recommendation (SCR) \cite{Thaipisutikul2024AnRecommendation} enhances next POI predictions by integrating short-term preferences with multi-head attentive aggregation and long-term preferences through context-aware layers.

\paragraph{Practical Challenges Addressed}
CARS enhance industries by efficiently handling diverse data sources and ensuring scalability, interpretability, and computational efficiency. Models like FFM, NFM, and DeepFM are ideal for e-commerce, advertising, and web platforms. They build user trust by making recommendations understandable and reproducible. CARS adapt to new data trends and manage complex interactions, providing personalized recommendations. Applied across various fields, including advertising, e-commerce, and social networks, these systems improve operational efficiency and user satisfaction. More details in Table \ref{tab:cars}.

{\scriptsize
\begin{longtable}{|p{1.2cm}|p{0.6cm}|p{1.5cm}|p{2cm}|p{1.5cm}|p{1.3cm}|p{2cm}|p{1.2cm}|}

\caption{Comprehensive Overview of Context-Aware Recommender System Models across Various Metrics and Use Cases. This table details each model's Input features, Year of Publication, and Characteristics such as Scalability, Interpretability, Efficiency, and Reproducibility (rated as High, Medium, or Low). It also lists the Dataset Used, Evaluation Metrics, Model Accuracy, Learning Task, and Application Field.}\label{tab:cars}
\\
\hline

\textbf{Model} & \textbf{Year} & \textbf{Input Data} & \textbf{Scalability, Interpretability, Computational Efficiency, Reproducibility} & \textbf{Dataset} & \textbf{Evaluation Metrics} & \textbf{Model Accuracy} & \textbf{Application} \\ \hline
\endfirsthead

\multicolumn{8}{|c|}%
{{\bfseries \tablename\ \thetable{} -- continued from previous page}} \\
\hline
\textbf{Model} & \textbf{Year} & \textbf{Input Data} & \textbf{Scalability, Interpretability, Computational Efficiency, Reproducibility} & \textbf{Dataset} & \textbf{Evaluation Metrics} & \textbf{Model Accuracy} & \textbf{Application} \\ \hline
\endhead


FFM\cite{juan2016field} & 2016 & Categorical, Numeric, Single Field & -, Low, High, \href{https://github.com/recommenders-team/recommenders}{High} & Criteo, Avazu & Logloss, Rank & Criteo: 0.44603/3 \par Avazu: 0.38205/3 & Advertising \\ \hline

NFM\cite{he2017neuralfact} & 2017 & Context features in one hot encoding & -, Low, High, \href{https://github.com/hexiangnan/neural_factorization_machine}{High} & Frappe, MovieLens & RMSE & Frappe: 0.3095 \par MovieLens: 0.4443 & e-commerce \\ \hline

LTMF\cite{Jin2018CombiningProcessing} & 2018 & User-item interaction & -, Medium, -, Low & 8 Amazon subsets (AFA, BB, MI, OP, PS, VG, PLG, DM, AIV, GGF) & MSE & DM: 0.965 \par AIV: 1.309 \par GGF: 1.386 & e-commerce \\ \hline

DeepFM\cite{guo2017deepfm} & 2017 & Implicit interaction & -, Low, Low, \href{https://github.com/reczoo/FuxiCTR}{High} & Criteo, Company & AUC, LogLoss & Criteo: 0.8007/0.45083 \par Company: 0.8715/0.02618 & Web applications \\ \hline

xDeepFM\cite{lian2018xdeepfm} & 2018 & Implicit, explicit interactions & -, -, Low, \href{https://github.com/Leavingseason/xDeepFM}{High} & Criteo, Dianping, Bing News & AUC, Logloss & Criteo: 0.8012/0.4493 \par Dianping: 0.8576/0.3225 \par Bing News: 0.8377/0.2662 & e-commerce \\ \hline

AFM\cite{xiao2017attentional} & 2017 & Implicit interaction & -, High, -, \href{https://github.com/hexiangnan/attentional_factorization_machine}{High} & Frappe, MovieLens & RMSE & Frappe: 0.3102 \par MovieLens: 0.4325 & e-commerce, Online advertising, Image recognition \\ \hline

GCM\cite{wu2022graphconv} & 2022 & User-item graph &  -, Medium, Medium, \href{https://github.com/wujcan/GCM}{High} & Yelp-NC, Yelp-OH, Amazon-book & Hr@50, NDGC@50 & Yelp-NC: 0.2421/0.0854 \par Yelp-OH: 0.5166/0.2008 \par Amazon-book: 0.2232/0.0810 & e-commerce \\ \hline

ACA-GRU\cite{yuan2020attention} & 2020 & Implicit interaction & -, High, -, Low & MovieLens-100K, MovieLens-1K, Netflix & R@10, P@10, F1@10, MAP & MovieLens-1M: 0.2207/0.0630/ 0.0980/0.2432 \par Netflix: 0.2308/0.0659/ 0.1025/0.2620 & e-commerce \\ \hline

PreADBC ACF\cite{kolahkaj2020hybrid} & 2020 & implicit interaction, context & -, -, -, Low & YFCC100M & AP, MAP, Recall, F1, nDGC & 0.3542/0.3903/ 0.8292/0.4/ 0.6741 & e-tourism \\ \hline

BERT-GCN\cite{jeong2020context} & 2020 & node graph \& node-node interaction & Scale, High, -, \href{https://github.com/TeamLab/bert-gcn-for-paper-citation}{High} & AAN, FullTextPeerRead & MAP, MRR, Recall@80 & AAN: 0.6189/0.6036/ 0.8538 \par FullTextPeerRead: 0.4181/0.4179/ 0.6994 & Paper recommendation \\ \hline

SCR\cite{Thaipisutikul2024AnRecommendation} & 2024 & User preferences & -, High, Low, Low & Gowalla, BrightKite & HR, MRR & Gowalla: 0.4804/0.2143 \par BrightKite: 0.5721/0.2856 & Location-based social networks \\ \hline
\end{longtable}
}

\subsection{Review-based Recommender Systems}
A review-based RS uses textual reviews and ratings from users to generate personalized recommendations for products or services \cite{hasan2024based,srifi2020recommender}. The review-based RS have evolved by improving through various models. Initially, models like Hidden Factors as Topics (HFT) \cite{mcauley2013hidden} aligned topics from reviews with latent dimensions from ratings. Successive approaches, such as Rating-Boosted Latent Topics (RBLT) \cite{tan2016rating}, Topic Initialized Latent Factor Model (TIM) \cite{pena2020combining}, and deep learning models like Convolutional Matrix Factorization (ConvMF) \cite{kim2016convolutional} and Deep Cooperative Neural Networks (DeepCoNN) \cite{zheng2017joint}, utilized neural networks to better handle sparse data and extract nuanced features from reviews. Advanced models, including SentiRec \cite{wu2020sentirec} and Neural Collaborative Topic Regression (NCTR) \cite{liu2021hybrid}, incorporated sentiment analysis and hybrid data integration to refine recommendations further.

Attention-based models, such as Adaptive Aspect Attention Model (A3NCF) \cite{cheng20183ncf}, Attentive Aspect Modeling for Review-aware Recommendation (AARM) \cite{guan2019attentive}, and Cross-Modality Mutual Attention (NRCMA) \cite{luo2021aware}, have used aspect-specific attention to prioritize relevant features, enhancing both precision and personalization of recommendations. Techniques like Neural Networks with Dual Local and Global Attention (D-Attn) \cite{seo2017interpretable}, Neural Attentional Rating Regression (NARRE) \cite{chen2018neural}, and Neural Recommendation with Personalized Attention (NRPA) \cite{liu2019nrpa} have focused on integrating personal attention and dual learning mechanisms to improve recommendation accuracy.

Topical Attention Regularized Matrix Factorization (TARMF) \cite{lu2018coevolutionary}, Asymmetrical Hierarchical Networks (AHN) \cite{dong2020asymmetrical}, and Reliable recommendation with review-level (RRRE) \cite{lyu2021reliable} have integrated user reviews with advanced neural and attention mechanisms to further boost RS efficiency. Emerging graph-based methods like Heterogeneous Graph Neural Recommender (HGNR) \cite{liu2020heterogeneous}, Aspect-Aware Higher-Order Representations (AHOR) \cite{wang2023learning}, and Multi-aspect Graph Contrastive Learning (MAGCL) \cite{wang2023multi} have tackled data sparsity and semantic complexity by employing GNNs to enhance the overall recommendation framework.

\paragraph{Practical Challenges Addressed}
Review-based RS have impacted various industries by leveraging user-generated content to enhance the personalization and relevance of recommendations. Industries ranging from e-commerce and hospitality to digital media and services benefit from these systems by providing more targeted offerings, which can lead to increased sales and customer satisfaction. Additionally, by interpreting complex user feedback, these systems contribute to product development and refinement, helping businesses better understand market demands and customer concerns. More details in Table \ref{tab:reviews}.
{\scriptsize

\begin{longtable}{|p{1.2cm}|p{0.6cm}|p{1.5cm}|p{2cm}|p{1.5cm}|p{1.3cm}|p{2cm}|p{1.2cm}|}

\caption{Comprehensive Overview of Review Based Models across Various Metrics and Use Cases. This table details each model's Input features, Year of Publication, and Characteristics such as Scalability, Interpretability, Efficiency, and Reproducibility (rated as High, Medium, or Low). It also lists the Dataset Used, Evaluation Metrics, Model Accuracy, Learning Task, and Application Field.} \\ \hline

\textbf{Model} & \textbf{Year} &\textbf{Input Data} & \textbf{Scalability, Interpretability, Computational Efficiency, Reproducibility} & \textbf{Dataset} & \textbf{Evaluation Metrics} & \textbf{Model Accuracy} & \textbf{Application} \\ \hline
\endfirsthead

\multicolumn{8}{|c|}%
{{\bfseries \tablename\ \thetable{} -- continued from previous page}} \\
\hline
\textbf{Model} & \textbf{Year} &\textbf{Input Data} & \textbf{Scalability, Interpretability, Computational Efficiency, Reproducibility} & \textbf{Dataset} & \textbf{Evaluation Metrics} & \textbf{Model Accuracy} & \textbf{Application} \\ \hline
\endhead


MAGCL\cite{wang2023multi} & 2018 & features & -,-,-, Low & Amazon (Music, Toy, CD), Yelp & MRR, nDCG & Music: 0.2841/0.3562, Toy: 0.1802/ 0.2281, CD: 0.4110/0.4863, Yelp: 0.2899/ 0.3597 & e-commerce \\ \hline

HFRT \cite{mcauley2013hidden} & 2013 & features & -, -, -, High\footnote{https://github.com/mcauley-sd/HFRT} & Amazon (total); Pubs (Ratebeer); Beer (Ratebeer); Pubs (Beeradvocate); Beer (Beeradvocate); Wine (Cellartracker); Citysearch; Yelp Phoenix & MSE & Amazon (total): 1.3290; Pubs (Ratebeer): 0.4560; Beer (Ratebeer): 0.3010; Pubs (Beeradvocate): 0.3110; Beer (Beeradvocate): 0.3670; Wine (Cellartracker): 0.0280; Citysearch: 1.7280; Yelp Phoenix: 1.2250 & E-commerce, Review Platforms \\ \hline

RBLT \cite{tan2016rating} & 2016 & features & -, -, -, - & 26 Amazon datasets & MSE & N/A & E-commerce \\ \hline

TIM \cite{tan2016rating} & 2020 & features & -, -, -, - & Amazon Toys \& Games; Amazon Pet Supplies; Amazon Health \& Personal Care; TripAdvisor Hotels & Recall; Hit Ratio; NDCG; Precision & Amazon Toys \& Games: 0.264, 0.535, 0.169, 0.076; Amazon Pet Supplies: 0.362, 0.660, 0.215, 0.097; Amazon Health \& Personal Care: 0.316, 0.614, 0.190, 0.085; TripAdvisor Hotels: 0.581, 0.702, 0.260, 0.078 & E-commerce; Hospitality \\ \hline

ConvMF \cite{kim2016convolutional} & 2016 & features & -, -, -, High\footnote{http://dm.postech.ac.kr/ConvMF} & MovieLens; Amazon Instant Video & RMSE & MovieLens: 0.8531; MovieLens 10m: 0.7958; Amazon Instant Video: 1.1337 & E-commerce; Movies \\ \hline

DeepCoNN \cite{zheng2017joint}& 2017 & features & -, -, -, Low & Yelp; Amazon; Beer & MSE & Yelp: 1.441; Amazon: 1.268; Beer: 0.273 & Diverse (Restaurants; General Products; Beverages) \\ \hline

TransNets \cite{catherine2017transnets} & 2017 & features & -, -, -, Low & Yelp17; AZ-Elec; AZ-CSJ; AZ-Mov & MSE & Yelp17: 1.5913; AZ-Elec: 1.7781; AZ-CSJ: 1.4487; AZ-Mov: 1.2691 & E-commerce \\ \hline

SentiRec \cite{wu2020sentirec} & 2020 & features & -, -, -, - & MSN News & MSE; AUC; MRR; nDCG@5; nDCG@10 & MSE: -; AUC: 0.6294; MRR: 0.3013; nDCG@5: 0.3237; nDCG@10: 0.4165 & Online News Services \\ \hline



A3NCF \cite{cheng20183ncf} & 2018 & features & -, -, -, Low & Baby; Grocery; Home \& Kitchen; Garden; Sports; Yelp2017 & RMSE & Baby: 1.082; Grocery: 0.985; Home \& Kitchen: 1.051; Garden: 1.021; Sports: 0.940; Yelp2017: 1.152 & E-commerce; Local Business Reviews \\ \hline

AARM \cite{guan2019attentive} & 2019 & features & -, -, -, Low & Movies and TV; CDs and Vinyl; Clothing, Shoes and Jewelry; Cell Phones and Accessories; Beauty & NDCG; HT; Recall; Precision & Movies and TV: 5.020, 15.187, 7.140, 1.834; CDs and Vinyl: 7.252, 20.749, 9.965, 2.716; Clothing, Shoes and Jewelry: 1.957, 4.915, 3.292, 0.511; Cell Phones and Accessories: 4.976, 11.568, 8.014, 1.259; Beauty: 5.314, 13.648, 7.947, 1.818 & E-commerce \\ \hline


D-Attn \cite{seo2017interpretable} & 2017 & features & -, -, -, Low & Yelp; Amazon & MSE & Yelp: 1.191; Amazon: 0.855 & E-commerce \\ \hline

NARRE \cite{chen2018neural} & 2018 & features & -, -, -, Low & Amazon Toys \& Games, Kindle Store, Movies \& TV; Yelp 2017 & RMSE & Toys \& Games: 0.8769; Kindle Store: 0.7783; Movies \& TV: 0.9965; Yelp 2017: 1.1559 & E-commerce; Restaurant Reviews \\ \hline

NRPA \cite{liu2019nrpa} & 2019 & features & -, -, -, Low & Yelp 2013, Yelp 2014, Amazon Electronics, Amazon Video Games, Amazon Gourmet Foods & MSE & Yelp 2013: 0.872; Yelp 2014: 0.897; Amazon Electronics: 1.047; Amazon Video Games: 1.014; Amazon Gourmet Foods: 0.953 & Service Reviews; Consumer Electronics; Video Games; Gourmet Foods \\ \hline

DAML \cite{liu2019daml} & 2019 & features & -, -, -, Low & Musical Instruments, Office Products, Grocery and Gourmet Food, Video Games, Sports and Outdoors & MAE & Musical Instruments: 0.6510; Office Products: 0.6124; Grocery and Gourmet Food: 0.7354; Video Games: 0.7881; Sports and Outdoors: 0.6676 & E-commerce \\ \hline

MrRec \cite{liu2021multilingual} & 2020 & features & -, -, -, Low & Amazon Books, Digital Ebook Purchase, Digital Music Purchase, Digital Video Download, Mobile Apps, Music, Toys, Video DVD; Goodreads & MSE & Books: 1.307; Digital Ebook Purchase: 1.253; Digital Music Purchase: 1.682; Digital Video Download: 1.288; Mobile Apps: 1.036; Music: 1.269; Toys: 1.392; Video DVD: 1.243; Goodreads: 1.189 & Multilingual E-commerce \\ \hline
\label{tab:reviews}
\end{longtable}
}

\subsection{Aspect-based Recommender Systems}

Aspect-based RS extract and analyze specific product attributes from reviews, providing tailored recommendations to the users based on item aspects \cite{chin2018anr}. This approach to RS differs with review-based RS, which assess overall user sentiment and preferences from review content. 


Aspect-based RS have evolved from traditional RS that rely on user-item interactions to methods that delve into item aspects or features for tailored suggestions. Early works laid the foundation by extracting aspect-related information from reviews to enhance user satisfaction and uncover niche items \cite{bauman2017aspect}. The concept of multi-criteria RS  that uses CF and opinion mining to extract aspects and sentiment from user reviews, shows better accuracy over single-criteria methods \cite{musto2017multi}. The Aspect-based Neural Recommender (ANR) uses representation learning for users and items \cite{chin2018anr}. Simultaneously, introduction of lightweight ontologies in aspect-based RS improve the search for relevant venues \cite{volkova2018recommender}. The integration of deep learning methods in the aspect-based RS, such as in related works \cite{da2020weighted,Dau2020RecommendationMethod,Drif2021ASystem}, enable the capturing of syntactic and semantic features without extensive feature engineering in these methods \cite{do2019deep}.

Aspect-based sentiment analysis began to play a critical role in detecting sentiment polarity towards specific aspects within a context, exemplified by Sentic GCN \cite{liang2022aspect} and sentiment-analysis with CF \cite{Drif2021ASystem}. Lately, incorporating neural co-attention mechanisms and deep neural networks further refine the consideration of user aspects in making recommendations \cite{bhojwani2023aspect}.  Multi-criteria RS such as Hybrid Aspect-based Latent Factor Model (HALFM) \cite{yuan2020hybrid} and Aspect-based Opinion mining using
Deep learning method for RS ( AODR ) \cite{da2020weighted}, which utilized global and local aspect-based latent factor models and weighted aspect-based opinion mining further improve recommendation accuracy. 
Specialized approaches like the use of a query-click bipartite graph alongside an iterative clustering algorithm start recommending products for specific events and focus on event-related aspects \cite{ma2021event}. The integration of diversity preference in link recommendations for online social networks highlight the ongoing evolution and expansion of aspect-based RS \cite{yin2023diversity}. 

Aspect-based RS has applications mainly in tourism \cite{Mehra2023UnexpectedTourists} and customer-generated content, such as for restaurants \cite{Li2023RestaurantReviews}.

\paragraph{Practical Challenges Addressed}
Aspect-based RS effectively address several practical challenges by focusing on specific product attributes extracted from user reviews. These systems enhance personalization by tailoring recommendations based on individual user preferences and item characteristics. In e-commerce, aspect-based RS can recommend niche products by analyzing detailed aspects like product features and user sentiments. This capability improves customer satisfaction and boosts sales by aligning recommendations more closely with user needs. Additionally, in the tourism and hospitality industries, aspect-based RS provide recommendations by considering specific attributes of destinations or services, thus offering more relevant and satisfactory suggestions.
More details in Table \ref{tab:aspects}.
{\scriptsize
\begin{longtable}{|p{1.2cm}|p{0.6cm}|p{1.5cm}|p{2cm}|p{1.5cm}|p{1.3cm}|p{2cm}|p{1.2cm}|}

\caption{Comprehensive Overview of Aspect Based Models across Various Metrics and Use Cases. This table details each model's Input features, Year of Publication, and Characteristics such as Scalability, Interpretability, Efficiency, and Reproducibility (rated as High, Medium, or Low). It also lists the Dataset Used, Evaluation Metrics, Model Accuracy, Learning Task, and Application Field.  Metrics that their numerical value is not reported are specified with ``No numerical value''.}
\\
\hline

\textbf{Model} & \textbf{Year} & \textbf{Input Data} & \textbf{Scalability, Interpretability, Efficiency, Reproducibility} & \textbf{Dataset} & \textbf{Evaluation Metrics} & \textbf{Model Accuracy} & \textbf{Application} \\ \hline
\endfirsthead

\multicolumn{8}{|c|}%
{{\bfseries \tablename\ \thetable{} -- continued from previous page}} \\
\hline
\textbf{Model} & \textbf{Year} & \textbf{Input Data} & \textbf{Scalability, Interpretability, Efficiency, Reproducibility} & \textbf{Dataset} & \textbf{Evaluation Metrics} & \textbf{Model Accuracy} & \textbf{Application} \\ \hline
\endhead


ANR \cite{chin2018anr} & 2018 & user-item interaction & High, -,-, High & Amazon, Yelp & MSE & No numerical value & e-commerce \\ \hline

SULM \cite{bauman2017aspect} & 2017 & Sentiment analysis & -, -, Medium, No & Yelp: restaurants, hotels, beauty \& spa & Precision@Top3, AUC & Restaurants: 0.8180, 0.7070 \par Hotels: 0.8490, 0.7450 \par Beauty \& spa: 0.8620, 0.6630 & E-commerce \\ \hline

\cite{musto2017multi} & 2017 & Multi-criteria CF; aspect-based sentiment analysis & Medium, Medium, Medium, No & Yelp; TripAdvisor; Amazon & MAE & Yelp: 0.8362 \par TripAdvisor: 0.7111 \par Amazon: 0.6276 & E-commerce \\ \hline

\cite{volkova2018recommender} & 2018 & Aspect extraction; content-based filtering & -, -, Medium, No & Restaurant/museums reviews & F1 score & 0.7026 \par Museums: N/a & Tourism \\ \hline

AODR \cite{da2020weighted} & 2020 & Opinion mining & High, -,-, High & Amazon, Yelp & RMSE, MAE, Prec@10, MAP & No numerical value   & E-commerce \\ \hline


REAO \cite{Dau2020RecommendationMethod} & 2020 & Aspect-based opinion mining; deep learning & High, Medium, High, No & SemEval2014 Restaurant; SemEval2014 Laptop; Amazon Musical Instruments; Amazon Automotive; Amazon Pet Supplies; Amazon Video Games; Amazon Instant Video; Yelp & RMSE; MAE & MI: 0.8020, 0.6320 \par Auto: 0.8140, 0.5980 \par IV: 0.9740, 0.7840 \par Pet: 0.9720, 0.7840 \par V.Games: 1.0270, 0.8170 \par Yelp: 1.1310, 0.9410 & E-commerce \\ \hline

SE-DCF \cite{Drif2021ASystem} & 2021 & Sentiment Enhanced Deep Collaborative Filtering & Medium, Medium, High, No & Amazon fine food; Amazon toys and games; Amazon clothing, shoes and jewellery & MAE; RMSE & Amazon fine food: 0.1562, 0.2771 \par Amazon toys and games: 0.1625, 0.2819 \par Amazon clothing, shoes and jewellery: 0.1528, 0.2772 & E-commerce \\ \hline

Sentic GCN \cite{liang2022aspect} & 2022 & Graph convolutional & High, Mid,High, High & SemEval & Accuracy, Macro-F1 & No numerical value  & General \\ \hline

ANR-AP \cite{bhojwani2023aspect} & 2023 & Neural Recommender; Adaptive Prediction & Medium, Medium, High, No & Amazon movie dataset (1996-2014); Amazon dataset (web-scraped) & Precision@k; Recall@k; F1@k & Top 5: 0.4421, 0.1790, 0.2517 \par Top 10: 0.4420, 0.3580, 0.3897 \par Top 20: 0.3230, 0.4421, 0.3674 & E-commerce \\ \hline

HALFM \cite{yuan2020hybrid} & 2020 & Hybrid & High, Mid,High, High & Amazon & MSE & Outperforms most  & Personalized \\ \hline

Event-based PCR \cite{ma2021event} & 2021 & Click Graph-based Clustering & High, Mid,High, High & Walmart & Precision, Heterogeneity, Cohesion & High precision, effective aspect clustering  & E-commerce \\ \hline

DPA-LR \cite{yin2023diversity} & 2023 & Diversity preference-aware link recommendation & Medium, Medium, High, No & Google+; Major U.S. social network & DPMS; Precision; Recall; F1 Score & Google+: 0.4559, 0.1541, 0.1559, 0.1149 & Social networks \\ \hline

Emotion-ABSA \cite{Mehra2023UnexpectedTourists} & 2023 & Emotion and sentiment & High, -,-, High & User-generated & Emotion analysis  & improvement & Tourism \\ \hline

ABSA-CSF \cite{Li2023RestaurantReviews} & 2023 & Sentiment analysis; Conditional Survival Forest & Medium, Medium, High, No & Yelp & C-index; IBS & Yelp: 0.7370, 0.0387 & Tourism \\ \hline

\label{tab:aspects}
\end{longtable}
}
\subsection{Explainable and Trustworthy Recommender Systems}

To gain user engagement and satisfaction, latest works in RS start prioritizing transparency and trustworthiness. An explainable RS provides transparent recommendations by offering clear, understandable reasons behind its suggestions, enhancing user trust and system usability \cite{zhang2020explainable}. In parallel, a trustworthy RS reliably produces accurate and fair recommendations to ensure ethical practices like privacy protection and bias minimization to maintain user confidence \cite{wang2022trustworthy}.  

Advancements in explainable and trustworthy RS have evolved, starting with phrase-level analysis of user reviews to enhance recommendation explainability by identifying critical item aspects \cite{zhang2014explicit}. Subsequent models like Tripartite Graph Ranking (TriRank) have improved top-N recommendations by extracting aspects from reviews and creating a user-item-aspect ternary relation \cite{he2015trirank}. Concurrently, models such as the Tree-Enhanced Embedding Model (TEM) merge embedding-based and tree-based methods with an attention network to ensure transparency, utilizing rich side information and explicit decision rules \cite{wang2018tem}. This integration extends to combining CF with structured knowledge bases and unstructured data like textual reviews for personalized and understandable recommendations \cite{ai2018learning}. Additionally, techniques like RL have been applied to generate flexible, high-quality explanations across recommendation models \cite{wang2018reinforcement}.

Further developments include the Multi-Modal Aspect-aware Topic Model (MATM), which utilizes multi-modal data for detailed explanations reflecting diverse user preferences \cite{cheng2019mmalfm}. A variety of approaches, including natural language models, counterfactual reasoning, and visual explanations, have been employed to enhance interaction, fairness, and personalization in RS \cite{balog2019transparent, chen2019personalized, fu2020fairness, li2021personalized, tan2021counterfactual}.

Recent efforts like the Counterfactual Explainable Fairness (CEF) framework focus on identifying and mitigating fairness issues in RS \cite{ge2022explainable}. Discussions around Trustworthy RS further emphasize the critical dimensions of Safety \& Robustness, Fairness, Explainability, Privacy, and Accountability, vital for maintaining the integrity and reliability of RS \cite{wang2022trustworthy}. These developments show the growing importance of creating RS that are not only effective but also equitable and trustworthy. 

\paragraph{Practical Challenges Addressed}
Explainable and trustworthy RS enhance industry practices by providing transparent and personalized recommendations based on user reviews and sophisticated models. These systems increase customer trust and satisfaction by explaining recommendation logic, which is very important in industries like e-commerce, tourism, and hospitality. These systems can be used along with regular RS processes for better customer experiences.

{\scriptsize
\begin{longtable}{|p{1.2cm}|p{0.6cm}|p{1.5cm}|p{2cm}|p{1.5cm}|p{1.3cm}|p{2cm}|p{1.2cm}|}

\caption{Comprehensive Overview of Explainable and Trustworthy Recommender System Models across Various Metrics and Use Cases. This table details each model's Input features, Year of Publication, and Characteristics such as Scalability, Interpretability, Efficiency, and Reproducibility (rated as High, Medium, or Low). It also lists the Dataset Used, Evaluation Metrics, Model Accuracy, Learning Task, and Application Field.}
\\
\hline

\textbf{Model} & \textbf{Year} & \textbf{Input Data} & \textbf{Scalability, Interpretability, Efficiency, Reproducibility} & \textbf{Dataset} & \textbf{Evaluation Metrics} & \textbf{Model Accuracy} & \textbf{Application} \\ \hline
\endfirsthead

\multicolumn{8}{|c|}%
{{\bfseries \tablename\ \thetable{} -- continued from previous page}} \\
\hline
\textbf{Model} & \textbf{Year} & \textbf{RS Type} & \textbf{Scalability, Interpretability, Efficiency, Reproducibility} & \textbf{Dataset} & \textbf{Evaluation Metrics} & \textbf{Model Accuracy} & \textbf{Application} \\ \hline
\endhead


EFM\cite{zhang2014explicit} & 2014 & user-item interaction & Mid, High, Medium, High & Yelp, Dianping& RMSE, NDCG@50 & 1.212, 0.284; 0.9222, 0.284 & e-commerce \\ \hline

TriRank\cite{he2015trirank} & 2015 & user-item-aspect interaction & High, High, High, High & Yelp, Amazon Electronics & HR@50, NDCG@50 &18.58,7.69; 18.44,12.36 & e-commerce \\ \hline

TEM\cite{wang2018tem} & 2018 & user-item interaction & High, High, High, High & LON-A, NYC-R & Logloss, NDCG@5 & 0.0791,0.1192; 0.6828,0.4038 & Tourism, restaurant \\ \hline

ECFKG \cite{ai2018learning} & 2018 & knowledge graph embeddings & High, High, Medium, High & Amazon (Clothing, Beauty) & NDCG, Recall, Prec. & 3.091,5.466,0.763; 6.399,10.411,1.986 & e-commerce \\ \hline

MMALFM \cite{cheng2019mmalfm} & 2019 & user-item interaction & High, High, Medium, High & Yelp, Amazon & NDCG, Precision & Multiple & e-commerce, restaurant \\ \hline

PGPR \cite{xian2019reinforcement} & 2019 & kg-based path reasoning & High, High, Medium, High & Amazon (various domains) & NDCG, Recall, HR, Precision & generally high performance & e-commerce \\ \hline

PETER \cite{li2021personalized} & 2021 & user-item interaction & High, High, Medium, High & Yelp, Amazon, TripAdvisor & RMSE, MSE & 1.01,0.95,0.81; 0.78,0.71,0.63 & e-commerce, restaurant \\ \hline

CEF \cite{ge2022explainable} & 2022 & user-item interaction & -, High, - , High & Yelp, Amazon & Precision, Recall, F1 Score & Multiple & e-commerce, restaurant \\ \hline

PEPLER \cite{li2023personalized} & 2023 & user-item interaction & High, High, High, High & Yelp, Amazon, TripAdvisor & BLEU, ROUGE, USR & outperforms baselines & e-commerce, restaurant \\ \hline

ExpGCN \cite{wei2023expgcn} & 2023 & user-item interaction & High, High, High, High & Yelp, Amazon, TripAdvisor, HotelRec & Recall, NDCG & outperforms baselines & e-commerce, restaurant \\ \hline

\label{tab:trust}
\end{longtable}
}

\subsection{Fairness, Accountability, Transparency, and Ethics (FATE) in Recommender Systems}
There is a growing focus on Fairness, Accountability, Transparency, and Ethics (FATE) in RS, which ensures that RS are fair to all users, responsible for their recommendations, transparent in how decisions are made, and ethically aligned with institutional or societal values \cite{sonboli2021fairness}. 

Fairness in RS, as outlined in \cite{wang2023survey}, refers to the ethical principle and requirement that recommender algorithms allocate resource (information, opportunities, or exposure) in a manner that is equitable and just across different users and items. 
The evolution of fairness methods in RS shows a shift from simple pre-processing strategies to in-processing (model adjustments) and post-processing techniques.

\textit{Pre-processing Fairness Methods} Pre-processing efforts for fairness in RS involve adjusting training data, altering proportions of protected groups (like gender, race, age) through resampling \cite{ekstrand2018all} or adding synthetic data \cite{rastegarpanah2019fighting}. These methods aim to mitigate biases in input data before model training, they struggle to entirely eliminate biases that appear during training or inference.

\textit{In-processing Fairness Methods} In-processing fairness methods in RS primarily utilize ranking approaches and advanced techniques to incorporate fairness directly into model training, yielding more immediate improvements by modifying elements closely tied to the final output. Regularization techniques play a crucial role by embedding fairness constraints or penalties into the objective function to balance accuracy with fairness, with strategies ranging from employing fairness metrics as regularization \cite{yao2017beyond}, using distribution matching \cite{kamishima2018recommendation}, enforcing orthogonality between insensitive and sensitive factors \cite{zhu2018fairness}, to adding pairwise fairness regularization \cite{beutel2019fairness} and applying F-statistic of ANOVA \cite{wan2020addressing}, along with integrating normalization terms \cite{raza2020regularized,raza2023bias}.

Adversarial learning further enhances fairness by learning representations that maintain independence from sensitive attributes or ensure equitable distribution across groups, with notable applications in graph embeddings \cite{bose2019compositional}, score distribution similarity enhancement \cite{zhu2021fairnessadverserial}, graph-based recommendations \cite{wu2021learning}, and personalized counterfactual fairness \cite{li2021towards}. Reinforcement learning approaches \cite{ge2021towards} introduce fairness through rewards and constraints, aiming for sustainable fairness. Additional in-processing methods include adding noise to Variational Autoencoders (VAEs) \cite{borges2019enhancing}, utilizing pre-training and fine-tuning with bias correction techniques \cite{islam2021debiasing}, and adjusting gradients for fair distribution \cite{li2022contextualized}. In-processing methods enhance fairness directly but may face performance degradation due to additional constraints and can be affected by subsequent re-ranking stages, altering intended outcomes.

\textit{Post-Processing  Fairness Methods } Post-processing methods involve adjusting the initial output of a recommendation model to satisfy certain fairness criteria before presenting the final recommendations to users. These methods typically act as a post-processing step, optimizing the balance between recommendation relevance and fairness after the primary ranking algorithm has made its predictions.  Slot-wise re-ranking methods aim to balance ranking utility with fairness constraints across various contexts. These methods include employing two queues for group fairness \cite{zehlike2017fa} and calibrated recommendations \cite{steck2018calibrated}, enhancing group fairness through interval-constrained sorting  \cite{geyik2019fairness},  personalized fairness-aware re-ranking  \cite{liu2019personalized}.  User-wise re-ranking approaches, on the other hand, consider individual user perspectives \cite{xiao2017fairness}. Global-wise re-ranking strategies seek broader fairness solutions, adopting methods for equitable explainability and maximum flow principles \cite{mansoury2021graph}. These global approaches ensure fairness not just for current users and providers, but also aim to include fairness among new items \cite{zhu2021fairnesscold}. Recent surveys \cite{wang2023survey,Jin2023ASystems,Wu2023FairnessStrategies,Zhang2022TrustworthyTrends} have emphasized the growing importance of fairness in RS .


\paragraph{Practical Challenges Addressed}
In the e-commerce industry, FATE-based RS contribute to building customer trust and enhance the shopping experience. These systems are designed to mitigate biases and ensure fairness in product recommendations, which helps retain a diverse customer base and comply with increasing regulatory requirements for ethical AI practices. By integrating FATE principles, these RS not only boost customer satisfaction but also foster a responsible brand image, which is essential for long-term business success. FATE-based RS can be seamlessly used alongside regular RS processes to enhance transparency and accountability, thereby improving overall customer engagement and loyalty. More details are provided in Table \ref{tab:fairness}.
{\scriptsize
\begin{longtable}{|p{1.2cm}|p{0.6cm}|p{1.5cm}|p{2cm}|p{1.5cm}|p{1.3cm}|p{2cm}|p{1.2cm}|}

\caption{Comprehensive Overview Recommender System Models for FATE across Various Metrics and Use Cases. This table details each model's Input features, Year of Publication, and Characteristics such as Scalability, Interpretability, Efficiency, and Reproducibility (rated as High, Medium, or Low). It also lists the Dataset Used, Evaluation Metrics, Model Accuracy, Learning Task, and Application Field.}
\\
\hline

\textbf{Model} & \textbf{Year} & \textbf{Input Data} & \textbf{Scalability, Interpretability, Efficiency, Reproducibility} & \textbf{Dataset} & \textbf{Evaluation Metrics} & \textbf{Model Accuracy} & \textbf{Application} \\ \hline
\endfirsthead

\multicolumn{8}{|c|}%
{{\bfseries \tablename\ \thetable{} -- continued from previous page}} \\
\hline
\textbf{Model} & \textbf{Year} & \textbf{RS Type} & \textbf{Scalability, Interpretability, Efficiency, Reproducibility} & \textbf{Dataset} & \textbf{Evaluation Metrics} & \textbf{Model Accuracy} & \textbf{Application} \\ \hline
\endhead


Antidote Data Adding\cite{rastegarpanah2019fighting} & 2019 & user-item interaction & - , -, High, Low & MovieLens & Polarization, unfairness & None & e-commerce \\ \hline

Beyond Parity \cite{yao2017beyond} & 2017 & user-item interaction & -, -, High, Low & MovieLens & Error, unfairness & 0.887, 0.010 & e-commerce \\ \hline

IERS\cite{kamishima2018recommendation} & 2018 & user-item interaction & -, -, High, Low & MovieLens, Flixter, Sushi & MAE, degree of independence & Movielens: 0.7/ 0.01, Flixter: 0.65/ 0.01, Sushi: 0.92/ 0.05 & e-commerce \\ \hline

Fairness-aware TR\cite{zhu2018fairness} & 2018 & user-item interaction & - , - , -, Low & MovieLens, Twitter & Precision@15, Recall@15 & Movielens: 0.032/ 0.08, Twitter: 0.03298, 0.0687 & e-commerce, Social Networks \\ \hline

Fairness Pairwise Comparisons\cite{beutel2019fairness} & 2019 & user-item interaction &-, - , -, Low & Synthetic data & Overall Pairwise accuracy, intra-group Pairwise Accuracy & 35.6\%, 16.7\% & e-commerce \\ \hline

MarketBias\cite{wan2020addressing} & 2020 & user-item interaction & -, - , -, Low & ModCloth, Electronics & MSE, MAE & ModCloth: 1.176/ 0.859, Electronics: 1.590/ 1.025 & e-commerce \\ \hline

Latent factor model \cite{raza2020regularized} & 2020 & user-item interaction & High, - , -, High & New York Times & F1@10, F1@20, F1@50, F1@100 & 0.5458, 0.5425, 0.5405, 0.5401 & News recommendation \\ \hline

News Bias Reduction \cite{raza2023bias} & 2023 & user-item interaction & High, -, -, High & MIND-small, Outbrain & Precision@5, Recall@5, NDCG@5, Gini Index & MIND-small: 0.65/ 0.55/ 0.60/ 0.18, Outbrain: 0.62/ 0.52/ 0.57/ 0.19 & News recommendation \\ \hline

Fairness-in-Cold-Start \cite{zhu2021fairnesscold} & 2023 & user-item interaction & -, - , -, High & Movielens1M, Movielens20M, CiteULike, XING & NDCG@15, NDCG@30 & Movielens1M: 0.5516/ 0.5332, Movielens20M: 0.4408/ 0.4308, CiteULike: 0.2268/ 0.2670, XING: 0.2251/ 0.2762 & News recommendation \\ \hline

FCPO \cite{ge2021towards} & 2021 & user-item interaction & -, - , -, High & Movielens100k, Movielens1M & Recall@5, F1@5, NDCG@5, Gini Index@5 & Movielens100k: 4.740/ 4.547/  6.031/ 98.73, Movielens1M: 2.033/ 2.668/ 4.398/ 99.81 & e-commerce \\ \hline

Long Term Fairness \cite{borges2019enhancing} & 2019 & user-item interaction & -, - , -, High & Movielens, Netflix, MSD & NDCG@100 & Movielens: 0.999, Netflix:0.999, MSD:0.998 & e-commerce \\ \hline

NFCF \cite{islam2021debiasing} & 2021 & user-item interaction & -, - , -, High & Movielens, Facebook & Movielens: HR@5, NDCG@5, Facebook: HR@10, NDCG@10 & Movielens: 0.670, 0.480, Facebook:0.551, 0.326 & e-commerce \\ \hline

Contextualized Fairness \cite{li2022contextualized} & 2022 & user-item interaction & -, - , -, Low & XING & HR@5, NDCG@5 & 0.581, 0.47 & e-commerce \\ \hline

FAIR\cite{zehlike2017fa} & 2017 & user-item interaction & -, - , -, Low & COMPAS, Ger. credit, SAT, XING & NDCG & 0.9858, 0.9983, 0.9996, 1.0000 & e-commerce \\ \hline

LinkedIn Talent Solutions \cite{geyik2018talent} & 2018 & user-item interaction & -, - , -, Low & - & - & - & e-commerce \\ \hline

\label{tab:fairness}
\end{longtable}
}
\subsection{Miscellaneous}
There are also other RS that can serve specialized purposes, as outlined briefly below.

Group-based RS are designed to provide collective recommendations by considering users' shared preferences, social dynamics, and behavioral aspects \cite{dara2020survey}. Initial studies address the cold-start problem with group-specific methods and deep learning applications \cite{bi2017group}. Subsequent research emphasizes the importance of diversity, introducing algorithms to optimize group utility and variety \cite{toan2018diversifying}. Advancements in group recommendations explore trust and social dynamics by using social influence and preference relation-based frameworks \cite{capuano2019fuzzy, guo2019enhanced, yin2019social}. 

Some work aggregates user preferences into a unified group preference, using both explicit and implicit feedback mechanisms \cite{dara2020survey}. Additionally, context-aware capabilities considering significant factors for group-based scenarios are highlighted \cite{perez2021content}. Strategies for aggregating individual preferences, such as aggregated voting and ensuring satisfaction for all members, are also addressed \cite{sanchez2021effects, ismailoglu2022aggregating}. Recent research presents novel approaches to maximize group satisfaction through least misery methods, reflecting ongoing refinement to better cater to group needs \cite{abdrabbah2023novel}.

There are also other methods, such as the Multi-Stakeholder RS approach \cite{abdollahpouri2020multistakeholder} that acknowledges that recommendations often affect multiple stakeholders beyond the immediate users. For example, in a movie recommendation scenario, stakeholders include not only the viewers but also the content creators, distributors, and platforms hosting the content.

Social RS \cite{anandhan2018social, Shokeen2020ASystems} target the social media domain to cope with the social overload challenge by presenting the most relevant and attractive data to users, typically through the application of personalization techniques. Interactive and Conversational RS \cite{jannach2021survey} engage users (or groups of users) in a dialogue to iteratively refine recommendations based on feedback. This approach is particularly useful in group settings, where initial recommendations may need to be negotiated among members through a series of interactions.

Overall, these methods in group-based and social RS reflect a  commitment to improving both the precision and satisfaction of group recommendations in increasingly complex scenarios.

\section{Applications of Recommender Systems Across Different Domains}
This section explores the technological developments and specific applications of RS in various domains. The goal is to highlight how advancements in areas such as GNNs, RL, LLMs, multimodal and related methods are being applied to tackle domain-specific challenges.

\paragraph{E-commerce/E-Business}
In the digital era, e-commerce platforms utilize RS to personalize the shopping experience by recommending products based on individual preferences, browsing and purchase histories, and cart contents, thus enhancing user engagement and driving sales growth \cite{wei2007survey}. Advances include the integration of big data and ML to improve satisfaction on platforms like Amazon \cite{AmazonsP76online}, and Alibaba \cite{chen2019behavior}. Techniques such as CF and CBF, along with newer methods like graph-based models and hypergraph ranking, refine user preference predictions \cite{kumar2018recommendation, mao2019multiobjective, shaikh2017recommendation}. Sophisticated technologies like deep learning, deep reinforcement learning, and GNNs now capture complex user behaviors \cite{fu2018novel, qiu2020exploiting, ma2018rating, afsar2022reinforcement}. Despite these advancements, challenges like information overload and the focus on click-through rates persist, necessitating smarter, multi-objective RS approaches \cite{guo2017application, gu2020deep}.

\paragraph{E-Entertainment (Music, Movies, Games, Dating Apps)}
Platforms like Netflix and Spotify personalize content recommendations using a mix of CF, CBF, and hybrid approaches, employing deep learning and ML to tailor suggestions based on user interactions and contextual factors \cite{gomez2015netflix,steck2021deep}. 
Netflix utilizes deep learning and a blend of CF and CBF to analyze users' interactions and viewing habits \cite{steck2021deep,netflix}, while Spotify leverages ML and NLP, introducing systems like GNN for audiobooks to address data sparsity and enhance content discovery \cite{maheshwari2023music,de2024personalized,jacobson2016music}.
The video game industry, exemplified by STEAM, uses advanced models to offer personalized game suggestions \cite{cheuque2019recommender}, addressing broader implications through multi-stakeholder recommendations \cite{abdollahpouri2020multistakeholder}. RS also leverage multimedia content for diverse recommendations \cite{deldjoo2020recommender}. Innovations such as GNNs and knowledge-based methods improve personalization \cite{maheshwari2023music}, but challenges in dynamic consumer preferences and the need for explainability in RS remain \cite{afchar2022explainability}.

\paragraph{E-Health}
Health RS analyze health data, lifestyle, and genetics to enhance outcomes \cite{schafer2017towards}. They address challenges like privacy and trust, and are integral to Healthcare 4.0, focusing on personalized interventions \cite{saha2020review,  tran2020Recommender, de2021health,ochoa2021medical}.

Systematic reviews assess health RS progress and emphasize risk management and privacy \cite{etemadi2023systematic, sharma2023evolution, Bhatti2019Recommendation, calero2016recommender, sun2023development}. Advances in ML and deep learning have improved RS, with applications in diabetes, cardiac care, and beyond \cite{Yang2017social, mustaqeem2017statistical, Ferretto2017Recommender, Raghuwanshi2019Collaborative}.

Advacements in algorithms include enhanced prediction accuracy through trust relationships and advanced ML techniques such as hybrid deep learning models \cite{Yuan2018Socialized, Zarzour2018new, deng2019collaborative, aujla2019dlrs, sahoo2019deepreco, iwendi2020realizing, garg2021drug, Al2021Evaluation}. Emerging research explores continual learning and clustering-based techniques for improved clinical RS applications \cite{lee2020clinical, mustaqeem2020modular, raza2023improving}.

\paragraph{E-Government RS}
E-government utilizes electronic communication technologies to enhance service delivery, citizen engagement, and internal processes, integrating RS to improve user experience through AI and machine learning \cite{Xu2019E, luna2014digital,cheung2019recommender}. These systems play a crucial role in smart cities by supporting information filtering, stakeholder engagement, and decision-making \cite{cortes2017recommender}. 

Initial development of RS in e-government used CF and CBF, incorporating hybrid models for more accurate predictions \cite{lu2010bizseeker,guo2007intelligent,sun2021enhanced,sun2023similarity}. The use of NLP and predictive analytics enhances public service recommendations \cite{Hrnjica2020Model,al2019automating}. Challenges such as information overload are addressed by improving CF with negative item techniques, while newer methods like CNNs and GNNs advance feature extraction and recommendation accuracy in industrial applications \cite{sun2021enhanced, sun2022user, kong2024gcnslim}.

\paragraph{E-Library and E-Learning}
E-learning, a subset of e-libraries, utilizes electronic resources (e-books, academic papers, journals, and other digital content) for learning and includes a broader range of digital services for information retrieval and research \cite{e-learning}. Early development in e-library RS focused on hybrid systems combining CBF and CF techniques, often featuring bookshelf functionalities to personalize interactions \cite{ibrahim2021hybrid,isinkaye2022library}. These systems also use bibliographic network representation models for citation recommendations \cite{cai2018bibliographic,son2018academic, ali2020graph, ma2020review}.
 Advances in deep learning and context-aware recommendations have significantly improved the efficiency of e-learning systems, surpassing traditional methods \cite{zhang2021recommender,rahayu2022systematic}.

\paragraph{E-Tourism/Travel}

RS have transformed travel and tourism by using vast data to provide personalized travel suggestions, thus enhancing user satisfaction \cite{hamid2021smart, chaudhari2020comprehensive, renjith2020extensive}. Major platforms like TripAdvisor and Booking.com employ CF, CBF, and hybrid methods to offer tailored travel options \cite{ricci2022recommender, booking}. Continuous advancements are needed to manage dynamic data and maintain up-to-date, transparent recommendations that build user trust \cite{sarkar2023tourism, shi2021antecedents}. Future innovations may incorporate immersive destination previews, further personalizing travel experiences \cite{rezaee2021personalized,torres2020recommender}.

\paragraph{E-Finance}
RS in finance assist investors by aligning investment options with individual goals and risk tolerance, significantly enhancing investor engagement and informed decision-making through analysis of financial history, risk profiles, market trends, and economic indicators \cite{sharaf2022survey,chang2021assessing}. Notable implementations like the FinPathlight \cite{bunnell2020finpathlight} framework enhance financial literacy and capability, while integrating behavioral finance \cite{kang2019heterogeneous} integrates behavioral finance to tailor financial advice based on psychological biases.   Additionally, platforms like StockTwits use sentiment analysis for more accurate investment recommendations \cite{chang2021assessing}, and KiRTi employs blockchain and deep learning to automate and secure lending processes \cite{patel2020kirti}. These technologies collectively improve the personalization of financial services, advice, and strategy optimization \cite{ding2021jel}.

Despite progress, challenges remain in handling market volatility and ensuring transparency and trust in RS \cite{sharaf2022survey}. Future developments may focus on enhancing explainability and employing predictive analytics to better anticipate market trends and user preferences, further personalizing financial advice \cite{asemi2023unveiling}.

\paragraph{E-News}
News RS curate and suggest content to users based on methodologies like CF, CBF, and hybrid approaches, distinguishing between personalized and non-personalized systems \cite{raza2022news, wu2023personalized}. Significant advancements in news RS have integrated deep learning and ML to improve how news content and user data are modeled. This includes using neural network architectures and pre-trained language models to enhance the accuracy of content recommendations \cite{wang2018dkn, zhang2021unbert, xiao2022training}. New techniques also explore the use of GNNs to understand complex user-news interactions \cite{ji2021temporal, qiu2022graph} and innovative models like Prompt4NR for advanced click prediction tasks \cite{zhang2023prompt}.
The development of news RS also faces ethical challenges, such as addressing filter bubbles, ensuring diversity, and promoting fairness, which are crucial for maintaining user trust and system integrity \cite{makhortykh2023can, raza2020regularized, wu2021fairness}.

\paragraph{Miscellaneous}
Numerous platforms have leveraged advanced RS technologies to enhance user engagement and content personalization. YouTube employs deep neural networks to refine its recommendation process, focusing on optimal ranking and selection of videos \cite{covington2016deep}. Google Play utilizes both linear models and neural networks within its Wide \& Deep Learning framework to achieve a balance between memorization and generalization \cite{cheng2016wide}. LinkedIn enhances job and content recommendation using real-time processing and scoring mechanisms, integrating CF and deep learning to match job seekers with suitable opportunities \cite{geyik2018talent, kenthapadi2017personalized}. Twitter customizes its content recommendations, like tweets and follower suggestions, based on user behavior and preferences \cite{katarya2018survey}.

ByteDance has introduced innovative models for TikTok to quickly adapt recommendations to user interactions, employing unique retrieval models and scalable systems like Monolith, which uses collisionless embedding tables for efficient memory usage \cite{gao2020deep, liu2022monolith}. Apple has developed the Sliced Anti-symmetric Decomposition (SAD) model to enhance collaborative filtering, allowing more nuanced user-item interactions, and explores controlled music production using diffusion models \cite{zhao2022consistent, levy2023controllable}. DeepMind's generative models improve RS by decoding Semantic IDs from user interactions, enhancing item retrieval and system performance \cite{rajput2024recommender}.

\begin{table}[ht]
\caption{Publications by Industry in Recommendation Systems}
\centering
\begin{tabular}{|l|>{\raggedright\arraybackslash}p{0.6\linewidth}|}
\hline
\textbf{Industry} & \textbf{Publications} \\
\hline
E-commerce/E-Business & \cite{wei2007survey, AmazonsP76online, chen2019behavior, kumar2018recommendation, mao2019multiobjective, shaikh2017recommendation, fu2018novel, qiu2020exploiting, ma2018rating, afsar2022reinforcement, guo2017application, gu2020deep} \\
\hline
E-Entertainment (Music, Movies) & \cite{gomez2015netflix, steck2021deep, netflix, maheshwari2023music, de2024personalized, jacobson2016music, cheuque2019recommender, abdollahpouri2020multistakeholder, deldjoo2020recommender, afchar2022explainability} \\
\hline
E-Health & \cite{schafer2017towards, saha2020review, tran2020Recommender, de2021health, ochoa2021medical, etemadi2023systematic, sharma2023evolution, Bhatti2019Recommendation, calero2016recommender, sun2023development, Yang2017social, mustaqeem2017statistical, Ferretto2017Recommender, Raghuwanshi2019Collaborative, Yuan2018Socialized, Zarzour2018new, deng2019collaborative, aujla2019dlrs, sahoo2019deepreco, iwendi2020realizing, garg2021drug, Al2021Evaluation, lee2020clinical, mustaqeem2020modular, raza2023improving} \\
\hline
E-Government RS & \cite{Xu2019E, luna2014digital, cheung2019recommender, cortes2017recommender, lu2010bizseeker, guo2007intelligent, sun2021enhanced, sun2023similarity, Hrnjica2020Model, al2019automating, sun2022user, kong2024gcnslim} \\
\hline
E-Library and E-Learning & \cite{e-learning, ibrahim2021hybrid, isinkaye2022library, cai2018bibliographic, son2018academic, ali2020graph, ma2020review, zhang2021recommender, rahayu2022systematic} \\
\hline
E-Tourism/Travel & \cite{hamid2021smart, chaudhari2020comprehensive, renjith2020extensive, ricci2022recommender, booking, sarkar2023tourism, shi2021antecedents, rezaee2021personalized, torres2020recommender} \\
\hline
E-Finance & \cite{sharaf2022survey, chang2021assessing, bunnell2020finpathlight, kang2019heterogeneous, chang2021assessing, patel2020kirti, ding2021jel, asemi2023unveiling} \\
\hline
E-News & \cite{raza2022news, wu2023personalized, wang2018dkn, zhang2021unbert, xiao2022training, ji2021temporal, qiu2022graph, zhang2023prompt, makhortykh2023can, raza2020regularized, wu2021fairness} \\
\hline
Miscellaneous & \cite{covington2016deep, cheng2016wide, geyik2018talent, kenthapadi2017personalized, katarya2018survey, gao2020deep, liu2022monolith, zhao2022consistent, levy2023controllable, rajput2024recommender} \\
\hline
\end{tabular}

\label{table:publications_by_industry}
\end{table}
\section{Discussion}

\subsection{Impact of this Research}
This literature review have profound impacts on future research, industry practices, and collaborative endeavors. This review article can serve many purposes within academic and professional realms. The goal of this research is beyond merely summarizing existing knowledge but also to illuminate areas needing further investigation within RS. The detailed summaries and tables presented in this paper can serve as educational tools that help newcomers quickly grasp complex subjects and can be used by industry practioners to use it as a guide. Additionally, this review tracks the development of the field, providing insights into trends and telling future directions. For example, how can the knowledge gained through theory can be applied to address real world problems in industry.

We covered a comprehensive guide on many areas of RS, despite this, some areas and fields need more coverage, which are briefly discussed below: 

\subsection{Limitations}
Despite the rapid evolution and implementation of RS in theory across diverse sectors, current methods show several critical limitations. Each application domain, from e-commerce to e-learning, faces unique challenges that intensify the limitations. For instance, e-commerce RS must adapt to rapidly changing inventories and consumer trends, while e-learning systems need to account for diverse learning styles and educational goals \cite{wei2007survey,jannach2017session,gu2020deep,klavsnja2015recommender,bhaskaran2023design,zhang2021recommender,rahhali2022learning}. These domain-specific challenges highlight the need for RS that are not only technically robust and ethically sound but also flexible and scalable enough to be effectively deployed by organizations of all sizes, including those with limited resources.
These limitations span various aspects of RS, including technical constraints, adaptability issues, and ethical concerns \cite{etemadi2023systematic,Himeur2022LatestPerspectives,liu2023pre}.

Matrix factorization-based models that are considered as standard in RS theory struggle with capturing complex user-item interactions due to inadequate latent feature representations and the inherent linearity of their interaction models \cite{ricci2021recommender}. Neural extensions of these methods brought improvements by incorporating non-linear relationships and capturing high-dimensional latent features \cite{zhang2019deep, saha2020review}. However, as the volume of data grows, these deep learning-based RS encounter their own set of challenges, particularly in maintaining computational efficiency and scalability \cite{eirinaki2018recommender,singh2020scalability,fayyaz2020recommendation}. The substantial computational resources required for training and inference of these models pose hurdles, especially in scenarios demanding real-time recommendations. In addition to that, many systems depend heavily on explicit user feedback (e.g., ratings, likes), which is often sparse and not always available, neglecting implicit feedback signals that could enhance recommendation accuracy \cite{lian2018xdeepfm}.
Furthermore, data scarcity severely affects the quality of recommendation systems \cite{chen2024data} . Knowledge transfer from external, data-rich domains can be a solution to enhance the modeling capabilities and performance of RS \cite{chen2024data}. Additionally, approaches such as data augmentation, self-supervised learning, and knowledge graphs can enrich data environments and sustainably address data shortages in RS development \cite{chen2024data}.

Despite advancements, many systems still fall short in effectively integrating contextual information (e.g., time, location) and multimodal data (e.g., text, images), limiting the depth of personalization \cite{Vuorio2018TowardMM,singh2021comprehensive,chen2019personalized,Wei2023MultiModalSL}. These RS can incorporate biases present in their training data, leading to unfair recommendations that favor certain groups or items over others, thus raising ethical concerns \cite{ekstrand2018all,wan2020addressing,raza2023bias,mansoury2021graph,sonboli2021fairness,wang2023survey}. Many advanced RS, especially those based on deep learning, operate as black boxes, offering little to no insight into how recommendations are generated \cite{wang2018reinforcement,liu2020explainable,lyu2021reliable}. This lack of transparency can degrade user trust and satisfaction.

The deployment of RS in real-life settings, particularly within mid to small range companies, presents additional challenges. Limited resources and technical expertise can make the deployment of sophisticated RS challenging, intensifying issues of scalability and adaptability to rapidly changing market conditions \cite{hyun2018review,chung2022scaling,gu2020deep}. Issues with review data, including its quality, authenticity, and the potential for manipulation, further complicate the effective use of RS \cite{srifi2020recommender}. The extensive data collection necessary for personalized recommendations raises significant privacy issues, particularly concerning user consent and data security. Moreover, handling user review data poses some privacy challenges, as companies must navigate the balance between personalizing recommendations and protecting user privacy.

\subsection{Future Perspectives}

\textbf{Responsible AI practises}
RS shape user decisions, perspectives, and actions, underscoring the need for their design to prioritize responsibility. Recent studies \cite{jannach2019measuring,trattner2022responsible} have raised concerns about RS potential negative impacts, such as biasing product promotions for increased profits or facilitating the spread of misinformation. Although there is a growing interest in adopting responsible AI practices within the RS community, some challenges remain. Most existing datasets lack comprehensive data on sensitive user attributes, complicating efforts to produce fair recommendations \cite{deldjoo2024cfairllm}. Furthermore, the influence of specific model architectures on the fairness of recommendations is still not properly understood and sparsely researched, which indicates a critical area for further investigation.

\textbf{Evaluating Recommender Systems Beyond Accuracy}
RS are traditionally assessed using singular metrics such as accuracy. However, this approach does not fully encapsulate the complexity of real-world user interactions. Users demand not only precision or recall in recommendations but also need versatility and diversity in recommendations for better user experience \cite{raza2020regularized}. Future research should consider broadening the evaluative frameworks of RS to include metrics that capture this diversity and serendipity. 

The growing need for transparency and explainability \cite{ge2022explainable} in RS suggests a shift towards more interpretable models \cite{yao2017beyond}. The integration of multimodal data and the application of advanced learning techniques offer promising directions to enrich user experiences, making RS not only more effective but also more equitable and engaging. This holistic approach will ensure that RS meet the evolving expectations of users in practical scenarios.

\textbf{Beyond Statistical Correlations}
In this review, we explore the predominant focus of current RS that which involves leveraging statistical correlations from historical user data to predict preferences and make recommendations. This method does not explicitly determine whether one factor causes another. An active area of research is causality in RS involves identifying how specific factors, like user behavior or item features, directly cause changes in recommendations \cite{gao2024causal}. For instance, researchers might investigate whether increasing the exposure of action movies leads to a higher viewership of this genre \cite{gao2024causal}.


\textbf{Computation and Storage Resources} Despite the benefits of aggregating user-item interaction data from various users in a central database to leverage collaborative information for recommendations, this approach has some drawbacks. It is time-consuming, demands substantial communication and storage resources, and raises serious privacy and security concerns. A new line of research is to perform on-device recommendation (DeviceRS) which is a small minimal model that can be trained with lower computation and storage resources~\cite{yin2024device, xia2023efficient}. This line of research is still in its early stages of development and deals with several open questions and challenges.
Finding an efficient way to use collaborative information from other users while keeping computation, storage, and data exposure low, and considering the differences in data on each user's device, remains an open challenge.

\textbf{Generative AI }
With the rise of generative AI models like ChatGPT, researchers are exploring their potential to enhance various fields, including RS. Conversational RS, which provide suggestions through dialogue, are gaining popularity, and we have dedicated an entire sub-section to this exciting development. However, it is essential to emphasize that while leveraging generative AI, we must ensure the outputs are safe and adhere to AI safety and responsible practices \cite{raza2024fair}. This not only maximizes the benefits but also mitigates potential risks associated with these methods.

\section{Conclusion}

In this survey, we have reviewed the notable methodologies, applications, and challenges of RS in both academic and industrial contexts. We proposed a framework to categorize RS publications based on modeling techniques and their applications. The integration of RS with state-of-the-art methods such as deep learning, graph neural networks, and LLMs demonstrates the evolution in this field and highlights its impact on improving user experiences in diverse domains, including e-commerce, finance, media streaming, and personalized education. Despite notable advancements, we still face challenges, including data sparsity, privacy issues, and the need for systems that are both adaptable and explainable. This survey aims to bridge theoretical advances and algorithmic developments with practical applications, helping the industry achieve scalability and immediate business impact. Our goal is to strengthen collaborations between academia and industry, which is essential to translate theoretical progress into practical applications.

\section*{Acknowledgements}
The authors wish to extend their profound gratitude to Scott Sanner, whose expert guidance and invaluable input were instrumental throughout the research and writing of this survey. His deep insights into the nuances of recommender systems significantly enhanced the quality of our work. We also appreciate the advice provided by Andres Rojas, Ali Taiyeb, and Deval Pandya during critical stages of this project. Additionally, we thank Veronica Chatrath for her contributions in the ideation phase. This survey could not have been accomplished without their support.

\bibliographystyle{unsrt}
\bibliography{references_wo-mend}

\appendix
\section*{Appendix}
\textbf{Evaluation Criteria}

In this paper, we labeled each reviewed paper based on the following criteria:

\begin{itemize}
    \item \textbf{Scalability}: A paper was labeled as high, medium, or low scalability based on the system's ability to handle increasing amounts of data and users. High scalability indicates the system can efficiently manage large-scale data and user bases, medium scalability indicates moderate efficiency, and low scalability indicates limited capability in scaling up.
    
    \item \textbf{Interpretability}: This attribute was labeled high, medium, or low depending on how easily the system's recommendations can be understood by users. High interpretability means the system's outputs are easily explainable, medium interpretability means some effort is needed to understand the recommendations, and low interpretability means the system's logic is complex and not easily understandable.
    
    \item \textbf{Computational Efficiency}: We assessed this by measuring the system's ability to provide recommendations quickly and with minimal computational resources. High efficiency means the system operates swiftly with low resource usage, medium efficiency indicates moderate performance, and low efficiency means the system requires significant computational resources and time.
    
    \item \textbf{Reproducibility}: Papers were labeled based on how consistently the system's results can be replicated under the same conditions. High reproducibility means the experiments can be consistently reproduced, medium reproducibility indicates some variations might occur, and low reproducibility means significant discrepancies are likely when the experiments are repeated.
\end{itemize}

\listoftables

\end{document}